%% file: main.tex
\documentclass{aa}
\usepackage{txfonts}

\usepackage{amssymb,epsfig,color}
\usepackage{amsmath}
\usepackage{graphicx}
\graphicspath{{figures/}}

\usepackage{makecell}  % For line breaks in table headers

\usepackage{siunitx}
\usepackage{booktabs}
\usepackage{multirow}
\DeclareSIUnit\arcsecond{arcsec}
\DeclareSIUnit\parsec{pc}
\DeclareSIUnit\milliarcsecond{mas}
\DeclareSIUnit\year{yr}
\usepackage{longtable} % Multi-page tables
\usepackage{lscape} % Landscape mode

 \usepackage{flushend} % Package used to balance the columns on the last page

\usepackage[hyperindex,breaklinks=true, colorlinks,citecolor=blue,draft=False]{hyperref}

\def\reff@jnl#1{{\rm#1\/}}

\def\aj{\reff@jnl{AJ}}                  % Astronomical Journal
\def\araa{\reff@jnl{ARA\&A}}            % Annual Review of Astron and Astrophys
\def\apj{\reff@jnl{ApJ}}                % Astrophysical Journal
\def\apjl{\reff@jnl{ApJ}}               % Astrophysical Journal, Letters
\def\apjs{\reff@jnl{ApJS}}              % Astrophysical Journal, Supplement
\def\ao{\reff@jnl{Appl.Optics}}         % Applied Optics
\def\apss{\reff@jnl{Ap\&SS}}            % Astrophysics and Space Science
\def\aap{\reff@jnl{A\&A}}               % Astronomy and Astrophysics
\def\aapr{\reff@jnl{A\&A\simRev.}}         % Astronomy and Astrophysics Reviews
\def\aaps{\reff@jnl{A\&AS}}             % Astronomy and Astrophysics, Supplement
\def\azh{\reff@jnl{AZh}}                        % Astronomicheskii Zhurnal
\def\baas{\reff@jnl{BAAS}}              % Bulletin of the AAS
\def\jrasc{\reff@jnl{JRASC}}            % Journal of the RAS of Canada
\def\memras{\reff@jnl{MmRAS}}           % Memoirs of the RAS
\def\mnras{\reff@jnl{MNRAS}}            % Monthly Notices of the RAS
\def\nar{\reff@jnl{New Astronomy Reviews}}            % Monthly Notices of the RAS
\def\pra{\reff@jnl{Phys.Rev.A}}         % Physical Review A: General Physics
\def\prb{\reff@jnl{Phys.Rev.B}}         % Physical Review B: Solid State
\def\prc{\reff@jnl{Phys.Rev.C}}         % Physical Review C
\def\prd{\reff@jnl{Phys.Rev.D}}         % Physical Review D
\def\prl{\reff@jnl{Phys.Rev.Lett}}      % Physical Review Letters
\def\pasa{\reff@jnl{PASA}}              % Publications of the ASA
\def\pasp{\reff@jnl{PASP}}              % Publications of the ASP
\def\pasj{\reff@jnl{PASJ}}              % Publications of the ASJ
\def\qjras{\reff@jnl{QJRAS}}            % Quarterly Journal of the RAS
\def\skytel{\reff@jnl{S\&T}}            % Sky and Telescope
\def\solphys{\reff@jnl{Solar\simPhys.}}    % Solar Physics
\def\sovast{\reff@jnl{Soviet\simAst.}}     % Soviet Astronomy
 \def\ssr{\reff@jnl{Space\simSci.Rev.}}    % Space Science Reviews
\def\zap{\reff@jnl{ZAp}}                % Zeitschrift fur Astrophysik
\def\nat{\reff@jnl{Nature}}             % Nature

\begin{document}
	
	% =========================
	% Title and authors
	% =========================
	\title{Spectral Properties of Anomalous Microwave\\Emission in 144 Galactic Clouds}
	
	% \subtitle{} % optional
		
	\author{
		Roke Cepeda-Arroita\inst{1,2}\thanks{\email{roke.cepeda@iac.es}}
		\and
		J. A. Rubi\~no-Mart\'{i}n\inst{1,2}
		\and
		R. T. G\'{e}nova-Santos\inst{1,2}
		\and
		C. Dickinson\inst{3}
		\and
		S. E. Harper\inst{3}
		\and
		\newline
		F. Poidevin\inst{1,2}
		\and
		M. W. Peel\inst{4}
		\and
		R. Rebolo\inst{1,2,5}
		% Alphabetical authors below
		\and
		D. Adak\inst{1,2}
		\and
		A. Almeida\inst{1,2}
		\and
		K. Aryan\inst{1,2}
		\and
		R. B. Barreiro\inst{6}
		\and
		F. J. Casas\inst{6}
		\and
		\newline
		J. M. Casas\inst{1,2}
		\and
		J. Chluba\inst{3}
		\and
		M. Fern\'{a}ndez-Torreiro\inst{7}
		\and
		D.\,Herranz\inst{6}
		\and
		G. A. Hoerning\inst{3}
		\and
		Michael E. Jones\inst{8}
		\and
		J. Leech\inst{8}
		\and
		\newline
		E. Mart\'{i}nez-Gonz\'{a}lez\inst{6}
		\and
		T. J. Pearson\inst{9}
		\and
		Angela C. Taylor\inst{8}
		\and
		P. Vielva\inst{6}
		\and
		R. A. Watson\inst{3}
		\and
		Z. Zhang\inst{3}
	}
		
	\institute{
		Instituto de Astrof\'{i}sica de Canarias, 38200 La Laguna, Tenerife, Canary Islands, Spain
		\and
		Departamento de Astrof\'{i}sica, Universidad de La Laguna (ULL), 38206 La Laguna, Tenerife, Spain
		\and
		Jodrell Bank Centre for Astrophysics, Alan Turing Building, Department of Physics and Astronomy, The University of Manchester, Oxford Road, Manchester M13 9PL, UK
		\and
		Imperial College London, Blackett Lab, Prince Consort Road, London SW7 2AZ, UK
		\and
		Consejo Superior de Investigaciones Cient\'{i}ficas, Spain
		\and
		Instituto de F\'{i}sica de Cantabria (IFCA), CSIC-Univ. de Cantabria, Avenida de los Castros s/n, 39005 Santander, Spain
		\and
		Laboratoire de Physique Subatomique et de Cosmologie, Universit\'e Grenoble Alpes, 53 Avenue des Martyrs, Grenoble, France
		\and
		Department of Physics, University of Oxford, Denys Wilkinson Building, Keble Road, Oxford, OX1 3RH, UK
		\and
		Cahill Centre for Astronomy and Astrophysics, California Institute of Technology, Pasadena, CA 91125, USA
	}
	
	\date{Received XXX; accepted XXX}

	\abstract{
		Anomalous Microwave Emission (AME) is a diffuse microwave component thought to arise from spinning dust grains, yet remains poorly understood. We analyze AME in 144 Galactic clouds by combining low-frequency maps from S-PASS (2.3\,GHz), C-BASS (4.76\,GHz), and QUIJOTE (10--20\,GHz) with 21 ancillary maps. Using aperture photometry and parametric SED fitting via MCMC methods without informative priors, we measure AME emissivity, peak frequency, and spectral width. We achieve peak frequency constraints nearly three times tighter than previous work and identify 83 new AME sources. AME spectra are generally broader than predicted by spinning dust models for a single phase of the interstellar medium, suggesting either multiple spinning dust components along the line of sight or incomplete representation of the grain size distribution in current models. However, the narrowest observed widths match theoretical predictions, supporting the spinning dust hypothesis. The AME amplitude correlates most strongly with the thermal dust peak flux and radiance, showing $\sim$30\% scatter and sublinear scaling, which suggests reduced AME efficiency in regions with brighter  thermal dust emission. AME peak frequency increases with thermal dust temperature in a trend current theoretical models do not reproduce, indicating that spinning dust models must incorporate dust evolution and radiative transfer in a self-consistent framework where environmental parameters and grain properties are interdependent. PAH tracers correlate with AME emissivity, supporting a physical link to small dust grains. Finally, a log-Gaussian function provides a good empirical description of the AME spectrum across the sample, given current data quality and frequency coverage.
	}

	% =========================
	% Keywords
	% =========================
	
	% =========================
	% Maketitle
	% =========================
	\maketitle
	
	% =========================
	% Sections
	% =========================
	
\section{Introduction}
\label{sec:intro}

Anomalous Microwave Emission (AME) is a significant component of Galactic diffuse emission in intensity in the $10 < \nu < 60\,\si{\giga\Hz}$ range. It was discovered in CMB observations as excess emission strongly correlated with far-infrared radiation that could not be attributed to either synchrotron or free-free mechanisms \citep{Kogut1996, Leitch1997, Oliveira1998}. This correlation extends to scales as small as $2\,$arcmin, though most emission remains diffuse at larger angular scales \citep{Watson2005, Casassus2008, Scaife2009, Dickinson2010, Tibbs2013, Battistelli2019, Arce2019, Harper2025}. AME is widespread in our galaxy, accounting for up to half of the emission at $30\,\si{\giga\Hz}$ \citep{Planck2015_XXV}, and is now recognized as a major CMB foreground in intensity, with even low levels of polarization potentially biasing cosmological constraints and challenging B-mode experiments \citep{Remazeilles2016, Armitage-Caplan2012, Dunkley2009a}.

Observations of extragalactic sources have shown that AME is more concentrated in localized bubbles \citep{Scaife2010a, Murphy2010}, where it can account for up to two-thirds of the total emission \citep{Hensley2015_ngc, Murphy2018}. In the Andromeda galaxy, the overall contribution of AME remains an area of active research, and it is not yet clear whether its integrated emission is less dominant than that observed locally in the Milky Way \citep{M31quijotemfi, Harper2023_Andromeda, Battistelli2019, Planck2015_andromeda}.

The electric dipole spinning dust mechanism, first proposed by \cite{Erickson1957} and developed by \cite{Draine1998b}, is the leading explanation for AME. In this model, dust grains with a component of an electric dipole moment in their spinning plane emit radiation at their rotational frequency. The emission is dominated by the smallest, fastest-spinning grains, which is why sub-nanometer-sized grains such as polycyclic aromatic hydrocarbons (PAHs) \citep{Draine1998a}, nanosilicates \citep{Hensley2016_pahs, Hensley2017}, hydrogenated fullerenes \citep{Iglesias-Groth2005}, and nanodiamonds \citep{Greaves2018} have been proposed as primary carriers, with no conclusive evidence for the preference of a single carrier to date. Several refinements to the original \cite{Draine1998a} model have been implemented in the IDL-based code \textsc{SpDust2}\,\footnote{\url{https://cosmo.nyu.edu/yacine/spdust/spdust.html}} \citep{Ali-Hamoud2009, Ysard2010a, Hoang2010, Hoang2011, Silsbee2011, Ali-Haimoud2010}. The most recent development is the Python-based model \textsc{SpyDust}\,\footnote{\url{https://github.com/SpyDust/SpyDust}} \citep{spydust2024}, which generalizes the treatment of grain shapes.

A compelling argument for the spinning dust hypothesis is the lack of observed polarization. Many alternative mechanisms would produce measurable polarization, particularly at CMB frequencies ($\sim$30--200\,GHz), yet current upper limits on polarization fraction are already $\approx1$ percent at $\ang{1}$ scales \citep{Dickinson2011, Lopez-Caraballo2011, Rubino-Martin2012a, Genova-Santos2017, GonzalezGonzalez2025} and a few percent at arcminute scales \citep{Mason2009, Battistelli2015}, with measurements primarily limited by instrumental systematics such as polarization leakage and residual background polarized synchrotron. Other potentially highly polarized mechanisms, such as magnetic dust emission \citep{Draine2013, Hoang2016_mde}, are thought to be present at higher frequencies and may be embedded in the low-frequency end of thermal dust emission, making them difficult to detect with current experiments. However, they could still pose a significant challenge in future B-mode experiments at the current limits \citep{Dunkley2009a, Remazeilles2016}.

The characteristic bump of the spinning dust spectrum results from competing factors: while emitted power increases with the fourth power of rotational speed, the distribution of random excitations combined with damping establish a distribution of angular momenta that makes very high rotation rates rare, and the rapid growth of damping processes, especially electric dipole radiation, prevents grains from sustaining very high-frequency rotation. For a single grain, the spectral shape is well understood. However, for an astrophysical source, the observed spectrum is the sum of all individual grain spectra and therefore depends directly on the distributions of rotation rates, dipole moments, and the alignment between dipoles and rotation axes. These distributions are, in turn, influenced by grain sizes, shapes, compositions, and by environmental conditions—specifically the intensity of the radiation field, gas temperature, and the abundances and ionization states of hydrogen and carbon, all of which are poorly known in practice.

Despite the success of spinning dust models in reproducing observed AME spectra, extracting grain parameters from observed spectral energy distributions (SEDs) remains a major challenge. This difficulty stems primarily from parameter degeneracy—many combinations of the eight required input parameters produce nearly identical spectral shapes. The problem is exacerbated by limited observational data, as current measurements have limited spectral coverage and are typically only sensitive to the upper part of spinning dust emission due to calibration systematics and limited sensitivity. Current theoretical models compound this problem by not accounting for the dynamical interplay between environmental parameters and grain characteristics. This limitation potentially allows for physically incompatible parameter combinations (such as a low gas temperature alongside a strong radiation field) and ignores processes like grain growth or fragmentation. Addressing these challenges would require a full radiative transfer model coupled with dust evolution code such as \textsc{DustEM} \citep{Compiegne2011}, as demonstrated in \cite{Ysard2011}. Additionally, the dimensionality of theoretical models would need to be reduced to directly fit SEDs, for example by using moment expansion methods as explored in recent work \citep{spydust2024} and originally introduced in \cite{Chluba2017_momentexpansion}.

% Paper Goals
Given these challenges, this paper adopts a simplified, phenomenological approach to modeling AME, as outlined in Section~\ref{sec:AME}. Our primary goal is to assess how changes in the AME spectrum correlate with changes in the physical environment. To this end, we measure SEDs for compact Galactic sources at 1-degree scales using aperture photometry, building upon two previous studies. \cite{Planck2014_AME} detected 42 significant AME sources out of 98, but the analysis was limited by sparse low-frequency data: in the southern sky there were no measurements between 2.3 and 22.8\,GHz, and in the northern sky none between 1.42 and 22.8\,GHz. More recently, \cite{Poidevin2023} identified 44 significant sources out of 52 and incorporated QUIJOTE data in the 10--20\,GHz range, which greatly reduced biases in peak frequency and width. However, a substantial gap between 1.42\,GHz and 11.1\,GHz remained.

This paper presents a major improvement in low-frequency coverage by combining multiple datasets from QUIJOTE \citep{mfiwidesurvey}, C-BASS (Taylor et al. in prep.), and S-PASS \citep{spass_release}, in the context of the \textit{RadioForegrounds+} project.\footnote{\url{https://research.iac.es/proyecto/radioforegroundsplus/}}. We incorporate C-BASS data at 4.76\,GHz in the northern hemisphere and S-PASS data at 2.3\,GHz in the southern hemisphere, enabling accurate characterization of low-frequency foregrounds. This eliminates the need to fully rely on older large single-dish surveys with poorly understood beam characteristics and calibration. Additionally, QUIJOTE data covering 11--19\,GHz uniquely captures the low-frequency slope of the AME bump, allowing for more precise constraints on the peak frequency and width.

Together, these datasets enable us to fully constrain the AME amplitude, peak frequency, and width without applying informative priors, as shown in \cite{LambdaOrionis}. By explicitly fitting for the width, we also reduce biases in peak frequency determination. The improved calibration of the new data allows us to measure much fainter AME sources, expanding our analysis to 144 sources—significantly more than previous studies—including many at mid and high Galactic latitudes. In contrast, both earlier studies primarily focused on sources near the Galactic plane.

Section~\ref{sec:obs} provides an overview of the datasets used in this analysis, and Section~\ref{sec:methods} outlines the aperture photometry, foreground modeling, and SED fitting methods, including a representative example. Results are discussed in Section~\ref{sec:discussion}, with main conclusions summarized in Section~\ref{sec:conclusions}.

%%%%%%%%%%%%%%%%%%%%%%%%%%%%%%%%%%%%%%%%%%%%%%%%%%%%%%%
%%%%%%%%%%%%%%%%%%%%%%%%%%%%%%%%%%%%%%%%%%%%%%%%%%%%%%%

\section{Maps}
\label{sec:obs}

The datasets used in this paper are listed in Table~\ref{tab:surveys}, and are smoothed to a common full width at half maximum (FWHM) of $1^{\circ}$ for analysis. Maps not already offered at this resolution are smoothed with a Gaussian kernel chosen to match the target resolution. The three main low-frequency surveys critical for characterizing AME are outlined in the following sections, with ancillary surveys and tracers discussed in Section~\ref{subsec:ancillary}.

\input{datasets}

\subsection{S-PASS}
The S-band Polarization All Sky Survey (S-PASS) is a southern-sky intensity and polarization survey conducted with the Parkes radio telescope at 2.303 GHz \citep{spass_release}, covering $\delta\lesssim\ang{-1}$ with an angular resolution of $8.9$ arcmin. The intensity data used in this paper plays the role of setting the combined baseline level of free-free and synchrotron in the southern sky, together with the \cite{Jonas1998} \mbox{HartRAO} 2.326\,GHz survey, shown in Table~\ref{tab:surveys}. The relatively low calibration uncertainty of 5\% makes S-PASS a valuable counterpart to \mbox{C-BASS} in the southern hemisphere for AME studies.

\subsection{C-BASS}
The C-Band All Sky Survey (C-BASS) is a full-sky survey at $4.76\,\si{\giga\hertz}$ with an angular resolution of approximately 44 arcmin \citep{project_paper}. This paper uses intensity data from the northern sky survey (Taylor et al., in prep.), adopting a conservative calibration uncertainty of 5\%. At this frequency, generally just below the spinning dust emission, C-BASS provides the most reliable anchor point for estimating the combined synchrotron and free–free emission. This in turn helps to disentangle free-free from AME, which are otherwise highly degenerate at low frequencies, and thus constrains the AME amplitude. In polarization, C-BASS also serves as a key reference point for synchrotron emission due to its high sensitivity and low Faraday depolarization. The C-BASS map used here will be made publicly available in the near future.

\subsection{QUIJOTE}
The QUIJOTE (Q-U-I JOint TEnerife) experiment is a multi-frequency microwave survey conducted at the Teide Observatory, covering the range $10$--$40\,\si{\giga\hertz}$ with angular resolutions between $\ang{0.9}$ and $\ang{0.3}$ \citep{Rubino-Martin2012b}. In this paper, intensity data from the Multi-Frequency Instrument (MFI) at $11.1$, $12.9$, $16.8$, and $18.8\,\si{\giga\hertz}$ are used \citep{mfiwidesurvey}, which are publicly available\footnote{\url{https://lambda.gsfc.nasa.gov/product/quijote/}}. It is the only dataset that directly observes the low-frequency downturn of AME, making it critical for determining its peak frequency and width. In combination with WMAP and \textit{Planck}, it enables the precise characterization of the AME spectrum across its full frequency range if the baseline level of free-free and synchrotron is already well-determined by C-BASS or S-PASS. 

\subsection{Ancillary Data}
\label{subsec:ancillary}

% General summary
In addition to the datasets presented, 21 additional maps spanning from 408\,MHz to 3\,THz are used for SED fitting, along with 6 additional maps that serve as tracers for later AME analyses.

% Low-frequency dataset limitations
At low frequencies, the maps from \cite{Haslam1982} and \cite{Reich2001} suffer from low beam efficiencies and an incomplete characterization of the main beam and sidelobes. This discrepancy creates a calibration mismatch between beam-scale and large-scale measurements, limiting the accuracy of these surveys. To mitigate this issue, we use the reprocessed Haslam map from \cite{Remazeilles2015} and apply a correction factor of 1.55 to the original Reich map, following \cite{Reich1988}, to adjust the full-beam to main-beam ratio. This ensures calibration compatibility with the predominantly compact sources studied in this paper. Since these scale-correction factors vary spatially and depend on scale \citep{Irfan2014, Wilensky2024}, an effective calibration uncertainty of 30\% is assigned to these surveys. Consequently, the most reliable low-frequency measurements come from S-PASS, HartRAO, and C-BASS, while the two lowest-frequency maps are primarily used to constrain the synchrotron component.

The HartRAO survey \citep{Jonas1998} characterized the beam out to $8^\circ$ scales and determined that point sources require a multiplicative correction factor of 1.45, which we adopt. Given the well-characterized beam properties and the consistency of observations from a single telescope, we conservatively assume an effective calibration uncertainty of 10\%.

We also incorporate data from the WMAP 9-year data release (DR5)\footnote{\url{https://lambda.gsfc.nasa.gov/product/wmap/dr5/}} \citep{Bennett2013}, \textit{Planck} PR3\footnote{\url{https://pla.esac.esa.int/pla/\#maps}} \citep{Planck2018_I}, COBE-DIRBE \citep{Hauser1998}, and the reprocessed IRAS maps by \cite{Miville2005IRIS}, which extend up to $3\,\si{\tera\hertz}$.

% Tracers
Additional datasets include the CO emission map from \cite{Ghosh2024}, which we use to subtract CO contamination from the 100, 217, and 353\,GHz \textit{Planck} HFI maps. Compared to the widely used predecessor map from \cite{Dame2001}, the new separation exhibits lower astrophysical contamination and noise. Several other datasets aid in assessing correlations with AME amplitude. These include the reprocessed IRAS 60, 25, and 12\,\textmu m bands \citep{Miville2005IRIS}, a dark gas map \citep{dark_gas}, and the WISE 12\,\textmu m dust map \citep{Meisner2014}. The latter, which removes artifacts and continuum emission, provides a full-sky map dominated by PAH emission. It also serves as the basis for a tracer of the PAH fraction, as described in \cite{Hensley2016_pahs}.

%%%%%%%%%%%%%%%%%%%%%%%%%%%%%%%%%%%%%%%%%%%%%%%%%%%%%%%
%%%%%%%%%%%%%%%%%%%%%%%%%%%%%%%%%%%%%%%%%%%%%%%%%%%%%%%

\section{Methods}
\label{sec:methods}

\subsection{Aperture Photometry}
\label{sec:photometry}

We use aperture photometry to measure the flux density of each source by integrating the brightness within a primary aperture and subtracting a median background estimated from a surrounding background annulus. This technique is effective for sources where the emission is localized and significantly brighter than the surrounding background. A key advantage of this approach is its insensitivity to the absolute zero levels of the maps. Furthermore, it makes no assumptions regarding the source's shape and brightness profile.

For most sources, the aperture configuration consists of a circular primary aperture with a radius of 60 arcmin, while the background is defined as a concentric circular annulus with inner and outer radii of 80 and 100 arcmin, respectively, as adopted in previous studies \citep{Planck2014_AME, Poidevin2023}, which selected these values to minimize scatter in recovered flux densities and avoid source flux over-subtraction. For sources with larger primary radii, the background annuli are scaled proportionally.

Each source is visually inspected by overlaying the aperture and background annulus on the corresponding maps listed in Table~\ref{tab:surveys}. In some cases, secondary sources contaminate the background region. If these do not contribute significantly to the primary aperture, an angular restriction is applied to the background annulus at all frequencies to exclude the contaminating source, ensuring an accurate background estimate. During aperture placement, we find that the central positions of several sources from previous studies are slightly offset at all frequencies. These offsets arise because the original positions were determined automatically via Gaussian fits and moment analysis, which may have introduced slight biases for Galactic plane sources with brighter backgrounds. To improve accuracy, we apply small manual adjustments of 0.1--0.5 degrees to the central coordinates to 17 of the sources that were already known from previous studies.

The uncertainty on the flux density is computed by combining, in quadrature, the calibration uncertainty and the random noise in the background, scaled by the square root of the number of beam areas in the primary aperture. The background noise is quantified using the median absolute deviation (MAD) in the background annulus, multiplied by 1.4826 to convert it to an equivalent standard deviation \citep{median_absolute_deviation}. This approach provides a robust estimate of the noise while mitigating the impact of individual contaminating sources. However, if significant large-scale structure is present in the background, the MAD-based estimate may still overestimate the noise, though to a lesser extent than a standard deviation.

\subsection{Spectral Modeling}
\label{sec:foreground_modelling}

Sky emission is modeled as the sum of five components: synchrotron (where applicable), free-free, AME, CMB, and thermal dust emission: 
 
\begin{equation}
	\begin{aligned}
		S_{\rm{total}}(\nu) &= S_{\rm{sync}}(A_{\rm{sync}}, \alpha) + S_{\rm{ff}}(\rm{EM}) \\
		&\quad + S_{\rm AME}(A_{\rm AME}, \nu_{\rm AME}, W_{\rm AME})
		\\
		&\quad + S_{\rm CMB}(\delta T_{\rm CMB})  + S_{\rm d}(\tau_{\rm 353}, T_{\rm d}, \beta_{\rm d}) \,.
	\end{aligned}
\end{equation}

This results in a total of 8 or 10 free parameters, which are detailed in the following subsections.

\subsubsection{Synchrotron Emission}
\label{sec:synchrotron}

Synchrotron emission is modeled as a power law in the optically thin regime above $\nu\gtrsim100$\,MHz:
\begin{equation}
	S_{\rm{sync}}(\nu) = A_{\rm{sync}} \cdot \left( \frac{\nu}{\nu_{0}} \right)^{\alpha} \,,
\end{equation}
\noindent where $\nu_{0}$ is the pivot frequency, set to 1\,GHz, $A_{\rm{sync}}(\nu_0)$ is the amplitude, and $\alpha$ is the flux density spectral index, typically ranging from $-0.7$ to $-1.1$ and related to the Rayleigh-Jeans brightness temperature spectral index $\beta$ via $\alpha=\beta+2$. Due to the limited precision of the lowest-frequency datasets, spectral curvature is not modeled. 

This emission is generally more diffuse than other components and is largely removed by background subtraction in aperture photometry. Thus, it is only included when the lowest-frequency residuals and $\chi^2_{\rm{red}}$ indicate statistical significance, which occurs for 37 sources ($\approx 25\%$). 

A hard prior enforces $\alpha < 0$ to exclude unphysical rising spectra, but this constraint does not truncate the posterior distribution, making it effectively uninformative. Omitting the synchrotron component has a negligible impact on AME parameters even in sources where synchrotron is significant, with effects well below the $1\sigma$ level. This suggests that the synchrotron contribution is largely absorbed by the free–free component, and confirms that the inclusion of synchrotron is not necessary to reproduce our results.

\subsubsection{Free-free Emission}
\label{sec:free-free}

Free-free emission flux density is modeled as
\begin{equation}
	S_{\rm ff}(\nu) = \frac{2 k_{\rm B} \nu^{2}}{c^{2}} \cdot \Omega_{\rm A} \cdot T_{\rm ff}(\nu)\,,
\end{equation}
where $\nu$ is the observing frequency, $k_{\rm B}$ is the Boltzmann constant, $c$ is the speed of light, $\Omega_{\rm A}$ is the primary aperture solid angle and $T_{\rm ff}$ is the free-free emission brightness temperature. We use the free-free brightness temperature model in \cite{Draine_book}:
\begin{equation}
	\begin{split}
		T_{\rm ff}(\nu) = T_{\rm e} \cdot \left \{ 1 - \exp \left [ -\tau_{\rm{ff}}(\nu) \right ] \right \}\,,
	\end{split}
\end{equation}
where $T_{\rm e}$ is the electron temperature and $\tau_{\rm{ff}}(\nu)$ is the free-free optical depth, given by
\begin{equation}
	\begin{split}
		\tau_{\rm ff}(\nu) = 5.468 \cdot 10^{-2} \cdot \rm EM \cdot \textit{T}_{\rm e}^{-\frac{3}{2}} \cdot \left ( \frac{\nu}{GHz} \right )^{-2} \cdot \textit{g}_{\rm ff}(\nu)\,,
	\end{split}
\end{equation}
where $\mathrm{EM} \approx \int n_{\rm e}^2\,dl$ is the emission measure along the line of sight, dependent on the electron density $n_e$, and $g_{\rm{ff}}(\nu)$ is the dimensionless free-free Gaunt factor
\begin{equation}
	\begin{split}
		\begin{aligned}
			\noindent \exp \Bigl [ g_{\rm{ff}}(\nu) \Bigl ] =  \\ \exp \Biggl \{ 5.960-\frac{\sqrt{3}}{\pi} \cdot \ln \left [ \frac{\nu}{\si{\giga\hertz}} \left ( \frac{T_{\rm e}}{10^4\,\si{\kelvin}} \right )^{-\frac{3}{2}} \right ] \Biggl \} + \rm{e} \,,
		\end{aligned}
	\end{split}
\end{equation}
where $\mathrm{e}\approx2.718$ is Euler's number. The spectrum follows a flux density spectral index of $\alpha \approx -0.10$, steepening above $\sim 100\,\si{\giga\Hz}$ to $\alpha \approx -0.14$ \citep{Planck2015_XXIII}. Due to the limited dependence of the model on the electron temperature, we use a fixed $T_{\rm e} = 7500\,\si{\K}$, as measured by \cite{Alves2012, Maddalena1987, Quireza2006}, leaving $\mathrm{EM}$ as the only free parameter.

\subsubsection{Anomalous Microwave Emission}
\label{sec:AME}

Given the challenges of directly fitting theoretical models (see Section~\ref{sec:intro}), a first-order approximation of the spinning dust spectrum is employed: the log-Gaussian distribution. This form is motivated by \cite{Stevenson2014_lognormal}, who proposed a more complex model consisting of a power law multiplied by a log-Gaussian core with a complementary error function cutoff at high frequencies, where the power law introduces an additional index that creates spectral tilt. In this paper, we adopt a simplified symmetric form that directly parametrizes the amplitude, peak frequency, and width:

\begin{equation}
	\label{eq:lognormal}
	S_{\rm AME}(\nu) = A_{\rm AME} \cdot \exp{\left\{ -\frac{1}{2} \cdot \left[ \frac{ \ln{(\nu/\nu_{\rm AME})}} {W_{\rm AME}} \right]^{2} \right\}}\,,
\end{equation}

\noindent where $A_\mathrm{AME}$ is the peak amplitude, $\nu_\mathrm{AME}$ is the peak frequency, and $W_\mathrm{AME}$ determines the width of the spectrum. Note that this function is log-Gaussian in the statistical sense, meaning that it is Gaussian in $\ln\nu$, and should not be confused with the log-normal probability distribution, which includes a characteristic $1/\nu$ normalization factor.

The connection between the log-Gaussian and theoretical models is examined by fitting a log-Gaussian model to typical \textsc{SpDust2} and \textsc{SpyDust} templates for various environments, including cold neutral media (CNM), dark clouds (DC), molecular clouds (MC), photodissociation regions (PDR), reflection nebulae (RN), warm ionized media (WIM), and warm neutral media (WNM), using the idealized set of parameters presented in \cite{Draine1998b}. No frequency shifts are applied to the fitted templates themselves, as the peak frequency naturally emerges from the superposition of grain populations, making such a transformation unphysical.

A key difference between log-Gaussian and theoretical spinning dust templates lies in asymmetry. Theoretical models predict a steady power-law rise at low frequencies, whereas the log-Gaussian exhibits a sub-exponential increase. While the sub-exponential decline of the log-Gaussian is a good approximation of the high-frequency cutoff, mismodeling results in excess emission at frequencies below the peak. Additionally, this asymmetry introduces a slight negative bias in the recovered peak frequency, causing log-Gaussian fits to slightly underestimate the peak relative to theoretical templates.

Since theoretical templates lack an explicit width parameter, we determine an effective log-Gaussian equivalent width through fitting. To reflect experimental sensitivity limits and calibration uncertainties, we fit only the portion of the template where the amplitude exceeds 10\% of its peak value, with uniform weighting. This choice effectively restricts us to the very top of the spinning dust spectrum, given typical calibration errors ($\sim5$\% at low frequencies) and the additional suppression from free-free and, to a lesser extent, synchrotron subtraction.

The weighting required to replicate real mismodeling depends on factors such as source brightness, the relative contributions of calibration and noise uncertainties, dataset correlations, and biases in component separation. Notably, the resulting bias in peak frequency from fitting with a log-Gaussian is sensitive to weighting: greater sensitivity to the fainter ends of the AME spectrum enhances the impact of asymmetry, increasing the bias. In contrast, the width remains largely unaffected by changes in weighting. Given these complexities, a simplified approach is adopted to estimate log-Gaussian equivalent widths and approximate peak frequency biases to first order.

Results for both the \textsc{SpDust2} and \textsc{SpyDust} models are presented in Table~\ref{tab:spdust_vs_lognormal}. The recovered widths remain nearly identical across different environments, with values of $\approx0.41$ for \textsc{SpDust2} and $\approx0.39$ for \textsc{SpyDust}. This indicates that theoretical spectra for a single component yield a consistent width, regardless of variations in peak frequency or emissivity. The slightly narrower width in the \textsc{SpyDust} models arises from using an ensemble of grain geometries, ranging from disk-like to spherical, rather than assuming a fixed disk-like shape as in \textsc{SpDust2}. This variation alters the relationship between rotational and observed frequencies, leading to increased damping at high frequencies and thus a narrower width.

\begin{table}
	\centering
	\begin{tabular}{l cc|cc} % Changed first column from "l" to "c"
		& \multicolumn{2}{c|}{\textsc{SpDust2}} & \multicolumn{2}{c}{\textsc{SpyDust}$^{\dagger}$} \\
		& $W_{\rm AME}$ & \makecell{$\nu_{\rm AME}$ \\ Bias (\%)} & $W_{\rm AME}$ & \makecell{$\nu_{\rm AME}$ \\ Bias (\%)} \\
		\hline
		\rule{0pt}{2.6ex}CNM & 0.41 & $-$6.3 & 0.39 & $-$5.8 \\
		DC  & 0.42 & $-$7.8 & 0.40 & $-$6.5 \\
		MC  & 0.42 & $-$5.6 & 0.39 & $-$3.9 \\
		PDR & 0.41 & $-$7.3 & 0.39 & $-$6.5 \\
		RN  & 0.41 & $-$7.0 & 0.39 & $-$6.5 \\
		WIM & 0.41 & $-$6.4 & 0.39 & $-$5.5 \\
		WNM & 0.41 & $-$5.9 & 0.39 & $-$4.9 \\
		\hline
		\rule{0pt}{2.6ex}Average & 0.41 & $-$6.6 & 0.39 & $-$5.7 \\
	\end{tabular}
	\vspace{0.5em} % small vertical space
	\caption{log-Gaussian equivalent widths, $W_{\rm AME}$, and biases in the fitted peak frequency for \textsc{SpDust2} and \textsc{SpyDust}$^{\dagger}$. The bias in the peak frequency is expressed as a percentage, where a negative value indicates that the peak frequency fitted by the log-Gaussian is lower than the original peak frequency of the theoretical template. $^{\dagger}$\,The SpyDust model used corresponds to an ensemble of grain geometries.}
	\label{tab:spdust_vs_lognormal}
\end{table}

The bias remains consistent across environments, at $\approx6\%$. For a typical source peaking at 22\,GHz, this corresponds to an underestimation of the peak frequency by $\approx1.3$\,GHz. It is important to note that while most theoretical templates exhibit spectra peaking in the 20–35\,GHz range, the idealized RN and PDR cases peak at approximately 80 GHz and 160 GHz, respectively—far higher than any observed peak frequency.

Since the log-Gaussian model is phenomenological, in the limit $W_{\rm{AME}} \to \infty$, $S_{\rm{AME}}(\nu) \to A_{\rm{AME}}$, meaning that a sufficiently broad model flattens out and can become degenerate with free-free emission. Excessive width may also lead to overfitting at low frequencies, while widths significantly narrower than single-phase theoretical widths can fit between individual data points, leading to overfitting of individual data points. To prevent unphysical solutions, limits are set at $0.05<W_{\rm{AME}}<1.2$, ensuring for each source that these constraints do not alter the distribution and are therefore effectively uninformative.

\subsubsection{Cosmic Microwave Background Anisotropies}
\label{sec:CMB}

The flux density of CMB temperature anisotropies is modeled as:
\begin{equation}
	\label{eq:CMB}
	S_{\rm CMB}(\nu) = \frac{2 k_{\rm B} \nu^2}{c^2} \cdot \Omega_{\rm A} \cdot \delta T_{\rm CMB} \cdot \eta_{\Delta T}(\nu)\,,
\end{equation}
\noindent where \( \delta T_{\rm CMB} \) is the anisotropy amplitude in thermodynamic temperature units. The factor $\eta_{\Delta T}(\nu)$ accounts for the frequency-dependent conversion from thermodynamic temperature to intensity units:  
\begin{equation}
	\label{eq:planckcorr}
	\eta_{\Delta T}(\nu) = \frac{x^2 e^x}{(e^x - 1)^2} \,,
\end{equation}
\noindent where $x = h\nu/ k_{\rm B} T_\mathrm{CMB}$ and $T_\mathrm{CMB} = 2.725$\,K. This can contribute significantly near $\sim100$\,GHz, and is best fitted explicitly rather than subtracted, since CMB maps often contain substantial residual foreground contamination, especially at low latitudes. Unlike the other components, the CMB anisotropy is constrained by the accurately known variance of temperature fluctuations as a function of angular scale. We thus impose a Gaussian prior, $\delta T_{\rm CMB} \sim \mathcal{N}(0, 32.3\,\text{\textmu K})$, corresponding to $1^\circ$ scales, derived from the standard deviation of aperture-photometry measurements on the SMICA map \citep{smica}.

\subsubsection{Thermal Dust Emission}
\label{sec:thermal_dust}

Thermal dust emission is modeled using a modified-blackbody spectrum
\begin{equation}
	\label{eq:thermaldust}
	S_{\rm d}(\nu) = \frac{2 h \nu^3}{c^2} \cdot \frac{    \left( \nu / 353\,\si{\giga\Hz} \right)^{\beta_{\rm d}}    }{    e^{h\nu / k_{\rm B} T_{\rm d}} - 1    } \cdot \tau_{\rm 353} \cdot \Omega_{\rm A}\,,
\end{equation}
where $\tau_{\rm 353}$ is the optical depth at $353\,\si{\giga\hertz}$, $\beta_{\rm d}$ is the emissivity index, and $T_{\rm d}$ is the dust equilibrium temperature \citep{Hildebrand1983, Planck2016_XXIX}.

At frequencies as low as $30\,\si{\giga\hertz}$, the thermal dust contribution is generally very small, assuming the power-law approximation of the emissivity extends down to these frequencies. The modified-blackbody model provides a good empirical description of the dust spectral energy distribution; however, fitted $T_{\rm d}$ tend to be biased towards colder values. This is both because the modified-blackbody curve itself is biased in shape, and because observations on the Rayleigh–Jeans tail are dominated by emission from the colder dust components along the line of sight \citep{PlanckGCC2016}.

\subsubsection{Carbon Monoxide Emission}
\label{sec:CO}

The primary CO rotational transition, \( J=1\rightarrow0 \) at $\nu_0\approx115.3$\,GHz, contributes approximately 50\% of the intensity in the \textit{Planck} 100\,GHz HFI channel, while \( J=2\rightarrow1 \) accounts for about 15\% of the 217 GHz channel. The \( J=3\rightarrow2 \) transition has a minor effect on the 353 GHz channel, at the 1\% level, while higher-order transitions affecting the 545 and 857 GHz bands are negligible \citep{Planck2013_CO}.

Corrections are applied by subtracting the \cite{Ghosh2024} CO emission maps from the three main affected bands, with no adjustments made for higher frequencies. However, since CO emission is significantly stronger in the 100 and 217 GHz bands, residuals from systematic uncertainties in CO map subtraction may be larger in these bands. As a result, they are excluded from SED fitting. A subsequent evaluation shows that, on average, the 100 and 217\,GHz bands exhibit a small excess of $0.7\pm0.4\%$ and $1.5\pm0.4\%$ relative to the best-fit model. However, this excess varies very significantly across individual sources due to uncertainties in CO characterization, justifying the exclusion of these two bands.

\subsection{Markov Chain Monte Carlo SED Fitting}
\label{sec:mcmc_fitting_method}

The spectral energy distributions are fitted using the \texttt{EMCEE} ensemble sampler backend by \cite{Foreman-Mackey2013} with a Gaussian likelihood function, using the author's publicly available implementation\footnote{\url{https://github.com/RokeCepedaArroita/mcmc}}. Markov Chain Monte Carlo (MCMC) fitting enables a comprehensive exploration of non-linear parameter spaces and inherently accounts for correlations between parameters by jointly fitting them, resulting in marginalised distributions that reflect both the direction and magnitude of any degeneracies. The converged chains can then be used to generate samples and calculate derived parameters and their uncertainties (e.g. dust radiance), as the walkers naturally incorporate all underlying parameter correlations. Unlike least-squares fitting, which can be susceptible to local minima or biased solutions, MCMC robustly samples skewed or multimodal distributions, allowing visualization of multiple modes in parameter space, accurate quantification of uncertainties, and direct verification of chain stability and convergence.

Some of the previous studies on which this work expands employed least-squares fitting, namely \cite{Planck2014_AME, Poidevin2023}. Original fits from these studies were obtained via private correspondence and compared with the sources shared with this work. While best-fit parameters were in agreement for most of the sources, a subset fell into local minima and many lacked sufficient low-frequency data to reliably determine the AME peak frequency and width (see Section~\ref{sec:comparison}).
 
To ensure results are driven entirely by the data, no informative priors are applied to any parameters except for $\delta T_{\rm{CMB}}$. Instead, weak, physically motivated limits are imposed to keep parameter exploration within reasonable bounds: $10<T_{\rm d}<100\,\si{\kelvin}$, $\alpha<0$, $5<\nu_{\rm AME}<100\,\si{\giga\hertz}$, and $0.05<W_{\rm AME}<1.2$. Posterior distributions are examined visually to confirm that these constraints do not truncate the remaining parameter distributions, making them effectively uninformative. No priors are set for component amplitude parameters $A_{\rm AME}$, $A_{\rm sync}$, $\mathrm{EM}$, and $\tau_{\rm 353}$ to ensure that detections are purely data-driven and not biased by positivity priors. This choice reflects the fact that we are not measuring absolute fluxes but net fluxes after a local background subtraction at each frequency. Such net fluxes can legitimately be negative due to background estimation or noise, and if the likelihood is centered near zero, enforcing positivity would truncate the posterior and artificially push probability to positive values, increasing the chance of a false detection. In this setting the amplitudes behave effectively as location parameters in the likelihood, for which a uniform prior is the standard non-informative choice. Notably, positivity priors on $A_{\rm AME}$ were imposed in most of the relevant previous studies \citep{Poidevin2023, Mateo2023, Planck2014_AME}, which may have led to false detections and/or an underestimation of uncertainties on $A_{\rm AME}$, particularly for sources with relatively low signal-to-noise. This would also bias other parameters, such as $\mathrm{EM}$, $\nu_{\rm{AME}}$ and $W_{\rm{AME}}$, leading to similarly underestimated uncertainties in low-S/N sources. In our case, all sources are detected at $\gtrsim2\sigma$ and the amplitude posteriors are strongly peaked away from zero, so the broadening of posteriors and parameter correlations that can occur in low-S/N regimes when negative amplitudes are allowed do not affect our results.
 
The MCMC implementation employs 300 walkers, each performing 10,000 steps. The initial 8,000 steps are discarded as ``burn-in'', leaving 600,000 posterior samples. All fits are visually confirmed to converge within 2,000 steps, with most reaching a steady state after approximately 200 steps, justifying the conservative choice of a 8,000-step burn-in period.

Three significant correlations or degeneracies are generally observed among the parameters. First, $A_{\rm AME}$ and $\mathrm{EM}$ exhibit a strong inverse linear correlation, highlighting the importance of data below 10\,GHz for detecting AME. Second, the well-known correlations between the three thermal dust parameters, particularly between $\beta_{\rm d}$ and $T_{\rm d}$, as seen in \cite{Planck2013_XI}. Third, when synchrotron and free-free components are fitted simultaneously, individual constraints on each are typically weak and highly correlated due to their degeneracy, though their combined amplitude remains well-constrained. Notably, correlations between AME and thermal dust and synchrotron parameters are weak due to their large frequency separation.

The best-fit model for each source is obtained from the median of the marginalized 1D posterior of each parameter.

%Overfitting: large widths and low frequencies and high amplitudes common theme!

\subsection{Color Corrections}
\label{sec:color_corrections}

Color corrections account for the effect of a finite instrument bandpass on measured flux densities, ensuring consistency with a monochromatic observation at the reference frequency \(\nu_0\). The measured flux density must be multiplied by a correction factor \(C(\nu_0,\alpha_{\rm src})\), which depends on the flux density spectral index \(\alpha_{\rm src}\) of the source. This factor is given by a ratio of integrals over the instrument bandpass \(g(\nu)\). For an instrument calibrated in intensity units:

\begin{equation}
	C(\nu_0,\alpha_{\rm src}) = \frac{\int g(\nu)(\nu/\nu_0)^{\delta_{\rm ref}}~d\nu}{\int g(\nu)(\nu/\nu_0)^{\alpha_{\rm src}} ~d\nu}.
\end{equation}

The magnitude of the correction increases with the fractional width of the bandpass, asymmetry in the bandpass and the difference between the source spectral index $\alpha_{\rm src}$ and the reference spectral index to which each survey has been calibrated, $\delta_{\rm ref}$. For our dataset shown in Table~\ref{tab:surveys}, the typical corrections are of the order of $\lesssim1\%$ for the C-BASS and QUIJOTE datasets, up to $\sim2\%$ for WMAP and \textit{Planck}, and up to $\sim10\%$ for \textit{Planck} HFI and DIRBE/IRAS bands above $100$\,GHz. No corrections are applied to datasets with a frequency lower than C-BASS, since their extremely narrow bandwidths make the corrections negligible relative to their overall calibration uncertainties.

The \textsc{FastCC}\,\footnote{\url{https://github.com/mpeel/fastcc}} package by \cite{FastCC} is used to efficiently calculate color corrections by precomputing their spectral index dependence and approximating it with a polynomial function. At high frequencies, \textsc{FastCC} also provides corrections in terms of dust temperature \(T_\mathrm{d}\) and spectral index \(\beta_\mathrm{d}\), which provide more accurate corrections than using a fixed source spectral index $\alpha_{\rm src}$, as this value varies significantly across the bandpass near the peak of thermal dust emission. The latter method is applied to datasets $\geq 353\,$GHz.

Since estimating $\alpha_{\rm src}$, $T_\mathrm{d}$, and $\beta_\mathrm{d}$ requires color corrections, an iterative process is necessary. In this work, we perform three iterations, refining the spectral index and therefore color corrections until convergence.

\subsection{Example SEDs}
\label{sec:sed_example}

\begin{figure}
	\begin{center}
		\includegraphics[width=\columnwidth]{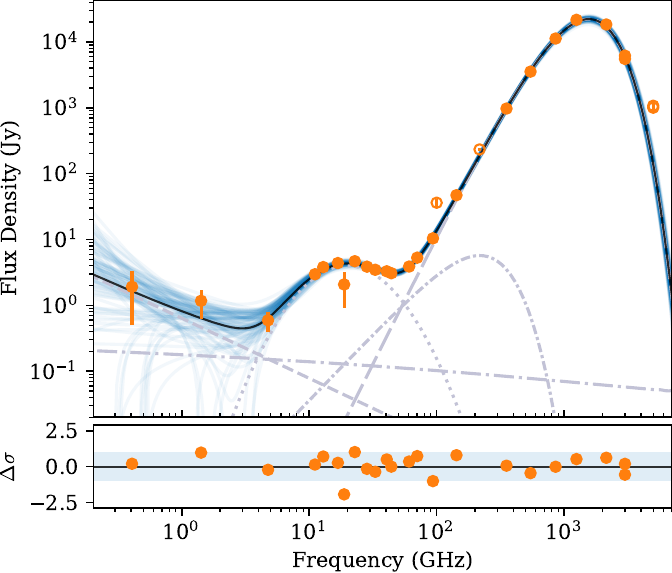}
		\vskip 0.3cm
		\includegraphics[width=\columnwidth]{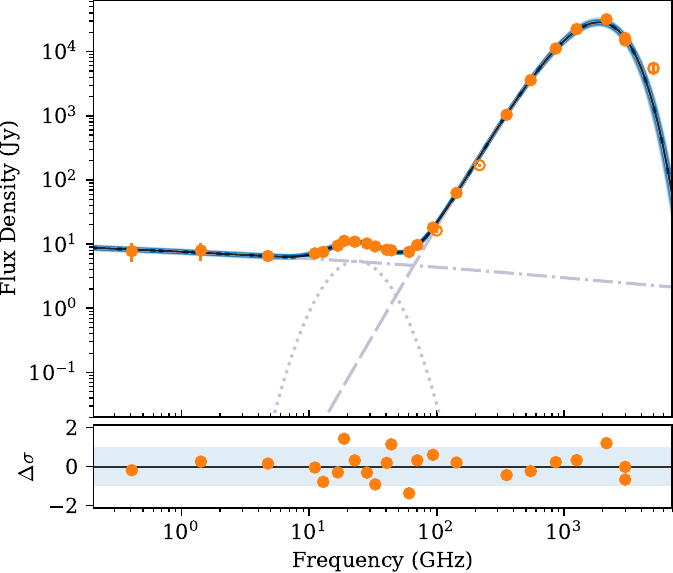}
		\caption{SEDs for two sources: \mbox{G159.02-33.88} (\textit{top}) and \mbox{G195.90-02.60} (\textit{bottom}). In each panel, the solid \textit{black} line shows the best-fit model, with $\chi^2_{\rm{red}} = 0.8$ and $0.7$, respectively. Individual realizations from the converged MCMC chain, illustrating model scatter, are shown in \textit{blue}. Color-corrected flux densities are plotted as \textit{orange} points, with hollow markers indicating data points excluded from the fit due to residual CO contamination or excessively high frequencies. Each individual best-fit model component is displayed in \textit{gray}, with the dotted line representing the AME component. The lower sub-panels display residuals in units of statistical deviation from the fit, with the $1\sigma$ region shaded in \textit{light blue}.}
		\label{fig:sed_fit}
	\end{center}
\end{figure}

\begin{figure*}
	\begin{center}
		\begin{minipage}{0.88\textwidth}
			\includegraphics[width=\textwidth,angle=0]{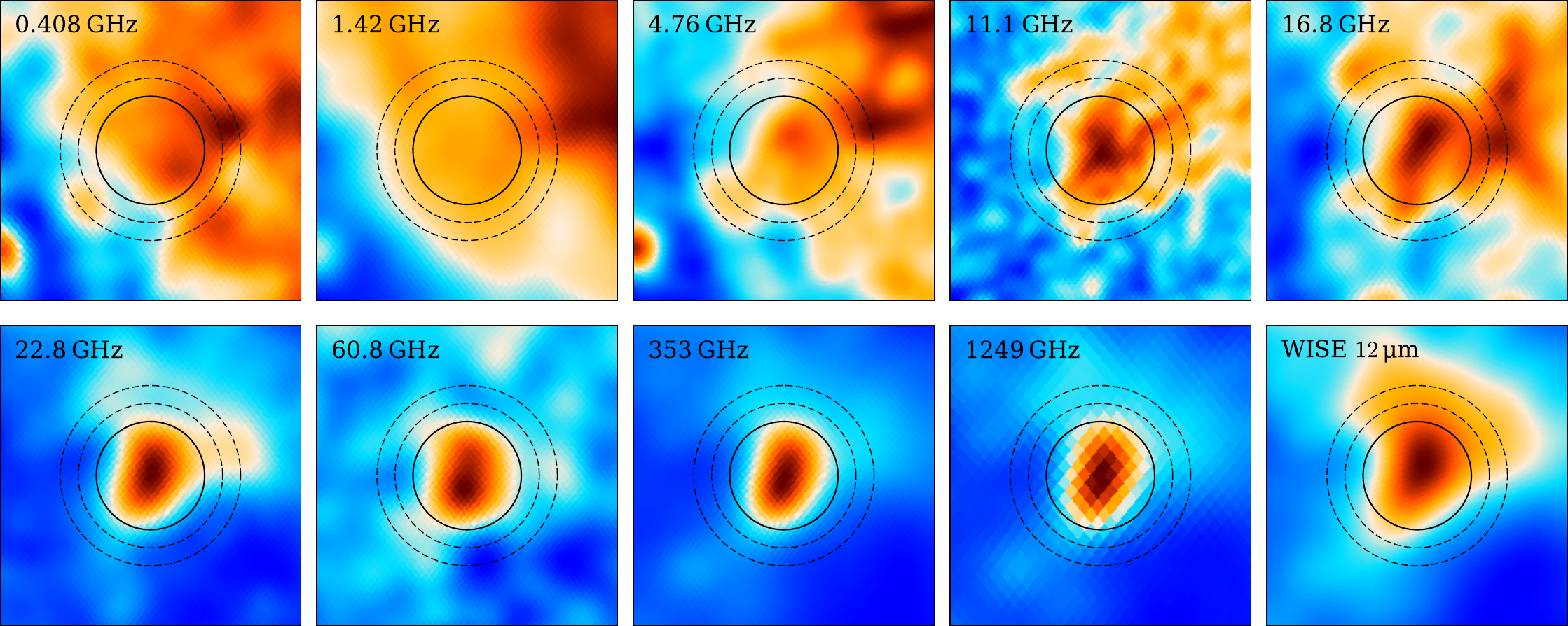}
		\end{minipage}%
		\hspace{0.2cm} % Adjust this value to control spacing
		\begin{minipage}{0.07\textwidth}
			\includegraphics[width=\textwidth]{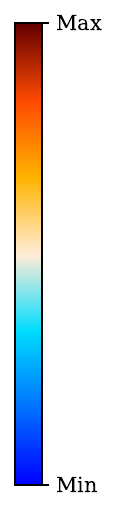}
		\end{minipage}
		\caption{Morphology of high-latitude source \mbox{G159.02-33.88} in selected maps, consisting of a collection of dark clouds including LDN 1454/53/58 and DOBASHI 4162. The solid circle ($72'$ radius) marks the primary aperture, while the dashed annulus ($96'$–$120'$) represents the background region. The grid spans $6.7^\circ$ per side. The AME component is visible above 5 GHz, with a strong correspondence between the WMAP K-band (22.8 GHz) near the AME peak and the DIRBE 240\,\textmu m (1249 GHz) map near the thermal dust peak. The reprocessed WISE 12\,\textmu m PAH-dominated map from \protect\cite{Meisner2014} is also shown. The color scale is linear, normalized to the pixel range of each image.}
		\label{fig:multifrequency_view}
	\end{center}
\end{figure*}

As an example, we present the high-latitude source \mbox{G159.02-33.88}. This cloud is slightly extended and comprises a collection of dark clouds, including LDN 1454, LDN 1453, LDN 1458, and DOBASHI 4162. It exhibits a peak flux density of $\approx4$\,Jy at its peak frequency of $\approx20$\,GHz. Its fitted SED is displayed in the top panel of Figure~\ref{fig:sed_fit}, while a corresponding multifrequency view is shown in Figure~\ref{fig:multifrequency_view}.

For comparison, the bottom panel of Figure~\ref{fig:sed_fit} shows the SED of \mbox{G195.90-02.60}, a high signal-to-noise source near the Galactic plane that includes LDN 1591, LDN 1592, and LDN 1593. Its AME level is similar at $\approx5$\,Jy, but the free-free emission is roughly 50 times brighter.

%%%%%%%%%%%%%%%%%%%%%%%%%%%%%%%%%%%%%%%%%%%%%%%%%%%%%%%
%%%%%%%%%%%%%%%%%%%%%%%%%%%%%%%%%%%%%%%%%%%%%%%%%%%%%%%

\section{Results and Discussion}
\label{sec:discussion}

\subsection{Sky Distribution and Source Selection}
\label{sec:source_selection}

\begin{figure*}
	\begin{center}
		\includegraphics[width=\textwidth,angle=0]{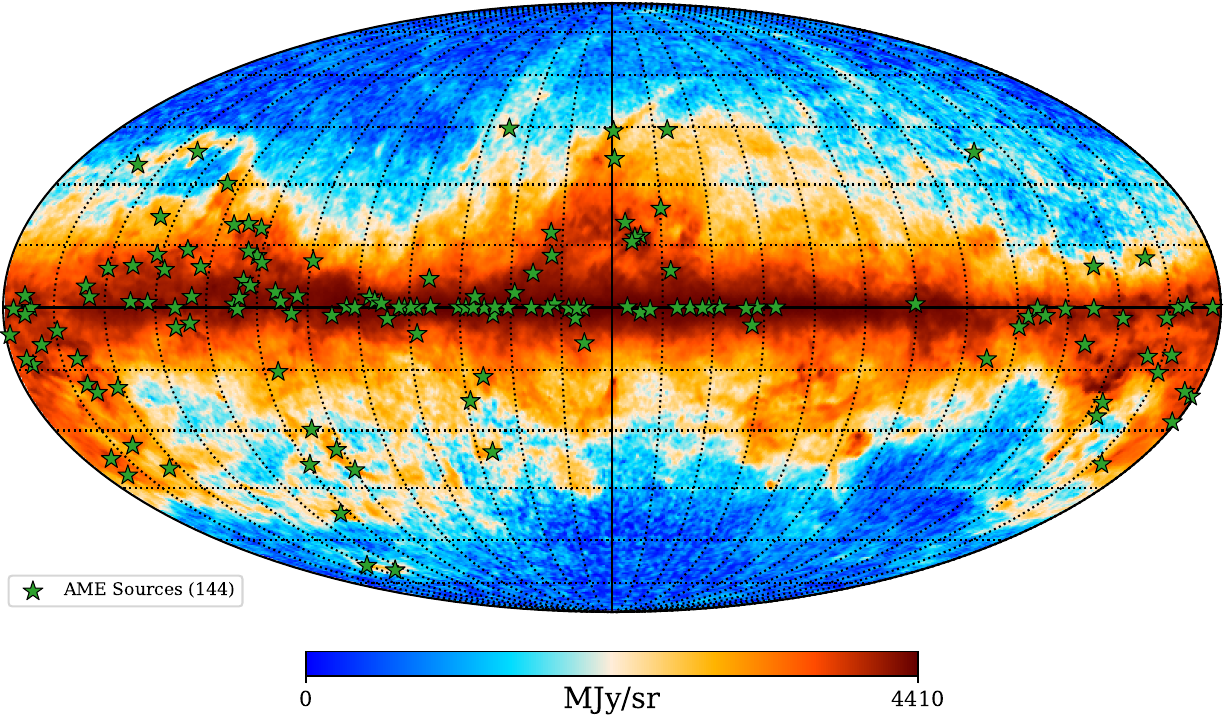}
		\caption{Locations of the 144 AME sources with $>2\sigma$ detection, represented by green stars, on top of the DIRBE 240\,\textmu m map.}
		\label{fig:sky_distribution}
	\end{center}
\end{figure*}

% General introduction
The sky distribution of the sources analyzed in this work is shown in Figure~\ref{fig:sky_distribution} against a background of thermal dust emission as seen by the DIRBE 240\,\textmu m band. Their properties and physical natures are listed in Table~\ref{tab:source_catalogue}. Most sources are concentrated along the Galactic plane, while a significant number trace thermal dust features at mid-latitudes; 40\% of sources are located at latitudes $|b| > 10^\circ$, with approximately 20\% of these having latitudes greater than $20^\circ$.

% Source selection
The sample includes 57 sources from \citet{Planck2014_AME} and 4 unique sources from \citet{Poidevin2023}. The remaining 83 new sources are identified through visual inspection of an updated reduction of the \citet{Planck2015_X} foreground model, derived using spectral parametric fitting with the \textsc{Commander} component separation code \citep{Eriksen2008}. This reduction incorporates C-BASS and S-PASS data, enabling a more reliable extraction of the AME amplitude (Hoerning et al., in prep.). The AME amplitude map is compared to the \textit{Planck} 857\,GHz emission at $1^\circ$ scales, and compact sources appearing in both maps are manually selected. Each source is then cross-checked in the SIMBAD \footnote{\url{https://simbad.u-strasbg.fr/simbad/}} database to confirm its Galactic origin and to characterize its nature and physical environment. Additionally, sources are verified against the Radio Fundamental Catalog \citep{Petrov2025} to exclude those with significant quasar or blazar contamination ($>0.5$\,Jy) in the primary or background apertures.

The use of the Hoerning et al. (in prep.) map is not essential for reproducing our source selection. A similar source list could be obtained through alternative methods, such as identifying clumps directly or extrapolating free-free emission from C-BASS maps to 22.8\,GHz and subtracting this estimate from WMAP K-band at the same frequency, though at significantly greater effort. However, our criterion of selecting AME sources only when they have a thermal dust emission counterpart inherently biases against any potential AME sources arising from a dust-independent emission mechanism.

Almost every source contains a mixture of dark and molecular clouds along with associated H\textsc{ii} regions. However, for some extended sources at high latitudes, molecular cloud and dark cloud catalogs are unavailable, with exceptions for well-known sources (e.g. $\rho$ Ophiuchi). For these high-latitude extended sources, the primary references are IRAS FIR sources \citep{IRAS1988} and \textit{Planck} Galactic Cold Clumps \citep{PlanckGCC2016}, which often encompass multiple individual dark and molecular clouds.

Most sources are compact relative to a $1^\circ$ beam, with 74\% having a primary aperture radius of $1^\circ$ and 24\% ranging from $1.0^\circ$ to $1.5^\circ$. Only three sources exceed this size, the largest being G170.60-37.30 (the translucent molecular cloud MBM16) with a radius of $2.5^\circ$. The sample has a mean radius of $1.1^\circ$, with a standard deviation of $0.2^\circ$. While these sources are not strictly compact in a morphological sense, their emission typically manifests as a single, centrally peaked structure without pronounced substructure, with the signal largely contained within a $2^\circ$ aperture; for an illustrative example, see Figure\,\ref{fig:multifrequency_view}.

% Selection biases
The selection criteria can introduce biases in the observed AME population, and as a result, we do not present a complete catalog—this is not the objective of our study. To be included in this analysis, sources must have a well-determined peak frequency and width, with AME detected at a significance level greater than $2\sigma$. Among the 144 sources that meet these criteria, 10 have significances in the $2$--$3\sigma$ range; the rest are $>3\sigma$. For comparison, \citet{Planck2014_AME} reported 42 significant AME sources out of 98, and \citet{Poidevin2023} found 44 out of 52. Our sample increases the number of significant sources by a factor of 3.3, with amplitude, peak frequency, and width constrained for each. However, in the southern hemisphere, the lack of QUIJOTE data limits width constraints, reducing the number of identified sources there. As a result, southern hemisphere sources tend to be the brightest, with the highest AME-to-free-free contrast. This strict requirement for fully constrained amplitudes, peak frequencies, and widths excludes many sources with potentially significant AME, particularly those with lower AME contrast for which the width cannot be well constrained.

A significant fraction of AME in the sky is diffuse on scales of several degrees or more, but our use of aperture photometry restricts the analysis to relatively compact sources ($\approx1^\circ$) that are substantially brighter than their surroundings. In some mid-latitude clouds, AME is so diffuse at $1^\circ$ scales that it becomes undetectable by aperture photometry techniques. Nevertheless, aperture photometry provides the advantage of isolating emission from individual physical sources rather than capturing the total emission along a given line of sight, as is the case in studies such as \cite{Mateo2023}. At high Galactic latitudes, detecting AME is particularly challenging because the background fluctuations within the aperture are often comparable to the source emission itself. This reduces the effective signal-to-noise ratio and can bias the derived spectrum if the background has a different spectral shape than the source. As a result, high-latitude detections are limited to sources with strong emission relative to the local background variations.

Another selection bias stems from the noise-limited sensitivity of the data, which prevents the detection of AME sources significantly fainter than $\lesssim1$\,Jy. Since the power emitted by spinning dust grains scales with the fourth power of their rotational speed, this sensitivity limit effectively imposes a constraint on peak frequency—sources with very low peak frequencies become too faint to detect below a certain threshold. This is supported by our ability to probe lower frequencies compared to previous studies due to improved low-frequency data (see Section~\ref{sec:peak_frequency_vs_temperature}).

Additionally, in bright regions dominated by calibration uncertainties, the relative contrast between AME and free-free emission introduces an additional limitation. If AME is brighter than free-free by an amount comparable to the calibration errors of low-frequency surveys (e.g., $\sim5\%$), it cannot be reliably detected. This bias favors sources with a higher contrast of AME relative to free-free emission. In fact, the minimum ratio of AME to total emission at the peak frequency for our sample is $\approx12\%$, consistent with the calibration uncertainties of the low-frequency data and AME significance levels of the sample.

Achieving completeness in flux density is only possible at the very brightest levels, given the combination of systematics and background complexity. Furthermore, completeness in luminosity, which would reflect the intrinsic emission of each source, is infeasible because of uncertain distances. Given these limitations, our sample is predominantly composed of compact sources with good data coverage, high contrast against free-free emission, and sufficient brightness to be detected above noise thresholds. As a result, the AME catalog presented in this paper is not complete to a well-defined flux-density threshold, although as a reference, the faintest AME detections in the sample are at the $\approx0.7$\,Jy level. Nevertheless, it represents the largest and most accurate set of AME detections and spectra to date, providing a robust basis for sub-sample analyses and tests of selection biases.

\subsection{AME Parameters}
\label{sec:ame_regions}

\subsubsection{Parameter Distributions}
\label{sec:parameter_distribution}

\begin{figure*}
	\begin{center}
		\includegraphics[width=\textwidth,angle=0]{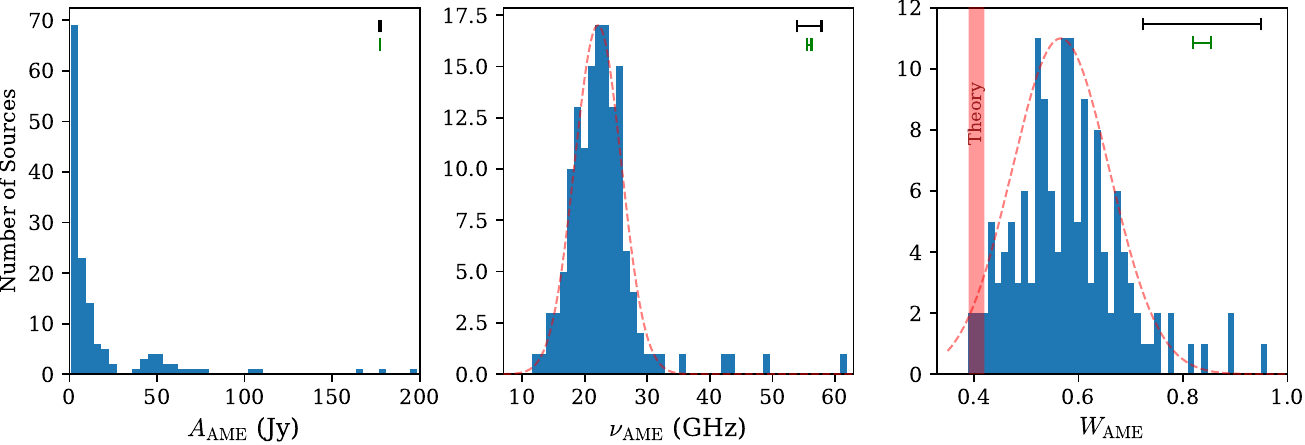}
		\caption{Distributions of AME parameters in the source sample. Each panel includes an inset in the \textit{top-right} showing the median (\textit{black}) and smallest (\textit{green}) 1$\sigma$ uncertainties for that parameter among all sources. For $A_{\rm{AME}}$, the median and smallest uncertainties are 0.6\,Jy and 0.1\,Jy, respectively; the corresponding uncertainty bars are not visually discernible, as their horizontal caps are unresolved at the adopted linewidth. A best-fit Gaussian is drawn as a \textit{red-dashed} line on the $\nu_{\rm{AME}}$ and $W_{\rm{AME}}$ distributions. The \textit{red} shaded region in the $W_{\rm{AME}}$ panel highlights the range of effective widths predicted by single-phase \textsc{SpDust2} and \textsc{SpyDust} templates (see Table~\protect\ref{tab:spdust_vs_lognormal}).}
		\label{fig:ame_parameter_distributions}
	\end{center}
\end{figure*}

Figure~\ref{fig:ame_parameter_distributions} shows the distributions of the fitted AME parameters: amplitude, peak frequency and width.

The left panel shows the AME amplitude distribution, which has a median value of 5.2\,Jy and a median uncertainty of 0.6\,Jy. These values translate to a median AME detection significance of $A_{\rm AME}/\sigma_{A_{\rm AME}}\approx7\sigma$ and a minimum significance of $\approx 2\sigma$. The ratios of AME to total emission at the peak frequency range from 12\% to nearly 100\%, with a median AME fraction of 64\% at the peak frequency.

The central panel presents the distribution of AME peak frequencies, predominantly between 10 and 35\,GHz. Few sources show peak frequencies above 35\,GHz, and these have relatively large uncertainties. The highest peak frequency in our sample is $\nu_{\rm AME}=62\pm13$\,GHz for the California Nebula, consistent with the \cite{Planck2014_AME} value of $50\pm17$\,GHz. The slightly smaller uncertainty arises from improved modeling of the low-frequency components, since the AME peak itself remains constrained entirely by WMAP/\textit{Planck} data. The median $\nu_{\rm AME}$ is 22.4\,GHz, with a median uncertainty of 1.9\,GHz. The distribution approximates a normal distribution with a mean of 21.9\,GHz and standard deviation of 3.7\,GHz (\textit{red-dashed} line). However, this Gaussian shape likely results from the central limit theorem rather than being an intrinsic property, as detection bias exists against very low peak-frequency sources (see Section~\ref{sec:source_selection}). This bias is evident in previous studies, which lacked detections below $\sim$17\,GHz due to fewer low-frequency datasets. Notably, no sources peak at the very high frequencies predicted by theoretical models for reflection nebulae and photodissociation regions, which the current data and source selection criteria would be sensitive to, at least at the low-frequency end ($<100$\,GHz) not embedded in thermal dust emission.

The peak frequency as a function of absolute latitude exhibits a small but statistically significant downward trend when fitted linearly, with higher-latitude sources tending to have lower peak frequencies. The best-fit line intercepts the Galactic plane at $23.7 \pm 0.6$ GHz with a slope of $-0.076 \pm 0.034$ GHz deg$^{-1}$ (2.2$\sigma$). This pattern parallels the general decrease in dust temperature at higher latitudes seen in our sample. The substantial scatter and high $\chi^2_{\rm red}=9.8$ suggest that local environmental factors strongly influence individual source peak frequencies, making latitude alone an inadequate predictor.

The right panel illustrates the distribution of the width parameter. The median width is 0.57, with a standard deviation of $0.10$ and a median uncertainty of 0.11 for any given source. We find no significant correlation between width and latitude, given that the uncertainties are large compared to the overall variation across the sample, although narrower widths generally have smaller associated uncertainties. Crucially, no source has a fitted width narrower than the theoretical effective widths predicted by single-component \mbox{\textsc{SpDust2}} or \mbox{\textsc{SpyDust}} models (red shaded region), and the observed distribution is generally much broader than these expectations. This result, consistent with findings for diffuse emission in the Galactic plane \citep{Mateo2023} and in $\lambda$ Orionis \citep{LambdaOrionis}, is a major conclusion of this analysis, and strongly indicates that single ISM phase spinning dust models cannot generally capture the observed AME width. Such broadness motivated the use of a two-component model in \cite{Planck2015_XXV}—with one fixed high-frequency and one varying low-frequency component—which better fit the then-limited data. It is now clear that this broadness is a prevalent feature, which we interpret as evidence for multiple spinning dust components within most compact sources or along most lines of sight, possibly also reflecting the fact that current theoretical models might not capture the full range of grain sizes encountered in the ISM, which would broaden the spectrum. Furthermore, widths narrower than 0.4 would have been observed if they existed, given that neither the sample selection nor fitting methodology is biased against narrower sources. The fact that theoretical predictions accurately reproduce the minimum observed width provides important new support for the spinning dust hypothesis.

Based on the observed distributions of peak frequencies and widths, a rough Gaussian approximation, $\nu_{\rm AME} \sim \mathcal{N}(21.9, 3.7\,\text{GHz})$ and $W_{\rm AME} \sim \mathcal{N}(0.56, 0.10)$, may serve as a practical starting point for defining informative priors in studies that lack direct constraints on these parameters, as shown by the \textit{red-dashed} lines in Figure\,\ref{fig:ame_parameter_distributions}. The derived distributions closely match those in \cite{Mateo2023}, suggesting broader applicability in diffuse regions. However, potential biases in our sample should be considered, such as reduced sensitivity to low $\nu_{\rm AME}$ values and the prior's inability to accommodate high-frequency sources such as the California nebula. These priors should therefore be used with caution.

The thermal dust emission (not shown) shows temperatures in the range $14<T_{\rm{d}}< 28$\,K, with a median of 19.0\,K and median uncertainty of 0.7\,K. The dust emissivity index is in the range $1.2<\beta_{\rm d}<2.2$, with a median of 1.6 and typical uncertainty of 0.09, consistent with values found by \cite{Planck2013_XI} over the full sky. The median value of the synchrotron spectral index for sources including this component is -1.1, consistent with expectations.

\subsubsection{The Impact of Low-Frequency Data: Comparison to Previous AME Studies}
\label{sec:comparison}

\begin{figure}
	\centering
	\includegraphics[width=0.9\columnwidth]{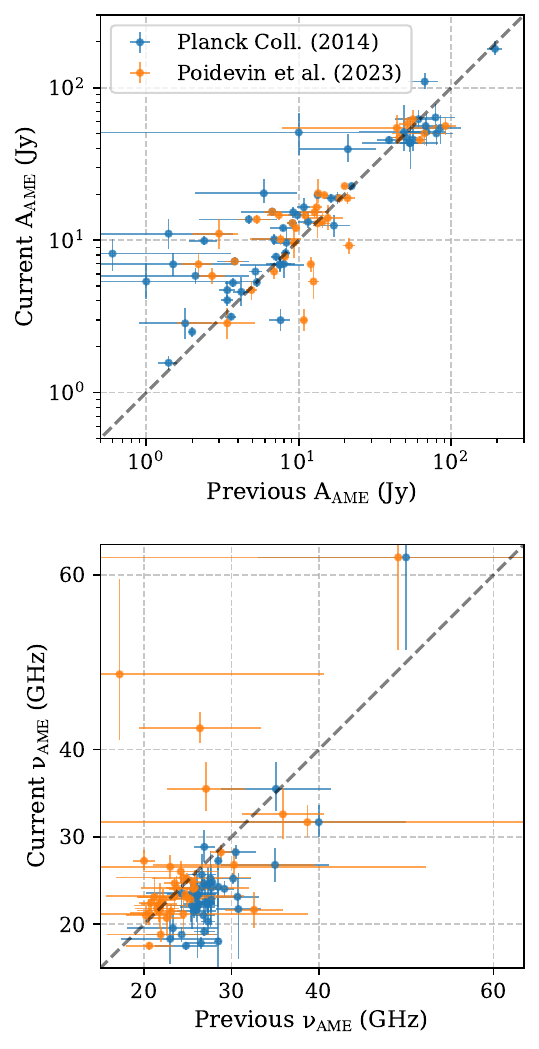}
	\caption{Comparison of this work (\textit{y}-axis) with previous studies (\textit{x}-axis) for $A_{\rm{AME}}$ (\textit{top} panel) and $\nu_{\rm{AME}}$ (\textit{bottom} panel). The dashed \textit{gray} line represents a 1:1 relation.}
	\label{fig:comparison}
\end{figure}

We compare our fitted parameters with those from previous studies for sources in common, focusing on AME amplitude and peak frequency (see Figure~\ref{fig:comparison}).

Our free-free emission estimates align well with previous studies, with uncertainties in $\rm{EM}$ about two-thirds those of \cite{Planck2014_AME}. However, \cite{Poidevin2023} reports $\rm{EM}$ uncertainties nearly half of ours, likely due to their assumption of 10\% effective calibration uncertainties for the lowest-frequency surveys. These surveys suffer from variable, scale-dependent calibration factors (see Section~\ref{subsec:ancillary}). The absence of C-BASS and S-PASS data in \cite{Poidevin2023} implies that the uncertainties in free-free and synchrotron emission may have been underestimated, potentially resulting in artificial high-significance AME detections in some sources.

The AME amplitudes (top panel) are generally consistent with \cite{Planck2014_AME}, with a systematic underestimation for sources in the 5--10 Jy range relative to our results. Compared to \cite{Poidevin2023}, our results show less scatter but some amplitudes are inconsistent, likely due to underestimated calibration uncertainties at low frequencies and least-squares fitting used in \cite{Poidevin2023} falling into local minima. In fact, AME significances in our study align better with those from \cite{Planck2014_AME}. The uncertainties in $A_{\rm{AME}}$ are, on average, 10\% smaller than \cite{Poidevin2023} and 30\% smaller than \cite{Planck2014_AME} for common sources thanks to additional data in the range 2.3--19\,GHz.

The bottom panel compares peak frequencies. \cite{Planck2014_AME} exhibits a systematic bias toward higher peak frequencies, predominantly falling to the right of the 1:1 line. A best-fit scaling factor of $0.87\pm0.02$ implies a peak frequency bias of $-13\pm2\,\%$ on average—substantially larger than the expected $\approx-6\%$ bias from modeling a theoretical template with a log-Gaussian function. We attribute the larger bias to the limited low-frequency data in \cite{Planck2014_AME}, which tends to push fitted peak frequencies higher, especially in cases where data do not cover both sides of the peak, as clearly demonstrated in \cite{Rennie_2022}. In addition, fixing the width to that of a theoretical template when the real widths are generally broader might also have biased the peak frequency. In contrast, \cite{Poidevin2023} shows a scaling factor of $1.04\pm0.04$, indicating consistency. Our peak frequency uncertainties are, on average, 2.7 times smaller than those in \cite{Poidevin2023}.

In H\textsc{ii} regions like the Rosette Nebula, where AME has very low contrast, the absence of QUIJOTE data leaves its width unconstrained and often biases it toward higher values. For this reason, the Rosette Nebula is not included in the catalog. This highlights the necessity of QUIJOTE observations in low AME contrast sources for fully constraining all AME parameters. More broadly, our comparison emphasizes the critical role of data below 20 GHz in accurately determining AME amplitude, frequency and width. Combining low-frequency datasets such as C-BASS and S-PASS to establish a reliable baseline for free-free and synchrotron emission—alongside QUIJOTE—remains essential for robust peak frequency measurements.

%\subsubsection{AME Detections at 5\,GHz}
%\label{sec:ame_5ghz}

%CGEM and other surveys might have AME at 5 GHz - need to get ratio of AME(5ghz)/(model(5ghz)) and print it with uncertainties for every source, then select the highest ones, give some sort of median figure

\subsubsection{Systematic Deviations from the Fitted Model and Calibration Systematics}  
\label{sec:reich_calibration_factor}

We assess potential miscalibration and foreground mismodeling by examining systematic deviations of flux densities from the best-fit SED models. This involves computing the ratio of fitted to measured flux density at each frequency and deriving a weighted mean to quantify any systematic offset. Our analysis focuses on point-like sources that fit within $1^\circ$-radius primary apertures, effectively excluding extended sources in order to assess calibration systematics.

Most datasets show good agreement with the fitted models, including C-BASS and the $1420\,\si{\mega\hertz}$ Stockert/Villa-Elisa survey \citep{Reich2001}. Our independent derivation of the point-source calibration factor is $1.57 \pm 0.06$, consistent with the $1.55 \pm 0.08$ reported by \cite{Reich1988} and applied in this study. Our factor is confirmed through a second iteration of the SED fits in which the data point is already corrected by the original \cite{Reich1988} factor, implying convergence. We find no significant latitude dependence, though uncertainties remain large for high-latitude bins due to the lower number of sources.

However, the S-PASS dataset systematically exceeds the model fit by a factor of $1.10 \pm 0.02$, while the HartRAO survey appears slightly low at $0.97 \pm 0.02$. A direct T-T plot comparison between the HartRAO 2.303\,GHz survey and S-PASS confirms that S-PASS is higher by a factor of $1.195 \pm 0.003$. Since we weight S-PASS more strongly than HartRAO in the fit and the two datasets frequently overlap, the higher discrepancy between the two datasets aligns with the deviations observed with respect to the line of best fit. Further validation comes from source W37, which lies in a region covered by S-PASS, HartRAO, C-BASS, and QUIJOTE. There, the S-PASS flux density is $\sim 13\%$ above the best-fit line, significantly larger than the quoted calibration uncertainty of 5\%. This potential miscalibration could lead to slightly inflated estimates of free-free emission in the Southern Hemisphere, reducing AME detections and biasing the sample toward regions with high AME/free-free contrast. 

A likely reason for the overcalibration of S-PASS is sidelobe integration when smoothing to $1^\circ$ FWHM from its native $8.9$ arcmin resolution. The Parkes telescope's main beam efficiency is approximately 63\% at the L-band \citep{Barnes2001}, and while it has not been publicly reported at the S-band, it is expected to be similar. \cite{spass_release} do not perform detailed beam modeling or give an explicit value. Proper beam characterization and deconvolution are therefore crucial for fully leveraging S-PASS data in spinning dust studies, particularly for AME analysis in future \textsc{Commander} runs. Similar overcalibration effects are also likely to impact polarized synchrotron studies at $1^\circ$ FWHM.  

The QUIJOTE MFI datasets exhibit a small but measurable trend relative to the line of best fit: the 11\,GHz band is slightly low at $0.972\pm0.009$, the 13\,GHz band is consistent with the fit, and the higher two bands are slightly high at $1.035\pm0.009$. Since this trend runs counter to the excess flux expected from modeling spinning dust spectra with a log-Gaussian distribution (see Section~\ref{subsec:ame_deviations}), it likely originates from excess atmospheric emission stripes, particularly at the higher 17 and 19\,GHz bands for high latitude and therefore lower signal-to-noise sources. These deviations are consistent with those found in \cite{mfiwidesurvey} through CMB cross-correlations, and remain within the quoted 5\% uncertainty.

% Talk about the thermal dust emission mismodelling??? 

\subsection{Spectral Variations and the Nature of AME}
\label{sec:nature_of_ame}

%This section describes and interprets...

\subsubsection{Correlation Analysis and Methodological Limitations}
\label{sec:spearman}

Spearman's rank correlation coefficients, $r_{\rm{s}}$, are a powerful tool for assessing monotonic but not necessarily linear relationships between parameters \citep{Spearman}. We compute $r_{\rm{s}}$ for every parameter combination across all datasets described in Section~\ref{sec:observables}, using the \textit{spearmanr} function from \textsc{SciPy} \citep{SciPy}. These correlations are robust to offsets and miscalibration, making them effective for identifying phenomenological trends when the underlying physical relationship is unknown.

However, Spearman correlation relies on rank ordering, and any noise that disrupts this ordering can reduce the measured $r_{\rm{s}}$ relative to the underlying true correlation. It may also fail to capture strong but non-monotonic relationships, such as those involving inflection points or more complex structures. Importantly, a high degree of correlation does not imply causation. Apparent correlations may arise from a shared dependence on a third variable or additional variables, while weak correlations may be statistically insignificant or spurious. Source selection for incomplete samples and fitting biases must also be considered.

In pairwise correlations, multidimensional relationships between variables may be obscured by confounding variables. As an analogy, a dataset describing the speed and mass of objects in relation to their kinetic energy will only show a tight pairwise correlation when one of the two variables is held relatively constant. Similarly, AME observables may depend on several environmental parameters simultaneously, so two-variable correlations alone may be inconclusive. This motivates the use of multivariate approaches, particularly machine learning methods. Random Forests \citep{Breiman2001}, for instance, can rank the relative importance of each parameter in predicting a given AME observable, highlighting the dominant physical processes at play. Symbolic Regression techniques, often based on genetic algorithms, extend this by producing compact analytical expressions that balance predictive power with interpretability. Sparse methods such as \textsc{SINDy} \citep{Brunton2016} assume that only a few terms govern most systems and have been successful in rediscovering known laws in complex regimes such as fluid dynamics. Similarly, Multivariate Adaptive Regression Splines \citep{MARS} model nonlinearities with piecewise linear fits, yielding explicit equations with clear interpretability. The exploration of multivariate AME correlations using these techniques is suggested as a future extension to this work.

To estimate the impact of observational uncertainties on $r_{\rm{s}}$, we perform a Monte Carlo randomization of each data point based on its Markov chain, neglecting systematic calibration errors and assuming Gaussian uncertainties for photometric data. We also verify that sources with and without QUIJOTE data follow consistent distributions, confirming the absence of spectral coverage bias. This is expected for the relatively high signal-to-noise sources studied in the Southern sky.

\subsubsection{AME Observables and Environmental Tracers}
\label{sec:observables}

The fitted SEDs provide three key AME observables: the AME amplitude, the peak frequency, and the spectral width, with no significant correlations observed between them. The latter two arise as emergent properties of the underlying dust grain population, while the amplitude must be normalized to derive a physically meaningful emissivity.

To quantify the AME emissivity, we adopt three normalization schemes for the fitted amplitude $A_{\rm{AME}}$. The first and most common approach is to normalize by the product of the dust optical depth at 353\,GHz, $\tau_{353}$, and the solid angle of the primary aperture, $\Omega$, using $A_{\rm{AME}} / (\tau_{353} \cdot \Omega)$. This method accounts for variations in source size across different regions. Although this normalization is standard, it introduces some scatter since the true source size is unknown and the aperture area is only an approximation.

The use of $\tau_{353}$ is motivated by the fact that it traces the dust column density and, under the assumption of a relatively constant dust-to-gas ratio, is often used as a proxy for the total hydrogen column density \citep{dark_gas}. This allows, in principle, for direct comparison with theoretical models of spinning dust emission by converting the measured AME emissivity into a per-hydrogen-atom basis. However, this assumption is only approximate: the $\tau_{353}$–$N_{\rm H}$ relation is known to vary by at least a factor of $\sim 2$ \citep{Planck2013_XI}, and potentially more at low Galactic latitudes where dust properties and radiation fields are more complex.

Furthermore, variations in density within an aperture introduce additional scatter: dense clumps may occupy only a small fraction of the beam, biasing the effective $\tau_{353}$ estimate when averaging over the entire region. Most critically, AME is thought to arise from very small grains or PAHs, whereas the total dust mass (and thus $\tau_{353}$) is dominated by large grains. This mismatch means that $\tau_{353}$ may not be an optimal tracer of the relevant grain population, particularly in regions where environmental conditions strongly affect the abundance of small grains.

As an alternative, we normalize the AME amplitude using the dust radiance, defined as

\begin{equation}
	\Re_{\rm{d}} = \int_{0}^{\infty} S_{\rm d}(\nu)\, \mathrm{d}\nu,
\end{equation}
\noindent
where $S_{\rm d}(\nu)$ is the spectral energy distribution of the thermal dust emission. We also compute the AME radiance, $\Re_{\rm{AME}}$, in an analogous way and normalize it by $\Re_{\rm{d}}$ to express the AME flux relative to the total thermal dust emission. This provides a direct comparison between the energy budgets of the two mechanisms.

Environmental parameters considered in our analysis include the dust temperature, optical depth, and emissivity index, as well as the emission measure of free-free emission. For a subset of sources, synchrotron parameters are also included. In addition to these, we derive composite quantities such as the peak frequency and flux density of the thermal dust emission, the dust and AME radiances, and a tracer of the interstellar radiation field (ISRF) strength, following \cite{Mathis1983}:

\begin{equation}
	\label{eq:G0}
	G_0 = \left( \frac{T_{\rm d}}{17.5\,\si{\kelvin}} \right)^{4+\beta_{\rm d}}.
\end{equation}
\noindent

Although $G_0$ is a widely used tracer of the ISRF, it assumes the average values produced by local stars in the diffuse ISM. It does not account for the specific stellar population or radiation field shape in each individual region, which can vary significantly. This makes $G_0$ a reasonable approximation at high Galactic latitudes, but at low latitudes greater scatter is expected.

To investigate the link between PAHs and AME, we use the reprocessed WISE 12\,\textmu m map from \citet{Meisner2014}, which removes instrumental artifacts and continuum starlight. Within this bandpass, PAH molecules are believed to dominate the emission, although their abundance and spectral features vary considerably across regions \citep{Boersma2010PAHs}. Following the method of \citet{Hensley2016_pahs}, we construct a tracer for the fractional PAH abundance by dividing photometry of the 12\,\textmu m map by the dust radiance.

\subsubsection{AME Tracers}

\begin{figure*}
	\centering
	\begin{minipage}{0.40\textwidth}
		\centering
		\includegraphics[width=\linewidth]{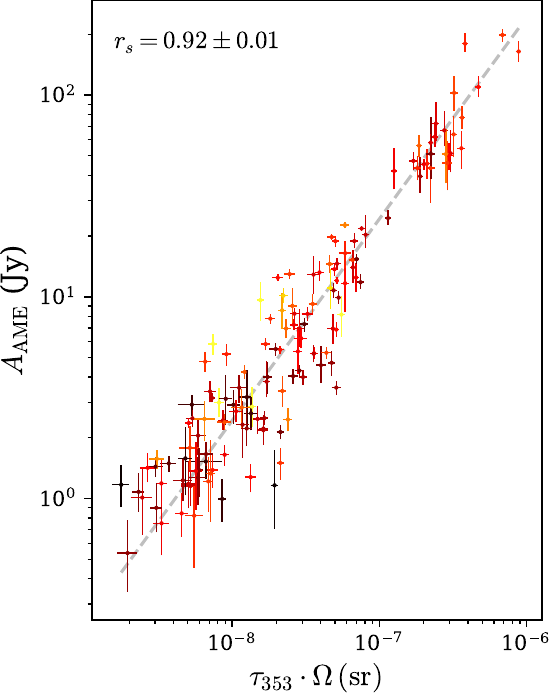}
	\end{minipage}%
	\hspace{0.8cm}
	\begin{minipage}{0.40\textwidth}
		\centering
		\includegraphics[width=\linewidth]{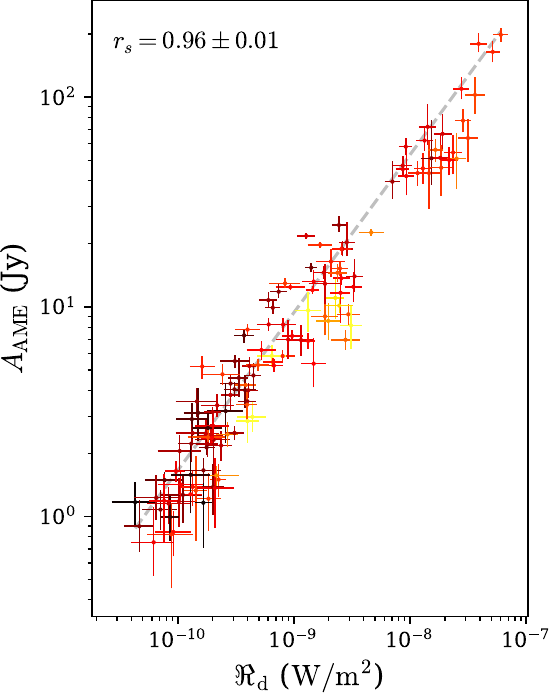}
	\end{minipage}%
	\hspace{0.8cm}
	\begin{minipage}{0.082\textwidth}
		\centering
		\includegraphics[width=\linewidth]{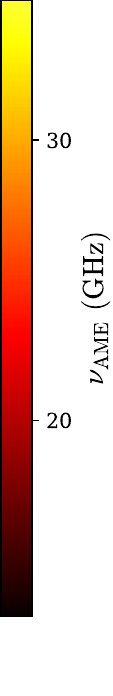}
	\end{minipage}%
	\caption{AME amplitude correlations against $\tau_{\rm{353}}\cdot\Omega$, where $\Omega$ is the primary aperture solid angle (\textit{left}), and dust radiance $\Re_{\rm{d}}$ (\textit{right}). The points are color-coded by peak frequency, with sources above 35 GHz truncated to the upper color limit. Best-fit models are shown by \textit{gray} dashed lines: linear for the optical depth correlation (\textit{left}) and power-law for the dust radiance correlation (\textit{right}).}
	\label{fig:a_ame_vs_tau_vs_dust_radiance}
\end{figure*}

\begin{figure*}
	\begin{center}
		\includegraphics[width=0.99\textwidth,angle=0]{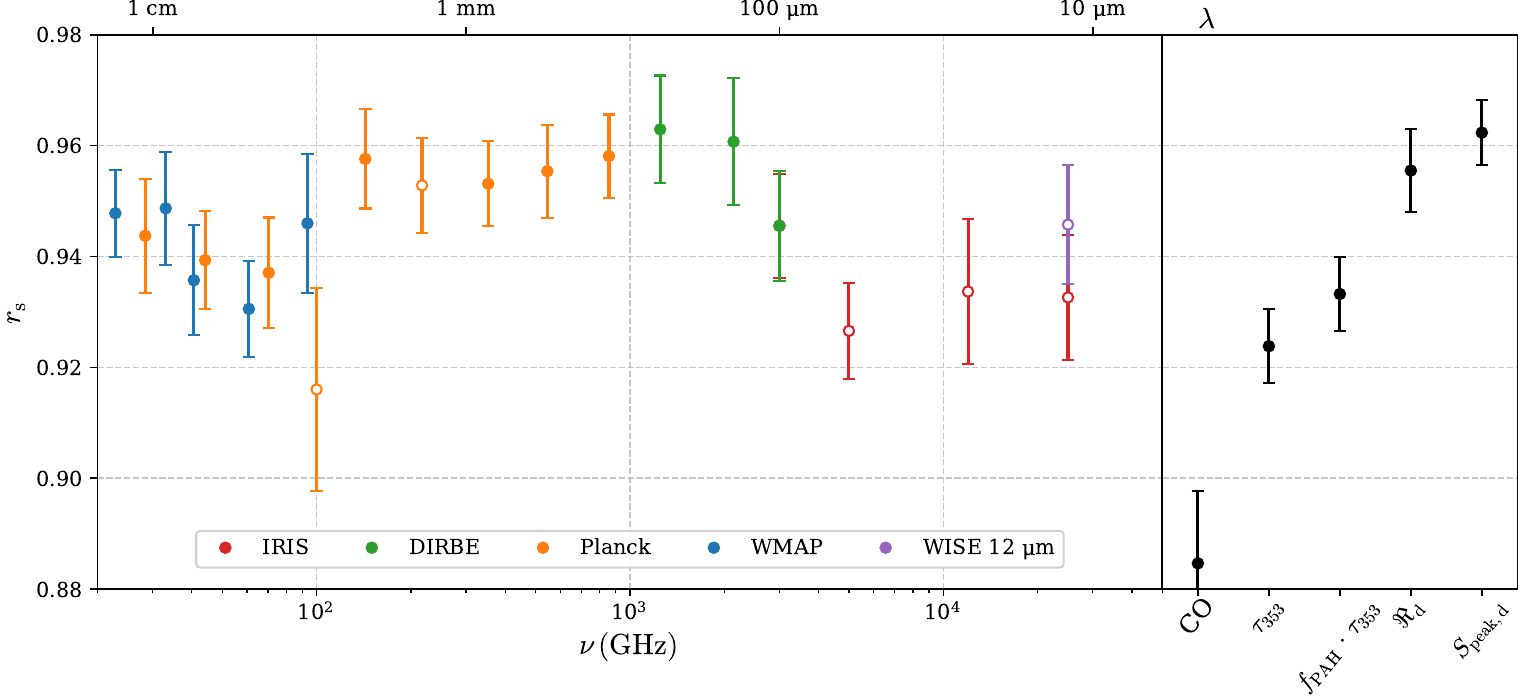}
		\caption{Spearman correlation coefficients between \protect{$A_{\rm AME}$} and individual frequency maps. Tracers are shown on the right, where CO denotes the 1-0 CO transition map by \protect\cite{Ghosh2024} and $S_{\rm{peak, d}}$ is the flux density at the peak of the thermal dust emission fitted. In both cases, $\tau_{\rm{353}}$ refers to the angular area-corrected optical depth $\tau_{\rm{353}} \cdot \Omega$. Hollow points denote frequencies that are excluded from the SED fitting. The 100 and 217\,GHz bands are CO-subtracted.}
		\label{fig:spearman}
	\end{center}
\end{figure*}

We analyze correlations between AME amplitude and various environmental parameters to identify the most effective predictor of AME at 1$^\circ$ scales. Figure~\ref{fig:a_ame_vs_tau_vs_dust_radiance} highlights the relationships between AME amplitude and both dust optical depth ($\tau_{353}$) and dust radiance ($\Re_{\rm{d}}$). We also correlate AME amplitude with the flux density from each frequency map in Table~\ref{tab:surveys}, following \cite{LambdaOrionis}, with results for datasets above 20\,GHz shown in Figure~\ref{fig:spearman}.

A strong correlation is found between AME amplitude and thermal dust emission. We specifically highlight the relationships with optical depth $\tau_{\rm{353}}$ and dust radiance. These correlations, previously reported in the literature, are now extended by our high-latitude sources to include fainter AME detections. Dust radiance exhibits a tighter correlation with AME amplitude than optical depth, suggesting it is a more effective tracer.

Fitting the $A_{\rm{AME}}$–$\tau_{\rm{353}} \cdot \Omega$ relation with a power law using orthogonal distance regression \citep{Boggs1990} yields an exponent of $0.99 \pm 0.03$, consistent with linearity. A linear fit returns an intercept of $0.25 \pm 0.18$\,Jy, marginally consistent with zero. The scatter is approximately 50\%, with an orthogonal $\chi^2_{\rm{red}} \approx 12$, even when forcing the fit through the origin (see Figure~\ref{fig:a_ame_vs_tau_vs_dust_radiance}). The best-fit relation is:
\begin{equation}
	\frac{A_{\rm{AME}}}{\mathrm{Jy}} = (2.4 \pm 0.1) \cdot 10^8 \cdot \left( \frac{\tau_{\rm{353}} \cdot \Omega}{\mathrm{sr}} \right)\,.
\end{equation}
\noindent
The high $\chi^2_{\rm{red}}$ reflects intrinsic environmental variations between regions, leading to $\approx$50\% deviations from this linear prediction. A linear relationship is expected physically, as $A_{\rm{AME}}$ scales with column density of small grains, which are usually well-mixed with the larger grains traced by $\tau_{\rm{353}}$. No significant dependence on peak frequency is observed in this correlation. For the typical $\nu_{\rm AME} \sim 22$\,GHz, the scaling yields an AME amplitude of $\sim16$\,K per $\tau_{353}$, although this value should be regarded as indicative only, given the angular dilution inherent in $\Omega$.

The $A_{\rm{AME}}$–$\Re_{\rm{d}}$ relation follows a power law:
\begin{equation}
	\frac{A_{\rm{AME}}}{\mathrm{Jy}} = (5.8 \pm 2.0) \cdot 10^7 \cdot \left( \frac{\Re_{\rm{d}}}{\si{\watt\per\metre\squared}} \right)^{0.75 \pm 0.02}
\end{equation}
\noindent
This fit exhibits a reduced scatter of 33\%, indicating that $\Re_{\rm{d}}$ is a more precise predictor than $\tau_{\rm{353}}$. However, the still-elevated $\chi^2_{\rm{red}} \approx 3$ suggests intrinsic source-to-source variability. A mild excess in dust radiance is observed around $A_{\rm{AME}} \approx 10$\,Jy, coinciding with sources that exhibit higher peak frequencies.

The improved correlation of $\Re_{\rm{d}}$ over $\tau_{\rm{353}}$ aligns with earlier findings \citep{Hensley2016_pahs,LambdaOrionis,Poidevin2023,Mateo2023}. Several factors likely contribute to this: (1) $\tau_{\rm{353}}$ is more model-dependent than $\Re_{\rm{d}}$, which can be directly calculated by integration with minimal modeling; (2) the use of primary aperture area as a proxy for source size introduces scatter, especially for more extended, lower-AME sources; and (3) $\tau_{\rm{353}}$ primarily traces large grains, whereas AME arises from very small grains—so variations in the ratio of large-to-small grains introduces additional scatter. A more in-depth discussion of small grain tracers, including PAHs, is provided in Section~\ref{subsec:pahs}.

A similarly tight correlation exists between AME amplitude and the peak flux of thermal dust emission:
\begin{equation}
	\frac{A_{\rm{AME}}}{\mathrm{Jy}} = (1.9 \pm 0.4) \cdot 10^{-3} \cdot \left( \frac{S_{\rm{d,peak}}}{\mathrm{Jy}} \right)^{0.79 \pm 0.02}
\end{equation}
\noindent
This relation has a typical scatter of 32\% and a Spearman coefficient comparable to that of $\Re_{\rm{d}}$. The corresponding AME emissivity, $A_{\rm AME}/S_{\rm d,peak}$, decreases with source brightness, ranging from $\sim 0.04\%$ for the faintest sources to $\sim 0.009\%$ for the brightest, changing by a factor of $\approx 5$ across our sample.

We also compare the total radiative energy outputs:
\begin{equation}
	\frac{\Re_{\rm{AME}}}{\si{\watt\per\metre\squared}} = (1.8 \pm 1.0) \cdot 10^{-8} \cdot \left( \frac{\Re_{\rm{d}}}{\si{\watt\per\metre\squared}} \right)^{0.75 \pm 0.03}
\end{equation}
\noindent
This relation yields $r_s = 0.94 \pm 0.03$ and a scatter of 52\%, likely due to larger uncertainties in deriving AME radiance. The corresponding AME radiative emissivity, $\Re_{\rm AME}/\Re_{\rm d}$, decreases with increasing source radiance, ranging from $\sim 3\cdot10^{-6}$ for the faintest sources to $\sim 1\cdot10^{-6}$ for the brightest, changing by a factor of $\approx 3$ across our sample.

A commonly used historical measure of AME emissivity is the ratio of the AME residual flux density at 28.4 GHz to the 100 µm flux density. In our sample, this ratio decreases from $\sim9\cdot10^{-4}$ for the faintest thermal dust emission sources to $\sim1\cdot10^{-4}$ for the brightest, a variation of nearly an order of magnitude. The median value, $3.7\cdot10^{-4}$, is similar to the values found in \citet{Planck2014_AME}. The ratio follows an approximate power-law dependence on the 100 µm flux density, scaling as $S_{100\,\mu\mathrm{m}}^{-0.27 \pm 0.03}$, indicating that a single average ratio poorly represents the sample. This trend mirrors the sublinear behavior of AME amplitude and radiance with respect to thermal dust emission. However, this emissivity definition is problematic because it relies on fixed reference frequencies that are highly sensitive to both the AME peak frequency and the thermal dust temperature, as evidenced by its large scatter. Peak amplitude ratios should therefore be used as a more fundamental and consistent basis for defining AME emissivities.

All in all, the sublinear indices for both AME amplitude and radiance indicate that AME increases more slowly than thermal dust emission. This suggests that brighter dust environments are less efficient at producing spinning dust emission. One possible cause is photodissociation of small grains in strong radiation fields, but if that were dominant, we would expect the brightest sources to have systematically hotter dust. We do not observe a clear trend, suggesting that photodissociation may not be the primary mechanism. A more likely explanation is grain evolution in dense, shielded environments, where the smallest grains may coagulate onto larger grains and be effectively depleted. \citet{Compiegne2008} show that in photodissociation regions the ratio of PAH to small-grain abundance declines by factors of 2--5 in dense zones compared to diffuse regions, while \citet{Arab2012} find strong PAH depletion in the Orion Bar consistent with coagulation processes. These results support the idea that small-grain depletion suppresses AME relative to thermal dust emission in denser dust environments.

Figure~\ref{fig:spearman} shows the frequency-dependent correlations of AME amplitude with individual maps. The strongest correlation occurs at DIRBE 240\,\textmu m ($\sim$1.3\,THz), near the peak of thermal dust emission for our sample, $1.8 \pm 0.2$\,THz. A secondary maximum appears at $\sim$20\,GHz, consistent with the AME peak. Correlations decline at CMB-dominated frequencies due to reduced AME contribution and lower signal-to-noise. The QUIJOTE frequencies, not shown, display reduced correlation coefficients due to lower signal-to-noise, precluding direct comparison with WMAP. The impact of noise is also evident in the \textit{Planck} 100\,GHz map, where the correlation is reduced due to scatter introduced by the CO correction.

Following the peak near the thermal dust maximum, the correlation coefficient declines steadily, reaching a minimum at the IRIS 60\,\textmu m band before rising again—an effect previously reported by \cite{Mateo2023,LambdaOrionis, Bell2019}. As thermal dust emission from large grains decreases above $\approx2$\,THz, the contribution from stochastically heated very small grains (VSGs), which are not in thermal equilibrium, becomes increasingly significant, particularly above $\approx3$\,THz ($100$\,\textmu m). This emission component typically dominates around the 60\,\textmu m band, where the observed correlation with AME is lowest \citep{Compiegne2011}. The dip in correlation between the thermal dust peak and the WISE 12\,\textmu m band suggests that VSGs are not co-located with AME, whereas PAHs appear to be well mixed with spinning dust—at least at $1^\circ$ resolution. The two points at 12\,\textmu m correspond to IRIS (\textit{red}) and WISE (\textit{violet}), with the reprocessed WISE\,12\,\textmu m map \citep{Meisner2014} displaying a significantly better correlation due to the removal of continuum emission. However, due to noise biases in Spearman coefficients and systematics in the WISE 12\,\textmu m data, the lower correlation coefficient relative to the thermal dust peak cannot be taken as evidence against PAHs as the primary carriers of spinning dust. Rather, it supports the interpretation that PAHs are better co-located with AME than the transiently heated VSG population.

Correlation coefficients for several tracers are also displayed on the \text{right} of Figure~\ref{fig:spearman}, with the correlation for the peak flux density of thermal dust emission being marginally better than $\Re_{\rm{d}}$, and comparable to the correlation coefficient of the DIRBE 240\,\textmu m band. In addition, a slightly stronger correlation is found through the product of $\tau_{\rm{353}}$ with a tracer for the fraction of PAHs, $f_{\rm PAH}$, to be discussed in Section~\ref{subsec:pahs}. The correlation coefficient with the dark gas map derived in \cite{dark_gas} (not shown) is $0.68\pm0.09$, though this is likely limited by systematics in its construction. In summary, the peak flux of thermal dust emission and dust radiance appear the best candidates for predicting the amplitude of AME at $1^\circ$ scales, potentially accurate to within $\approx30$\%.

\subsubsection{Correlation Between AME Peak Frequency and Thermal Dust Temperature}
\label{sec:peak_frequency_vs_temperature}

\begin{figure}
	\begin{center}
		\includegraphics[width=\columnwidth,angle=0]{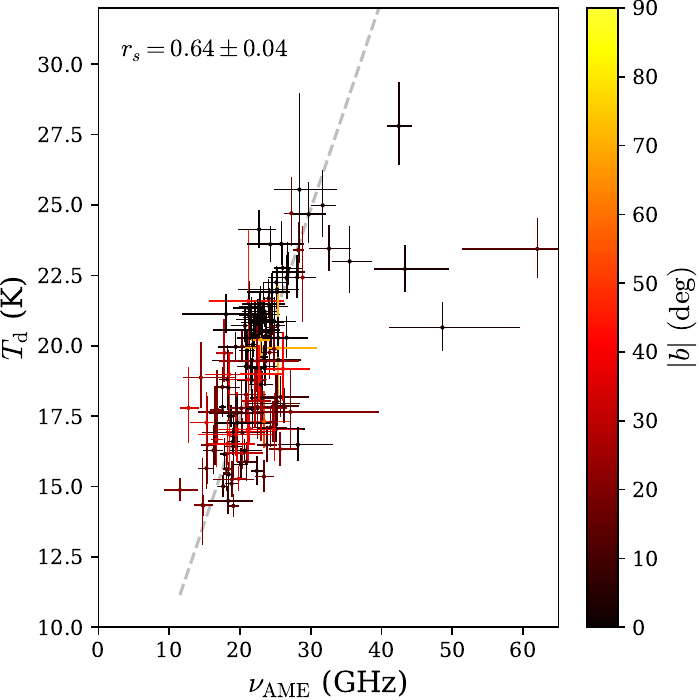}
		\caption{Increase in AME peak frequency with thermal dust temperature. Color encodes the absolute Galactic latitude of each source. A linear best-fit model is shown in \textit{gray}.}
		\label{fig:nu_ame_vs_td}
	\end{center}
\end{figure}

The AME peak frequency is observed to positively correlate with the temperature of large dust grains, as shown in Figure~\ref{fig:nu_ame_vs_td}. This observed trend is not reproduced by current theoretical models. In nearly all the idealized environments listed in Table~\ref{tab:spdust_vs_lognormal}, increasing the temperature or radiation field over the ranges relevant to our sample leads to only negligible changes in the predicted AME peak frequency, as shown in \cite{Ali-Hamoud2009} for a CNM template.

Specifically, varying the dust temperature between 10 and 30\,K produces nearly constant peak frequencies for most environments. Two notable exceptions are PDRs, which already predict very high peak frequencies ($\gtrsim$100\,GHz), and dark clouds, which exhibit a quasi-logarithmic increase, with predicted peak frequencies rising from $\sim29$ to $\sim37$\,GHz over the same temperature range. This latter trend is particularly relevant, as most of the sources in our sample are identified as containing dark clouds. This logarithmic behavior would manifest as an exponential function in Figure~\ref{fig:nu_ame_vs_td}; however, a linear fit provides a better match to the observed data, yielding $\chi^2_{\rm red} \approx 2.4$, a typical scatter of 14\%, and a slope of $1.35\pm0.09$\,GHz\,K$^{-1}$. 

Similarly, increasing the radiation field intensity $G_0$ while holding $T_{\rm d}$ constant produces a significant increase in peak frequency only for dark cloud environments, with values rising from $\sim30$ to $\sim80$\,GHz over the range of $G_0$ of the sample. When both $T_{\rm d}$ and $G_0$ are increased simultaneously by coupling the variables through Equation~\ref{eq:G0}, the models predict a quasi-linear rise in the peak frequency from 30 to $\sim80$\,GHz—about four times steeper per unit temperature than the trend observed in our data.

We interpret this disconnect between models and data as a consequence of the simplified treatment of environmental factors in theoretical frameworks, which employ fixed grain size distributions. The lack of dynamical modeling of how local conditions affect the grain size distribution likely prevents these models from capturing the full complexity of the phenomenon. In realistic environments, varying a single parameter—such as temperature or radiation field—cannot occur in isolation without also affecting other interrelated quantities like density, grain size distribution, and dipole moments. This interdependence underscores the need to integrate dust evolution and radiative transfer processes into spinning dust models in a self-consistent way.

While it is conceivable that a multi-parameter minimization algorithm—tuning three or four environmental variables simultaneously—could reproduce the observed trend, this approach would be computationally intensive and inefficient, and would risk producing solutions that lack a solid physical basis. A more effective strategy is to aim to reproduce the observed correlation from first principles, guided by a more realistic and self-consistent treatment of environmental interactions.

Because the proxy used for $G_0$ is closely derived from $T_{\rm d}$, the correlation between AME peak frequency and $G_0$ yields a similarly strong coefficient and scatter. Nonetheless, we hypothesize that the observed phenomenological correlation is ultimately driven by the local radiation field, which both raises the dust temperature and produces more rapid spinning dust rotation through a reduction in grain sizes, increased collisions and stronger radiative torques.

Although our sample is biased against sources with very low $\nu_{\rm AME}$ due to detectability limits, there is no comparable bias up to $\sim100$\,GHz. At peak frequencies $>100$\,GHz, faint AME would be increasingly difficult to detect, as it becomes embedded within the thermal dust emission and is excluded by the limits we imposed on $\nu_{\rm AME}$ to keep parameters physically reasonable. Nevertheless, the complete absence of peak frequencies near $100$\,GHz, expected in reflection nebulae and PDRs, indicates that they are intrinsically rare or suppressed rather than simply overlooked. A plausible explanation is reduced emissivity caused by photodissociation of small grains or rapid radiative energy loss from grains spinning at very high frequencies.

Four notable outliers with relatively large uncertainties in AME peak frequency are evident: the California Nebula ($\ell, b = 160.27, -12.36$), with $\nu_{\rm{AME}} = 62 \pm 13$\,GHz; Sharpless 280 ($208.80, -2.65$), at $49 \pm 9$\,GHz; W40 ($28.79, +3.49$), at $43 \pm 5$\,GHz; and IC 410 (the Tadpole Nebula, $173.56, -1.76$), at $42 \pm 2$\,GHz. All are predominantly H\textsc{ii} regions characterized by low AME contrast relative to the underlying free-free emission, contributing only 15--22\% of the total flux density at the peak frequency. While similarly low AME fractions are observed in a few sources with peak frequencies between 20 and 35\,GHz, none surpass 22\% at higher frequencies. For comparison, the median AME fraction at peak in the full sample is 64\%.

Importantly, these sources show significantly poorer fits when modeled with a purely free-free component, exhibiting clear residual bumps in the WMAP and \textit{Planck} frequency bands—strongly indicating the presence of an AME component. The observed AME amplitudes are also well above the level of calibration systematics in the data, and the derived peak frequencies are largely insensitive to the choice of background subtraction aperture—further supporting their reliability. Their deviation from the main population may reflect intrinsically distinct physical conditions or environmental factors. Follow-up studies at higher angular resolution will be critical for probing the origin of their elevated peak frequencies and understanding the conditions that give rise to these outliers. Furthermore, if extremely high peak frequency ($\gtrsim$100\,GHz) sources imply very low AME fractions (lower than $\sim5\%$), calibration uncertainties in the current datasets would present a bias against their detection.

We also test the analogy to Wien's displacement law for thermal dust emission (i.e., $\nu_{\rm{peak,d}} \propto T_{\rm d}^1$), which is effectively built into the parametric form of a modified-blackbody. Fitting Figure~\ref{fig:nu_ame_vs_td} with a power law $\nu_{\rm{AME}} \propto T_{\rm d}^\gamma$, we find a best-fit exponent of $\gamma = 1.17 \pm 0.08$. Excluding the four outliers reduces this to $\gamma = 1.10 \pm 0.07$, placing it marginally above unity. While this suggests that AME peak frequency increases more steeply with temperature than predicted by a direct Wien's-law-like proportionality, the proximity of the exponent to 1 implies that the analogy may still be physically informative. However, we do not expect AME to strictly follow a Wien's law relation. The grains responsible for thermal dust emission are not the same as those producing AME, and the fastest-spinning grains are likely in an out-of-thermal-equilibrium state due to rapid energy loss from radiative damping. The AME may originate from a ``Goldilocks'' region within the line of sight—where the local conditions favor efficient spinning dust emission—resulting in an effective temperature that differs from the beam-averaged $T_{\rm d}$. However, the fact that we see a relation implies that the local and beam-averaged $T_{\rm d}$ must be linked at large scales.

Nevertheless, due to sample selection effects and low-$\nu_{\rm{AME}}$ detection biases—as well as the presence of poorly understood outliers—this relation warrants re-evaluation in a full-sky component-separated dataset where completeness can be better quantified.

Overall, the positive correlation between AME peak frequency and thermal dust temperature likely reflects the influence of the radiation field on the grain size distribution—an interpretation that might explain prior observations \citep[e.g.,][]{Arce2019, LambdaOrionis}. The relatively tight relation observed for the majority of sources suggests that a fundamental, and potentially simple, physical mechanism governs the peak frequency of spinning dust, one that should eventually be reproduced by theoretical models that couple dust evolution, radiative transfer, and environmental conditions.

\subsubsection{PAHs}
\label{subsec:pahs}

\begin{figure*}
	\centering
	\begin{minipage}{0.40\textwidth}
		\centering
		\includegraphics[width=\linewidth]{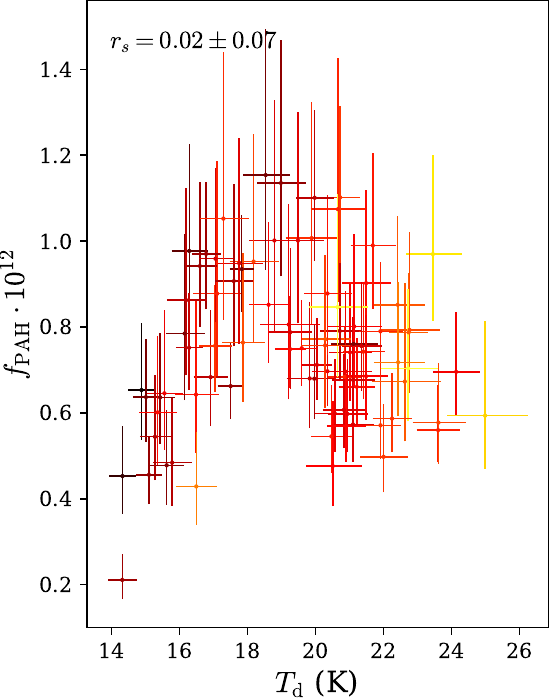}
	\end{minipage}%
	\hspace{0.8cm}
	\begin{minipage}{0.40\textwidth}
		\centering
		\includegraphics[width=\linewidth]{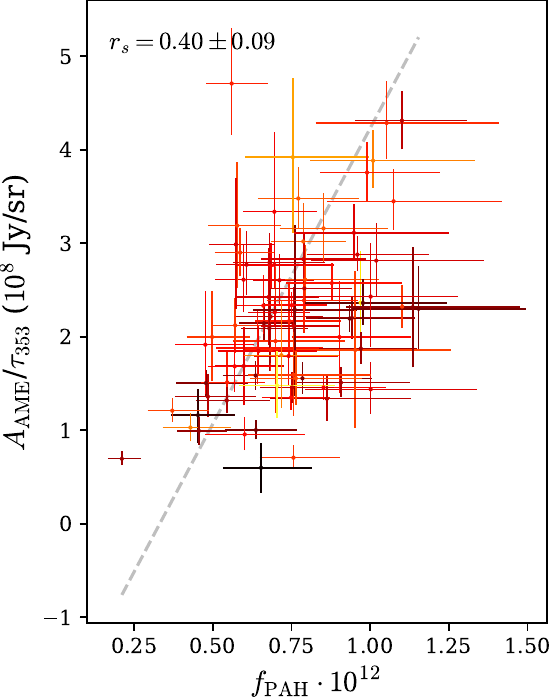}
	\end{minipage}%
	\hspace{0.8cm}
	\begin{minipage}{0.082\textwidth}
		\centering
		\includegraphics[width=\linewidth]{colorbar_nu_ame.pdf}
	\end{minipage}%
	\caption{\textit{Left}: Relation between the fraction of PAHs traced at 12\,\textmu m, \protect{$f_{\rm{PAH}}$}, and thermal dust temperature, showing an inflection point. \textit{Right}: AME emissivity relative to $\tau_{353}$ (a tracer of total column density) as a function of $f_{\rm{PAH}}$, with a linear best-fit model shown by the \textit{grey} dashed line. The color encodes AME peak frequency.}
	\label{fig:pahs}
\end{figure*}

Polycyclic aromatic hydrocarbons are planar organic molecules composed of multiple fused benzene rings, which are hexagonal and around $1$\,nm across. Their out-of-plane vibrational energy levels (bending modes) lead to distinct emission lines in the mid-IR \citep{Tielens2008}. Their ubiquity and small size make them strong candidates for spinning dust emission. The smallest PAHs, such as naphthalene, have an effective rotational diameter of $\approx0.72$\,nm and are believed to dominate the microwave emission, as they can reach the fastest rotational speeds and therefore radiate significant power. Nevertheless, in principle, any small, abundant molecule with a dipole moment could emit spinning dust radiation.

To investigate the link between PAHs and anomalous microwave emission, we follow the methodology of \citet{Hensley2016_pahs}, but take advantage of more accurate AME amplitude estimates enabled by new low-frequency data. As a tracer of PAH emission, we use the reprocessed WISE\,12\,\textmu m map \citep{Meisner2014}, which captures the strongest mid-infrared PAH features and is largely dominated by PAH emission. However, mid-IR observations at 12\,\textmu m are biased toward slightly larger PAHs than those thought to dominate spinning dust. Since rotational emission is highly sensitive to grain size, the smallest PAHs---only a few benzene rings in size---contribute disproportionately to the microwave signal, even if they represent only a minor fraction of the total PAH mass. Shorter-wavelength bands, such as DIRBE 3.5\,\textmu m, provide greater sensitivity to these smaller PAHs \citep{Chuss_2022}, but suffer from low sensitivity and strong stellar contamination, making them difficult to use without heavy processing. We therefore implicitly make the assumption that large and small PAHs are spatially co-located. Very recently, \citet{Sponseller2025} reported PAH–AME correlations using the 3.3\,\textmu m maps, though their conclusions are limited by two systematics evident in regions such as $\lambda$ Orionis \citep{LambdaOrionis}: (i) reliance on a parametric \textsc{Commander} AME separation lacking low-frequency data, biased by up to a factor of $\sim2$, and (ii) possible stellar contamination, since many point sources visible at 3.3\,\textmu m are absent in the 12\,\textmu m AKARI data shown in \citet{LambdaOrionis}. These challenges underscore why we adopt the reprocessed WISE 12\,\textmu m map as our main tracer, while highlighting the need for cleaner AME templates (Hoerning et al., in prep.) and future spectroscopy from SPHEREx \citep{SPHEREx2024} to robustly isolate PAH emission at 3.3\,\textmu m.

Unlike the full-sky pixel-based analysis in \citet{Hensley2016_pahs}, we perform aperture photometry on individual sources, meaning that we cannot use the fine-grained mask used in their study. Instead, we manually mask nine prominent stripes caused by Moon contamination in the WISE\,12\,\textmu m map, affecting $\approx5$\% of the sky. We then perform photometry at $1^\circ$ resolution on each source in our sample to obtain the 12\,\textmu m flux density, from which we derive a quantity proportional to the PAH fraction, $f_{\rm{PAH}} \propto S_{\rm{12\,\textmu m}} / \Re_{\rm d}$, following \citet{Hensley2016_pahs}. The dominance of PAHs in the reprocessed WISE 12\,\textmu m data is supported by the stronger correlation in Figure~\ref{fig:spearman} relative to the IRIS 12\,\textmu m map, which isn't corrected for unrelated emission mechanisms.

% Temperature dependence
Across our sample, this PAH fraction tracer varies by a factor of $\sim6$, implying that the fractional abundance of PAHs differs by less than an order of magnitude among the observed regions, as shown in Figure~\ref{fig:pahs}. Despite the relative large uncertainties in deriving $f_{\rm PAH}$, a clear relationship with $T_{\rm d}$ emerges in the \textit{left} panel. The low Spearman rank correlation coefficient, consistent with zero, is unsurprising given the non-monotonic nature of the relationship, to which the Spearman statistic is largely insensitive. Notably, $f_{\rm{PAH}}$ exhibits a peak around $T_{\rm d} \sim 18$--$20$\,K. The reduction in the fraction of PAHs at both high and low temperatures might be indicative of two physical processes: (1) At temperatures below 18\,K, the decrease in the observed gas-phase PAH fraction might be attributable to the adsorption or freeze-out of PAH molecules onto the surfaces of the much larger cold dust grains in dense molecular clouds \citep{Michoulier2018_pah_adsorption}. This process depletes the population of free PAHs in the gas phase, inhibiting vibrational modes and therefore IR emission. (2) Conversely, at temperatures above this 20\,K peak, the reduction in PAH fraction may be caused by the increased efficiency of PAH destruction through photodissociation by more intense UV radiation fields, which correlate with higher dust temperatures. Laboratory and modeling studies suggest that small PAHs are readily dissociated under such conditions, while larger PAHs are significantly more stable \citep[e.g.,][]{Leger1984, Montillaud2013}. This parabolic behavior is not mirrored in the $T_{\rm d}$–$\Re_{\rm d}$ relation, which shows a broadly positive trend with significant scatter, suggesting the observed maximum is not a sample-selection artifact.

% Correlation with tau353 times fpah
To examine the link between AME and PAHs, we compare the AME amplitude $A_{\rm AME}$ against both $\tau_{353}$ and $\tau_{353} \cdot f_{\rm PAH}$, the latter serving as a proxy for the column density of PAHs. This connection was sought in \cite{Hensley2016_pahs}, reporting a lack of correlation between AME and PAH emission and interpreting this as evidence against PAHs as AME carriers. However, their analysis also relied on AME amplitudes derived from the outdated \cite{Planck2015_X} \textsc{Commander} reduction, making the separation of AME from free-free emission highly degenerate. We argue that the lack of correlation in their study may be primarily driven by these systematics, which can easily mask underlying trends.

In our analysis, even though the product $\tau_{353} \cdot f_{\rm PAH}$ introduces additional noise, we obtain a correlation coefficient of $0.933 \pm 0.007$—slightly higher ($1.0\,\sigma$) than the $0.924 \pm 0.006$ found for $\tau_{353}$ alone, as seen in Figure\,\ref{fig:spearman}. The correlation also yields a best-fit power-law index of $1.02 \pm 0.13$, consistent with linearity. If $f_{\rm PAH}$ were uncorrelated with $A_{\rm AME}$, introducing it should degrade the correlation. We test this by scrambling the $f_{\rm PAH}$ values and repeating the analysis; the resulting correlations drop systematically to $r_s = 0.90 \pm 0.01$ on average, significantly ($2.7\,\sigma$) below the observed value. This is further supported by the direct correlation between $f_{\rm PAH}$ and AME emissivity (Figure~\ref{fig:pahs}, right panel), where a statistically significant positive trend is observed despite large scatter.

Taken together, these results provide circumstantial evidence for a connection between PAHs and AME, with spatial co-location at $1^\circ$ scales. That said, our tracer is imperfect: the WISE\,12\,\textmu m band suffers from systematic uncertainties and may include non-PAH contributions. Moreover, $f_{\rm PAH}$ traces only the subset of PAHs emitting in the 12\,\textmu m band, whereas rotational emissivity is expected to depend sensitively on molecular size, charge state, and structure. Different PAH sub-populations (e.g., neutral vs.\ ionized, small vs.\ large) therefore contribute unequally to AME, but these cannot be disentangled with broadband WISE data. In addition, \citet{Hensley_2022_cnm_pah} argue that PAHs are systematically depleted in warmer gas, and that their IR emissivity per hydrogen atom varies with environment---possibly explaining why the mid-IR--AME correlation is weaker than expected. All of these effects would obscure the relation between PAHs and AME.

% Conclusions
A definitive answer will require high-resolution spectroscopy to isolate PAH features from continuum and line emission emission mechanisms and to compare AME correlations against both PAH and thermal dust tracers. Such analyses would be particularly valuable at higher angular resolution, as the small spatial offsets seen at $1^\circ$ between the PAH and AME (e.g. Fig.\,\ref{fig:multifrequency_view}) could be amplified, enabling better identification of the specific carriers responsible for AME.

\subsubsection{Systematic Deviations from a log-Gaussian Model}
\label{subsec:ame_deviations}

\begin{figure*}[!htbp]
	\begin{center}
		\includegraphics[width=0.93\textwidth,angle=0]{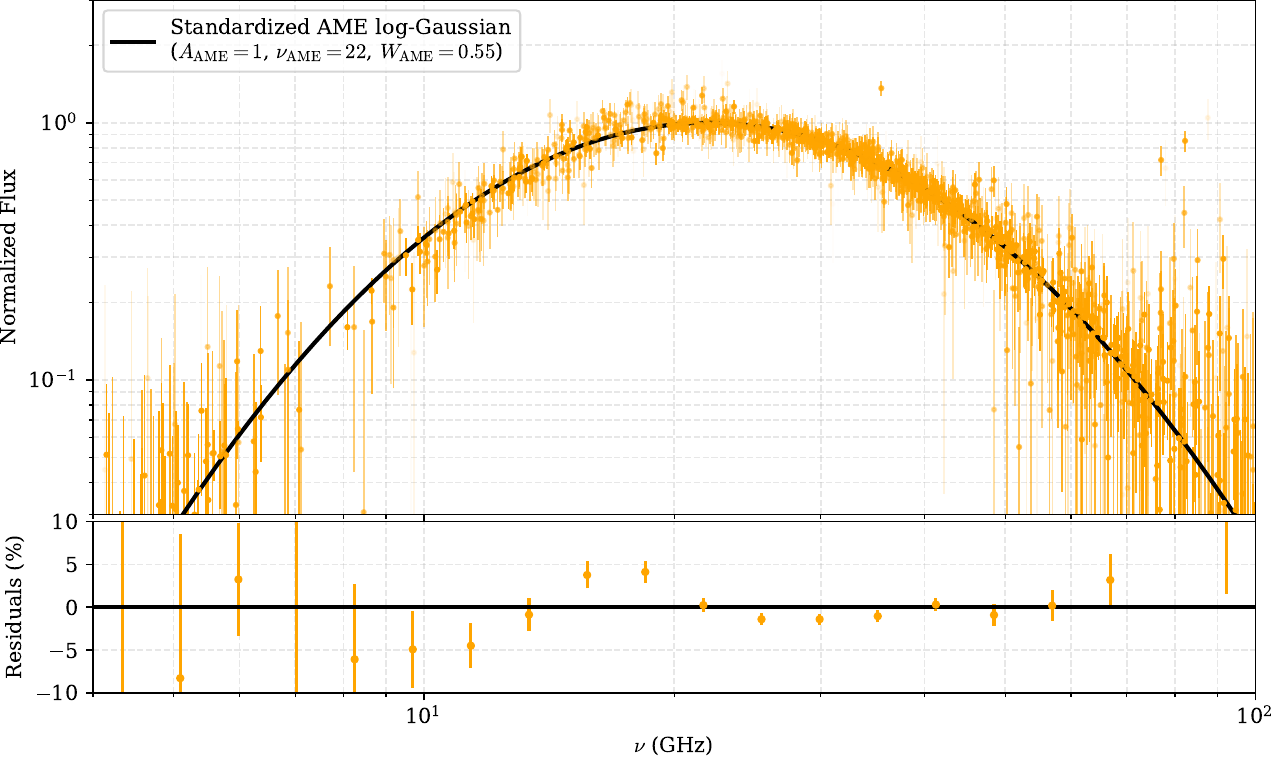}
		
		\vspace{0.1cm}
		
		\includegraphics[width=0.93\textwidth,angle=0]{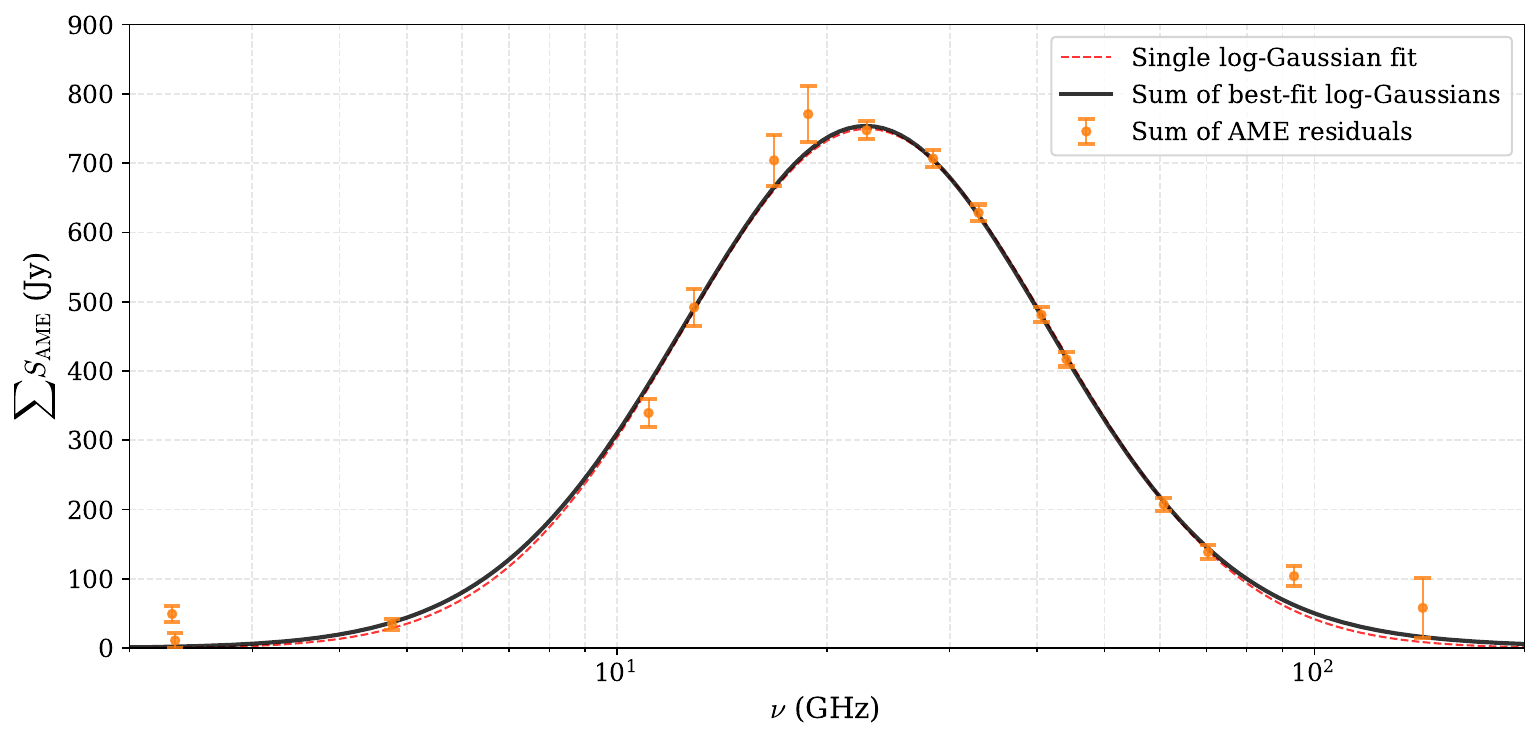}
		\caption{Residuals of the fitted data after subtracting all non-AME components. \textit{Top panel}: residuals relative to a normalized log-Gaussian distribution. A transparency gradient is applied to de-emphasize points with large error bars and enhance visual clarity. The lower sub-panel shows the binned, weighted mean residuals in percentage units relative to the standardized model, with bins evenly spaced in logarithmic scale. \textit{Bottom panel}: total fitted AME flux densities (\textit{orange}) in a linear \textit{y}-scale, compared with the sum of best-fit log-Gaussian components (\textit{black}). The sum only includes the 85 sources with simultaneous QUIJOTE and C-BASS coverage. A single log-Gaussian distribution fitted to the AME residuals is shown as a \textit{red} dashed line, with $\nu_{\mathrm{AME}} = 22.8$ GHz, $W_{\mathrm{AME}} = 0.62$ and $\chi^2_r = 3.1$.}
		\label{fig:ame_deviations}
	\end{center}
\end{figure*}

As noted in Section~\ref{sec:AME}, the log-Gaussian approximation for the AME spectrum differs from physically motivated spinning dust templates. Consequently, if spinning dust emission is the true underlying process, we might expect systematic deviations of the photometry from the log-Gaussian best fit. To search for such deviations in the AME residuals, we use two approaches: (i) we normalize the photometric data relative to a single, standardized log-Gaussian function, and (ii) we examine the spectrum of the summed residual AME flux densities. The results of both methods are shown in Figure~\ref{fig:ame_deviations}.

To standardize the photometric data relative to a reference log-Gaussian model, we first subtract the best-fit contributions from other foreground components. The residual AME emission for each source is then rescaled linearly so that its log-Gaussian parameters match a common reference shape and normalization: we adopt an amplitude of $A_{\rm target} = 1$\,Jy purely for convenience, along with a peak frequency $\nu_{\rm target} = 22$\,GHz and a width $W_{\rm target} = 0.55$. In this context, primed variables ($\nu'$, $S'$) refer to the transformed quantities, while unprimed variables represent the original photometric measurements. The transformation of the frequencies is given by:

\begin{equation}
	\frac{\nu'}{\nu} = \frac{\nu_{\rm target}}{\nu_{\rm AME}}.
\end{equation}

For flux densities, using the log-Gaussian function defined in Equation~\ref{eq:lognormal}, the transformation can be expressed as the ratio:
\begin{equation}
	\frac{S'(\nu')}{S(\nu)} = \frac{S_{\rm AME}(\nu'; A_{\rm target}, \nu_{\rm target}, W_{\rm target})}{S_{\rm AME}(\nu; A_{\rm AME}, \nu_{\rm AME}, W_{\rm AME})}.
\end{equation}

This scaling adjusts both the amplitude and shape of the distribution. Uncertainties in flux density are scaled by the same factor. This transformation allows direct comparison of the systematic deviations of the photometry from the fitted model.

The standardized residuals (top panel of Figure~\ref{fig:ame_deviations}) display the normalized deviations for all sources. Given that most sources in our sample peak near $\sim 22$\,GHz, one might expect a consistent excess of flux below 10\,GHz—corresponding to the slowly rising low-frequency tail of AME—if deviations from the log-Gaussian shape were present. However, this frequency range is poorly constrained due to the gap between C-BASS and the lowest QUIJOTE channel, which is visible in both figures.

Although the standardized log-Gaussian plot (\textit{top panel}) displays a large number of points, only $\approx13$ photometric data points are fitted per source. These are not independent across the sample, as they are all residuals relative to each source's fitted model. As such, the residuals are influenced both by observational systematics (e.g., calibration uncertainties) and by potential mismodeling of the AME spectrum, making it difficult to conclusively isolate intrinsic deviations from the log-Gaussian shape. Consequently, with such a limited number of points per source, the model can nearly always reproduce the data, potentially masking any underlying deviations. This is especially true in the $5$--$11$\,GHz range, where additional data near $\approx7$\,GHz might be required to robustly identify low-frequency excesses.

The second approach examines the summed residual AME flux densities for the 85 sources with both C-BASS and QUIJOTE coverage (bottom panel of Figure~\ref{fig:ame_deviations}). Remarkably, the summed spectrum of fitted AME components is well modeled by a single log-Gaussian function (\textit{red} dashed line). This suggests that any individual source deviations from the log-Gaussian shape are either small relative to observational uncertainties or tend to average out across the sample, supporting the validity of the log-Gaussian parametrization as a reasonable empirical description of the ensemble AME behavior.

Both analyses reveal consistent features: a minor excess appears between 17 and 19\,GHz with a corresponding deficit near 11\,GHz. This feature aligns with the deviations discussed in Section~\ref{sec:reich_calibration_factor}, which we attribute to residual atmospheric signal in the higher QUIJOTE bands rather than intrinsic AME spectral deviations. A small excess is also seen just below $100$\,GHz, which might correspond to an unaccounted component or mismodeling of the foregrounds.

Overall, the analysis shows that the log-Gaussian model provides a remarkably good approximation across the available data, serving as a reasonable empirical description of the current datasets given their spectral coverage and systematics. To quantitatively assess whether theoretical spinning dust templates fit the data better, dedicated simulations would be required—particularly ones in which the width of the spinning dust spectrum is varied based on physical considerations, to reflect the broader distributions observed. We leave this as an illustrative case study of how such systematic deviations might be identified and interpreted in future work.

%%%%%%%%%%%%%%%%%%%%%%%%%%%%%%%%%%%%%%%%%%%%%%%%%%%%%%%
%%%%%%%%%%%%%%%%%%%%%%%%%%%%%%%%%%%%%%%%%%%%%%%%%%%%%%%

\section{Conclusions}
\label{sec:conclusions}

This paper represents the most largest sample of compact AME sources to date, with AME amplitudes, peak frequencies and widths derived with improved accuracy thanks to the merger of new low-frequency datasets, namely S-PASS, \mbox{C-BASS} and QUIJOTE. We summarize the paper into 5 key findings:

\begin{enumerate}
	\item The observed widths of AME spectra are generally broader than those predicted by spinning dust models that assume a single ISM phase. This suggests contributions from multiple spinning dust phases within most of the sources studied or along the line of sight, and may also reflect that theoretical models do not fully capture the range of grain sizes present in the ISM. The narrowest observed widths align with theoretical predictions for idealized templates, providing a new piece of evidence supporting the spinning dust hypothesis.
	\item AME amplitude is most effectively traced by thermal dust emission. Dust radiance and the peak flux of thermal dust exhibit the tightest correlations, with typical scatter around 30\%, outperforming dust optical depth as predictors. The sublinear scaling of AME radiance with thermal dust radiance suggests that AME becomes less efficient in denser, more shielded environments, consistent with depletion of small grains via coagulation onto larger ones.
	\item We find a tight positive correlation between the AME peak frequency and the temperature of large dust grains, which current theoretical models fail to reproduce. The observed trend likely reflects the role of the local radiation field on grain properties, suggesting that spinning dust models must incorporate self-consistent radiative transfer and dust evolution treatment of the environmental parameters. Notably, we observe no sources with the extremely high peak frequencies expected for reflection nebulae and PDRs, suggesting the presence of a suppressive mechanism or inherent detection bias such as extremely low AME contrast or emissivity.
	\item PAH tracers exhibit characteristic environmental dependencies—likely driven by grain growth and photodissociation—and show a statistically significant but scattered correlation with AME emissivity. While systematic uncertainties in mid-IR tracers remain, our results support a physical connection between PAHs and AME, consistent with spatial co-location on degree scales.
	\item The log-Gaussian model provides a good empirical approximation for the AME spectrum given the current spectral coverage and systematics of the available data. 
\end{enumerate}

\noindent Going forward, this work highlights several directions for further study. The relationships identified across our sample could inform improved Python Sky Model \citep{Thorne_2017} templates for spinning dust, particularly for better AME amplitude estimation, improving the separation of low-frequency components. Ongoing work (Hoerning et al., in prep.) and forthcoming 8--10\,GHz observations from CGEM \citep{VillalbaGonzalez2023} and improved 10--20\,GHz observations with QUIJOTE MFI2 \citep{MFI2_eng} will further refine these templates. To validate and extend these findings, a new \textsc{Commander} component separation incorporating \mbox{C-BASS}, S-PASS, and QUIJOTE should be conducted to test the observed trends and improve sample completeness. The few sources with unusually high peak frequencies deserve detailed follow-up studies, as they may help isolate the physical conditions responsible for producing such high frequencies. More broadly, extending low-frequency datasets to full-sky coverage would be highly beneficial, particularly given that the absence of QUIJOTE coverage in the southern hemisphere visibly reduces detections in our study. Additionally, S-PASS would benefit from complete beam and sidelobe characterization, which is crucial for studying compact sources in both intensity and polarization at large angular scales. While we found no clear deviations from the log-Gaussian model, simulations could test whether current data are sensitive enough to detect asymmetries at lower frequencies and determine what level of systematic control would be required. Ultimately, establishing a definitive link between AME and specific molecules will require clean spectroscopic separation of PAHs and direct comparison with AME at finer angular scales. Upcoming SPHEREx all-sky survey data releases \citep{SPHEREx2024}, which probe the 3.3\,\textmu m PAH feature, will be particularly valuable in this regard.

Finally, theoretical models must evolve to explain the trends observed, including the broader spectral widths, the apparent absence of extremely high peak frequency sources, and the correlations with AME amplitude and peak frequency. This may require incorporating dynamical environmental conditions in spinning dust models. Data-driven tools such as symbolic regression could help uncover deeper multivariate patterns beyond simple pairwise correlations, providing insights that inform first-principles modeling.

%%%%%%%%%%%%%%%%%%%%%%%%%%%%%%%%%%%%%%%%%%%%%%%%%%%%%%%
%%%%%%%%%%%%%%%%%%%%%%%%%%%%%%%%%%%%%%%%%%%%%%%%%%%%%%%
\begin{acknowledgements}
The authors thank Jean-Philippe Bernard for providing the dark gas map used in the AME correlations, and Brandon Hensley for his insightful discussions and valuable input on theoretical aspects of this work. We also express our gratitude to David Chuss for organizing the ``AME and its Connections to the Interstellar Medium'' workshop at Villanova University, which greatly facilitated the exchange and development of ideas presented in this work, and for his helpful feedback on this paper. We further thank Alan Kogut and Ari Cukierman for useful discussions during the workshop. We also thank the anonymous referee for for constructive comments that improved the clarity of this work.
We thank the staff of the Teide Observatory for invaluable assistance in the commissioning and operation of QUIJOTE. The QUIJOTE experiment is being developed by the Instituto de Astrofisica de Canarias (IAC), the Instituto de Fisica de Cantabria (IFCA), and the Universities of Cantabria, Manchester and Cambridge.
Partial financial support was provided by the Spanish Ministry of Science and Innovation 
under the projects AYA2007-68058-C03-01, AYA2007-68058-C03-02, AYA2010-21766-C03-01, AYA2010-21766-C03-02, AYA2014-60438-P, ESP2015-70646-C2-1-R, AYA2017-84185-P, ESP2017-83921-C2-1-R, PGC2018-101814-B-I00, PID2019-110610RB-C21, PID2020-120514GB-I00, 
IACA13-3E-2336, IACA15-BE-3707, EQC2018-004918-P, the Severo Ochoa Programs SEV-2015-0548 and CEX2019-000920-S, the Maria de Maeztu Program MDM-2017-0765, and by the Consolider-Ingenio project CSD2010-00064 (EPI: Exploring the Physics of Inflation). We acknowledge support from the ACIISI, Consejeria de Economia, Conocimiento y Empleo del Gobierno de Canarias and the European Regional Development Fund (ERDF) under grant with reference ProID2020010108, and 
Red de Investigaci\'on RED2022-134715-T funded by MCIN/AEI/10.13039/501100011033.
This paper uses pre-publication data from the C-BASS project which is a collaboration between Oxford and Manchester Universities in the U.K., the California Institute of Technology in the U.S., Rhodes University, UKZN and the South African Radio Astronomy Observatory in South Africa, and the King Abdulaziz City for Science and Technology (KACST) in Saudi Arabia.
The work at Oxford was supported by funding from STFC, the Royal Society and the University of Oxford. The work at the California Institute of Technology and Owens Valley Radio Observatory was supported by National Science Foundation (NSF) awards~AST-0607857, AST-1010024, AST-1212217, and AST-1616227, and by NASA award NNX15AF06G. The work at Manchester was supported by several STFC Consolidated Grants and a UKSA grant (ST/Y005945/1) funding LiteBIRD foreground activities. A.C. Taylor and M.E. Jones also acknowledge support from the Horizon Europe Project RadioForegroundsPlus (GA 101135036) which is supported in the UK by UKRI grant number 10101603.
This project has received funding from the Horizon Europe research and innovation program under 
GA 101135036 (RadioForegroundsPlus).
CD and SEH acknowledge funding from an STFC (Consolidated Grant ST/P000649/1) and UKSA (LiteBird UK ST/Y005945/1).
FP acknowledges support from the MICINN under grant number PID2022-141915NB-C21.
DH acknowledges support from the MICINN under grant number PID2022-140670NA-I00.
GAH acknowledges the funding from the Dean's Doctoral Scholarship by the University of Manchester.
MFT acknowledges support from the Enigmass+ research federation (CNRS, Université Grenoble Alpes, Université Savoie Mont-Blanc).
JC acknowledges support from the Fundación Occident and the Instituto de Astrofísica de Canarias under the Visiting Researcher Programme 2022-2025 agreed between both institutions.
EMG acknowledges support from Plan Complementario AstroHEP funded by the "European Union NextGenerationEU/PRTR" and the Government of the Autonomous Community of Cantabria.
\end{acknowledgements}

%\section*{Data Availability}

%%%%%%%%%%%%%%%%%%%%%%%%%%%%%%%%%%%%%%%%%%%%%%%%%%%%%%%
%%%%%%%%%%%%%%%%%%%%%%%%%%%%%%%%%%%%%%%%%%%%%%%%%%%%%%%

%%%%%%%%%%%%%%%%%%%%

\bibliographystyle{aa}
% Use the LaTeX power, use bibtex properly.
\bibliography{paper_refs}

%%%%%%%%%%%%%%%

\begin{appendix}

\section{AME Source Catalog}
\label{sec:catalogue}

The full catalog of sources and best-fit parameters is presented in Table~\ref{tab:source_catalogue}.
\input{source_catalogue.tex}

\end{appendix}

%%%%%%%%%%%%%%%%%%%%%%%%%%%%%%%%%%%%%%%%%%%%%%%%%%%%%%%
%%%%%%%%%%%%%%%%%%%%%%%%%%%%%%%%%%%%%%%%%%%%%%%%%%%%%%%

\label{lastpage}

\end{document}

%% file: datasets.tex
\begin{table*}
	\caption{Summary of multi-frequency data, where $\nu$ is the effective frequency, $\Delta\nu$ denotes bandwidth, $\delta$ is declination in \mbox{J2000} celestial coordinates and $\sigma_{\rm cal}$ is the effective calibration uncertainty. The effective calibration uncertainty accounts for scale-calibration uncertainties due to incomplete beam characterization, which primarily affect the $408\,\protect\si{\mega\hertz}$ and $1.42\,\protect\si{\giga\hertz}$ maps. The FWHM and sky coverage values correspond to the native survey properties prior to convolution to $1^\circ$. $^\dagger$\,Frequencies excluded from fitting due to residual CO contamination. \label{tab:surveys}}
	\centering
	\resizebox{0.95\textwidth}{!}{
		\begin{tabular}{lcccccl}
			\toprule
			\multirow{ 2}{*}{\textbf{Survey}} & \boldmath{$\nu_0$} & \boldmath{$\Delta\nu$} & \textbf{FWHM} & \textbf{Sky} & \boldmath{$\sigma_{\rm cal}$} & \multirow{ 2}{*}{\textbf{Reference}} \\
			& \textbf{(GHz)} & \textbf{(GHz)} & \textbf{(arcmin)} & \textbf{Coverage} & \textbf{(\%)} & \\  \midrule
			Haslam & 0.408 & 0.0035 & 51 & Full Sky & 30 & \cite{Remazeilles2015} \\
			Stockert/Villa-Elisa & 1.420 & 0.014 & 35 & Full Sky & 30 & \cite{Reich2001} \\
			\mbox{S-PASS} & 2.303 & 0.168 & 8.9 & $\delta\lesssim\ang{-1}$ & 5 & \cite{spass_release} \\
			HartRAO & 2.326 & 0.040 & 20 & $\ang{-83}<\delta<\ang{12}$ & 10 & \cite{Jonas1998} \\
			\mbox{C-BASS} & 4.76 & 0.49 & 44 & $\delta>-15.\!^{\circ}6$ & 5 & Taylor et al. (in prep.) \\
			\mbox{QUIJOTE} MFI & 11.1 & 2.2 & 55 & $\delta\gtrsim\ang{-30}$ & 5 & \cite{mfiwidesurvey} \\
			\mbox{QUIJOTE} MFI & 12.9 & 2.2 & 56 & $\delta\gtrsim\ang{-30}$ & 5 & \cite{mfiwidesurvey} \\
			\mbox{QUIJOTE} MFI & 16.8 & 2.2 & 39 & $\delta\gtrsim\ang{-30}$ & 5 & \cite{mfiwidesurvey} \\
			\mbox{QUIJOTE} MFI & 18.8 & 2.3 & 40 & $\delta\gtrsim\ang{-30}$ & 5 & \cite{mfiwidesurvey} \\
			\textit{WMAP} K & 22.8 & 5.5 & 49 & Full Sky & 1 & \cite{WMAP2013_RESULTS} \\
			\textit{Planck} LFI & 28.4 & 6 & 32.4 & Full Sky & 1 & \cite{Planck2018_I} \\
			\textit{WMAP} Ka & 33.0 & 7 & 40 & Full Sky & 1 & \cite{WMAP2013_RESULTS} \\
			\textit{WMAP} Q & 40.6 & 8.3 & 31 & Full Sky & 1 & \cite{WMAP2013_RESULTS} \\
			\textit{Planck} LFI & 44.1 & 8.8 & 27.1 & Full Sky & 1 & \cite{Planck2018_I} \\
			\textit{WMAP} V & 60.8 & 14 & 21 & Full Sky & 1 & \cite{WMAP2013_RESULTS} \\
			\textit{Planck} LFI & 70.4 & 14 & 13.3 & Full Sky & 1 & \cite{Planck2018_I} \\
			\textit{WMAP} W & 93.5 & 21 & 13 & Full Sky & 1 & \cite{WMAP2013_RESULTS} \\
			\textit{Planck} HFI & 100\,$^\dagger$ & 33 & 9.7 & Full Sky & 1 & \cite{Planck2018_I} \\
			\textit{Planck} HFI & 143 & 47 & 7.3 & Full Sky & 1 & \cite{Planck2018_I} \\
			\textit{Planck} HFI & 217\,$^\dagger$ & 72 & 5.0 & Full Sky & 1 & \cite{Planck2018_I} \\
			\textit{Planck} HFI & 353 & 100 & 4.8 & Full Sky & 1.3 & \cite{Planck2018_I} \\
			\textit{Planck} HFI & 545 & 180 & 4.7 & Full Sky & 6.0 & \cite{Planck2018_I} \\
			\textit{Planck} HFI & 857 & 283 & 4.3 & Full Sky & 6.4 & \cite{Planck2018_I} \\
			DIRBE 240 & 1249 & 500 & 42 & Full Sky & 13.5 & \cite{Hauser1998} \\
			DIRBE 140 & 2141 & 620 & 38 & Full Sky & 10.6 & \cite{Hauser1998} \\
			DIRBE 100 & 2998 & 970 & 39 & Full Sky & 11.6 & \cite{Hauser1998} \\
			IRAS 100 & 2998 & $\sim1100$ & 4.7 & Full Sky & 13.5 & \cite{Miville2005IRIS} \\  
			\midrule
			\multicolumn{7}{l}{\textbf{Other Datasets and Tracers}} \\
			IRAS 60 & $60\,\si{\micro\meter}$ & $40\,\si{\micro\meter}$ & 3.6 & Full Sky & 10.4 & \cite{Miville2005IRIS} \\
			IRAS 25 & $25\,\si{\micro\meter}$ & $11\,\si{\micro\meter}$ & 3.5 & Full Sky & 15.1 & \cite{Miville2005IRIS} \\
			IRAS 12 & $12\,\si{\micro\meter}$ & $6.5\,\si{\micro\meter}$ & 3.5 & Full Sky & 5.1 & \cite{Miville2005IRIS} \\
			WISE 12 & $12\,\si{\micro\meter}$ & $\sim10\,\si{\micro\meter}$ & 0.1 & Full Sky & -- & \cite{Meisner2014} \\
			CO Emission & -- & -- & -- & Full Sky & -- & \cite{Ghosh2024} \\
			Dark Gas & -- & -- & -- & Full Sky & -- & \cite{dark_gas} \\
			\bottomrule
	\end{tabular}}
\end{table*}

%% file: source_catalogue.tex
\onecolumn
    \begin{landscape}
	\renewcommand{\arraystretch}{0.5} % Slightly increased for better readability

	% Center the table with appropriate margins
	\setlength{\LTleft}{0pt plus 1fil}
	\setlength{\LTright}{0pt plus 1fil}

	% Calculate appropriate caption width based on table width
	\setlength{\LTcapwidth}{1.0\linewidth} % Slightly less than full line width

	\fontsize{8.5}{15.0}\selectfont
	\begin{longtable}{@{\extracolsep{\fill}}l@{\hspace{0.8em}}c@{\hspace{0.8em}}c@{\hspace{0.8em}}c@{\hspace{0.8em}}c@{\hspace{0.8em}}c@{\hspace{0.8em}}c@{\hspace{0.8em}}c@{\hspace{0.8em}}c@{\hspace{0.8em}}c@{\hspace{0.8em}}c@{\hspace{0.8em}}c@{\hspace{0.8em}}p{7.5cm}@{}}
    \caption{Catalog of 144 AME regions with best-fit parameters, identified by their central coordinates. The primary physical environment of each source is shown in parentheses: DC for dark cloud, MC for molecular cloud, RNe for reflection nebula and FIR for infrared source. A \textit{plus} sign indicates the presence of multiple additional environments of the same type, none of which are dominant. In pre-existing sources whose central coordinates were adjusted, the listed name corresponds to the new coordinates and the original name appears in the final column as a superscript. Sources from the Dobashi dark cloud catalogue are denoted by DB. $\theta$ is the radius of the primary aperture; all other variables are defined in Section\,\ref{sec:foreground_modelling}. $^\dagger$\,Regions that include a synchrotron component.}
	\label{tab:source_catalogue} \\
	\toprule
	\multirow{2}{*}{Name} & $\theta$ & $A_{\mathrm{AME}}$ & $\nu_{\mathrm{AME}}$ & $W_{\mathrm{AME}}$ & $\sigma_{\mathrm{AME}}$ & $\delta T_{\mathrm{CMB}}$ & $\rm{EM}$ & $\tau_{\rm{353}}$ & $T_{\rm{d}}$ & $\beta_{\rm d}$ & $\chi^2_{\rm r}$ & \multicolumn{1}{l}{Main Sources (+Environment)} \\
	& (deg) & (Jy) & (GHz) &  &  & ($\mu\mathrm{K}_{\mathrm{CMB}}$) & (pc cm$^{-6}$) & $\cdot10^{6}$  &  (K) &  &  & \\
	\midrule
	\endfirsthead

	\caption[]{\parbox{\LTcapwidth}{\raggedright Catalog of 144 AME regions with best-fit parameters -- continued}} \\
	\toprule
	\multirow{2}{*}{Name} & $\theta$ & $A_{\mathrm{AME}}$ & $\nu_{\mathrm{AME}}$ & $W_{\mathrm{AME}}$ & $\sigma_{\mathrm{AME}}$ & $\delta T_{\mathrm{CMB}}$ & $\rm{EM}$ & $\tau_{\rm{353}}$ & $T_{\rm{d}}$ & $\beta_{\rm d}$ & $\chi^2_{\rm r}$ & \multicolumn{1}{l}{Main Sources (+Environment)} \\
	& (deg) & (Jy) & (GHz) &  &  & ($\mu$K) & (pc cm$^{-6}$) & $\cdot10^{6}$  &  (K) &  &  & \\
	\midrule
	\endhead

	\midrule
	\multicolumn{13}{r}{{Continued on next page}} \\
	\endfoot

	\bottomrule
	\endlastfoot

    G008.22$-$08.53 & 1.0 & $1.3 \pm 0.6$ & $25.7 \pm 3.8$ & $0.59 \pm 0.25$ & 2.3 & $-4 \pm 4$ & $6 \pm 8$ & $7.5 \pm 0.5$ & $18.2 \pm 0.7$ & $1.56 \pm 0.10$ & 1.7 & LDN 248 (DC) \\ 
    G008.37$-$00.10 & 1.0 & $43.4 \pm 16.6$ & $24.8 \pm 3.0$ & $0.58 \pm 0.21$ & 2.6 & $2 \pm 30$ & $1480 \pm 540$ & $232.4 \pm 16.6$ & $20.9 \pm 0.8$ & $1.79 \pm 0.08$ & 0.51 & DB 391/402/384, RAFGL 5436 \,\textsuperscript{G008.51$-$00.31}\, (DC, H\textsc{ii})\,$\dagger$ \\ 
    G010.36$-$00.21 & 1.0 & $50.1 \pm 7.7$ & $23.2 \pm 1.5$ & $0.39 \pm 0.07$ & 6.5 & $-10 \pm 29$ & $2666 \pm 83$ & $311.1 \pm 17.1$ & $20.9 \pm 0.6$ & $1.84 \pm 0.05$ & 0.92 & SRBY 15, DB 454, W31 \,\textsuperscript{G010.19$-$00.32}\, (DC, MC, H\textsc{ii}) \\ 
    G010.84$-$02.59 & 1.0 & $12.9 \pm 2.9$ & $21.3 \pm 1.5$ & $0.53 \pm 0.15$ & 4.5 & $10 \pm 22$ & $200 \pm 60$ & $37.3 \pm 3.5$ & $21.1 \pm 1.2$ & $1.63 \pm 0.12$ & 0.98 & GGD 27 (MC)\,$\dagger$ \\ 
    G012.80$-$00.19 & 1.0 & $54.5 \pm 11.6$ & $24.2 \pm 2.0$ & $0.42 \pm 0.10$ & 4.7 & $-2 \pm 30$ & $2788 \pm 115$ & $376.5 \pm 21.9$ & $20.5 \pm 0.6$ & $1.85 \pm 0.05$ & 0.65 & W33 (H\textsc{ii} + MC, DC) \\ 
    G017.00+00.85 & 1.0 & $56.1 \pm 7.2$ & $26.8 \pm 1.9$ & $0.58 \pm 0.10$ & 7.8 & $20 \pm 29$ & $2956 \pm 85$ & $194.2 \pm 8.1$ & $22.7 \pm 0.6$ & $1.69 \pm 0.04$ & 1.0 & M16, W37 (H\textsc{ii} + DC, MC) \\ 
    G018.02+12.45 & 1.0 & $5.2 \pm 0.7$ & $24.3 \pm 2.5$ & $0.78 \pm 0.14$ & 8.0 & $6 \pm 6$ & $-6 \pm 9$ & $9.6 \pm 0.7$ & $17.1 \pm 0.7$ & $1.69 \pm 0.13$ & 0.89 & TGU H189, DB 0768 (MC)\,$\dagger$ \\ 
    G018.53+18.05 & 1.0 & $2.2 \pm 0.5$ & $18.9 \pm 2.9$ & $0.46 \pm 0.10$ & 4.7 & $4 \pm 4$ & $4 \pm 3$ & $13.1 \pm 1.2$ & $15.3 \pm 0.6$ & $1.77 \pm 0.13$ & 0.84 & TGU H198, PGCC G018.23+18.06 (MC) \\ 
    G019.00$-$00.20 & 1.0 & $46.1 \pm 12.7$ & $25.4 \pm 5.0$ & $0.43 \pm 0.15$ & 3.6 & $-8 \pm 30$ & $1988 \pm 96$ & $302.3 \pm 22.3$ & $20.3 \pm 0.7$ & $1.87 \pm 0.07$ & 1.1 & GAL 019.07-00.28, LDN 399 (H\textsc{ii} + many DC/MC) \\ 
    G023.47+08.19 & 1.0 & $4.6 \pm 1.0$ & $18.3 \pm 3.2$ & $0.58 \pm 0.14$ & 4.6 & $22 \pm 8$ & $-4 \pm 9$ & $41.9 \pm 3.1$ & $14.5 \pm 0.5$ & $1.82 \pm 0.10$ & 0.38 & LDN462 (DC) \\ 
    G023.85$-$00.01 & 1.3 & $198.6 \pm 14.9$ & $25.1 \pm 1.7$ & $0.45 \pm 0.07$ & 13.3 & $10 \pm 30$ & $3731 \pm 89$ & $423.5 \pm 21.0$ & $22.2 \pm 0.6$ & $1.75 \pm 0.04$ & 1.8 & [WAM82] 023.955+0.152 (H\textsc{ii} + DC/MC) \\ 
    G028.79+03.49 & 1.0 & $8.2 \pm 1.9$ & $43.3 \pm 5.3$ & $0.57 \pm 0.16$ & 4.3 & $-45 \pm 23$ & $413 \pm 24$ & $57.7 \pm 3.6$ & $22.7 \pm 0.8$ & $1.45 \pm 0.06$ & 0.68 & W40 (H\textsc{ii}) \\ 
    G030.77$-$00.03 & 1.0 & $179.8 \pm 19.7$ & $24.3 \pm 2.2$ & $0.68 \pm 0.11$ & 9.2 & $12 \pm 31$ & $4694 \pm 234$ & $399.2 \pm 20.2$ & $23.6 \pm 0.6$ & $1.65 \pm 0.04$ & 1.1 & W43 (SFR) \\ 
    G034.62+00.07 & 1.0 & $72.2 \pm 16.5$ & $21.7 \pm 3.3$ & $0.62 \pm 0.17$ & 4.4 & $4 \pm 30$ & $1436 \pm 343$ & $252.7 \pm 13.5$ & $21.1 \pm 0.6$ & $1.69 \pm 0.05$ & 0.99 & GAL 034.4+00.23, MSXDC G034.43+00.24 (H\textsc{ii}, DC)\,$\dagger$ \\ 
    G035.20$-$01.74 & 1.0 & $39.6 \pm 8.4$ & $19.4 \pm 2.8$ & $0.61 \pm 0.15$ & 4.7 & $3 \pm 30$ & $873 \pm 169$ & $197.8 \pm 9.9$ & $20.0 \pm 0.5$ & $1.62 \pm 0.05$ & 0.86 & W48 (H\textsc{ii}, SNR)\,$\dagger$ \\ 
    G037.50+44.68 & 1.4 & $1.2 \pm 0.4$ & $22.7 \pm 3.0$ & $0.62 \pm 0.17$ & 2.8 & $-3 \pm 2$ & $-1 \pm 3$ & $1.8 \pm 0.1$ & $19.0 \pm 1.0$ & $1.50 \pm 0.14$ & 0.88 & S73, MBM 40 (H\textsc{ii}, MC) \\ 
    G037.79$-$00.11 & 1.0 & $62.1 \pm 7.8$ & $22.1 \pm 2.6$ & $0.57 \pm 0.11$ & 8.0 & $13 \pm 29$ & $1740 \pm 74$ & $248.7 \pm 14.2$ & $20.9 \pm 0.6$ & $1.69 \pm 0.05$ & 0.63 & W47 (H\textsc{ii}) \\ 
    G039.07$-$16.71 & 1.2 & $3.1 \pm 0.9$ & $16.5 \pm 3.2$ & $0.70 \pm 0.16$ & 3.5 & $-4 \pm 4$ & $6 \pm 8$ & $6.6 \pm 0.6$ & $17.6 \pm 0.9$ & $1.55 \pm 0.14$ & 1.2 & TGU H334 (DC)\,$\dagger$ \\ 
    G040.52+02.53 & 1.0 & $5.4 \pm 1.5$ & $23.1 \pm 1.2$ & $0.47 \pm 0.10$ & 3.6 & $25 \pm 14$ & $160 \pm 38$ & $29.2 \pm 1.9$ & $20.5 \pm 0.8$ & $1.73 \pm 0.09$ & 0.58 & W45 (H\textsc{ii}, MC)\,$\dagger$ \\ 
    G040.10$-$35.50 & 2.0 & $3.2 \pm 1.1$ & $15.3 \pm 3.4$ & $0.59 \pm 0.19$ & 3.0 & $3 \pm 2$ & $1 \pm 2$ & $3.3 \pm 0.4$ & $17.3 \pm 1.0$ & $1.75 \pm 0.16$ & 0.59 & MBM 47 (MC) \\ 
    G041.03$-$00.07 & 1.0 & $58.1 \pm 5.7$ & $21.8 \pm 0.7$ & $0.57 \pm 0.05$ & 10.3 & $7 \pm 30$ & $998 \pm 144$ & $233.1 \pm 9.9$ & $20.0 \pm 0.4$ & $1.66 \pm 0.04$ & 0.84 & SDC G41.003-0.097 (DC)\,$\dagger$ \\ 
    G043.20$-$00.10 & 1.0 & $47.2 \pm 4.9$ & $21.6 \pm 1.3$ & $0.60 \pm 0.09$ & 9.6 & $6 \pm 28$ & $1457 \pm 60$ & $177.9 \pm 11.0$ & $20.8 \pm 0.6$ & $1.66 \pm 0.06$ & 0.31 & W49 (H\textsc{ii}, MC) \\ 
    G043.97$-$22.62 & 1.2 & $1.6 \pm 0.6$ & $14.5 \pm 3.7$ & $0.53 \pm 0.17$ & 2.6 & $-2 \pm 3$ & $1 \pm 4$ & $3.5 \pm 0.4$ & $18.9 \pm 1.2$ & $1.61 \pm 0.17$ & 1.3 & IRAS 20286-0110 (FIR)\,$\dagger$ \\ 
    G045.47+00.06 & 1.0 & $45.4 \pm 2.7$ & $22.3 \pm 0.8$ & $0.53 \pm 0.04$ & 16.8 & $4 \pm 23$ & $1079 \pm 32$ & $209.7 \pm 9.2$ & $20.3 \pm 0.5$ & $1.64 \pm 0.04$ & 0.59 & CORNISH G045.4545+00.0591 (H\textsc{ii} + MC) \\ 
    G053.63+00.19 & 1.0 & $12.5 \pm 2.1$ & $22.5 \pm 2.0$ & $0.55 \pm 0.12$ & 5.9 & $-17 \pm 17$ & $678 \pm 27$ & $72.6 \pm 3.9$ & $21.0 \pm 0.6$ & $1.59 \pm 0.06$ & 0.44 & S82 (H\textsc{ii} + MC) \\ 
    G054.25+06.87 & 1.0 & $1.7 \pm 0.2$ & $18.2 \pm 1.1$ & $0.64 \pm 0.15$ & 7.5 & $-4 \pm 4$ & $4 \pm 3$ & $6.9 \pm 0.8$ & $18.8 \pm 1.2$ & $1.58 \pm 0.16$ & 0.88 & TGU H386 (DC) \\ 
    G057.66+00.09 & 1.0 & $24.6 \pm 2.1$ & $18.7 \pm 0.6$ & $0.66 \pm 0.06$ & 11.6 & $22 \pm 16$ & $94 \pm 38$ & $119.6 \pm 5.2$ & $17.5 \pm 0.4$ & $1.73 \pm 0.05$ & 0.57 & DB 2003/05/06/07/09/12, TGU H398 (DC)\,$\dagger$ \\ 
    G057.82$-$06.32 & 1.0 & $1.2 \pm 0.4$ & $25.4 \pm 2.9$ & $0.53 \pm 0.19$ & 3.2 & $-3 \pm 5$ & $9 \pm 7$ & $7.2 \pm 0.6$ & $19.5 \pm 0.9$ & $1.50 \pm 0.12$ & 0.98 & TGU H397, IRAS 19598+1823 (DC, FIR)\,$\dagger$ \\ 
    G059.48$-$00.02 & 1.0 & $18.9 \pm 1.6$ & $21.0 \pm 1.1$ & $0.62 \pm 0.09$ & 11.6 & $7 \pm 19$ & $367 \pm 20$ & $70.6 \pm 4.6$ & $19.8 \pm 0.7$ & $1.66 \pm 0.08$ & 0.38 & W55 \,\textsuperscript{G059.42$-$00.21}\, (H\textsc{ii}, MC) \\ 
    G061.34$-$00.02 & 1.0 & $13.7 \pm 1.0$ & $22.9 \pm 0.9$ & $0.56 \pm 0.05$ & 13.9 & $-3 \pm 11$ & $380 \pm 13$ & $51.8 \pm 2.6$ & $21.5 \pm 0.6$ & $1.57 \pm 0.05$ & 0.48 & S88 \,\textsuperscript{G061.47$+$00.11}\, (H\textsc{ii} + MC) \\ 
    G062.98+00.05 & 1.0 & $14.6 \pm 1.0$ & $21.5 \pm 0.8$ & $0.60 \pm 0.06$ & 14.7 & $18 \pm 13$ & $191 \pm 12$ & $53.9 \pm 3.4$ & $19.3 \pm 0.6$ & $1.70 \pm 0.08$ & 0.74 & S89 (H\textsc{ii}) \\ 
    G066.45$-$02.73 & 1.0 & $8.2 \pm 0.6$ & $21.7 \pm 0.9$ & $0.68 \pm 0.07$ & 15.0 & $12 \pm 9$ & $-6 \pm 7$ & $27.7 \pm 1.9$ & $17.8 \pm 0.7$ & $1.72 \pm 0.10$ & 1.1 & PGCC G066.53-02.87, DB 2140/41 (DC) \\ 
    G068.16+01.02 & 1.0 & $2.5 \pm 0.6$ & $28.4 \pm 4.5$ & $0.57 \pm 0.16$ & 4.4 & $-10 \pm 12$ & $197 \pm 6$ & $6.8 \pm 1.2$ & $25.5 \pm 3.0$ & $1.40 \pm 0.24$ & 0.38 & S98 (H\textsc{ii}) \\ 
    G070.14+01.61 & 1.0 & $14.5 \pm 1.5$ & $26.6 \pm 1.3$ & $0.62 \pm 0.09$ & 10.0 & $8 \pm 18$ & $536 \pm 19$ & $48.1 \pm 2.7$ & $22.4 \pm 0.8$ & $1.44 \pm 0.06$ & 0.46 & NGC 6857, S100 (H\textsc{ii}) \\ 
    G071.75+02.47 & 1.0 & $15.2 \pm 1.3$ & $25.3 \pm 1.0$ & $0.52 \pm 0.05$ & 11.7 & $-26 \pm 15$ & $572 \pm 17$ & $68.7 \pm 3.2$ & $20.7 \pm 0.6$ & $1.51 \pm 0.05$ & 0.58 & S101, DB 2225/2235 \,\textsuperscript{G071.59$+$02.85}\, (H\textsc{ii}, DC) \\ 
    G076.00$-$00.06 & 1.0 & $14.0 \pm 2.6$ & $21.2 \pm 2.1$ & $0.63 \pm 0.14$ & 5.3 & $0 \pm 16$ & $727 \pm 33$ & $69.0 \pm 3.2$ & $21.3 \pm 0.6$ & $1.58 \pm 0.05$ & 0.41 & S106 \,\textsuperscript{G076.38$-$00.62}\, (H\textsc{ii}) \\ 
    G078.44+00.17 & 1.2 & $42.0 \pm 10.4$ & $22.7 \pm 2.7$ & $0.66 \pm 0.18$ & 4.1 & $-9 \pm 25$ & $2037 \pm 94$ & $91.4 \pm 4.1$ & $24.1 \pm 0.7$ & $1.42 \pm 0.04$ & 0.21 & DB 2403/04/09/15/23/27 (DC) \\ 
    G082.81$-$01.94 & 1.0 & $11.7 \pm 3.4$ & $23.3 \pm 3.1$ & $0.54 \pm 0.19$ & 3.4 & $-17 \pm 23$ & $931 \pm 42$ & $61.1 \pm 3.5$ & $21.5 \pm 0.7$ & $1.47 \pm 0.06$ & 0.25 & LDN 914, TGU H497, DB 2650/56/60/64 (DC) \\ 
    G089.33+11.17 & 1.4 & $4.0 \pm 0.7$ & $17.5 \pm 0.8$ & $0.57 \pm 0.09$ & 5.6 & $0 \pm 4$ & $6 \pm 6$ & $9.3 \pm 0.7$ & $18.5 \pm 0.7$ & $1.55 \pm 0.09$ & 0.95 & PGCC G089.54+11.28 (MC)\,$\dagger$ \\ 
    G089.88$-$40.26 & 1.0 & $1.2 \pm 0.2$ & $21.1 \pm 3.2$ & $0.52 \pm 0.15$ & 4.9 & $-2 \pm 3$ & $1 \pm 3$ & $5.5 \pm 0.8$ & $16.9 \pm 1.2$ & $1.66 \pm 0.18$ & 1.1 & IRAS 23058+1546 (FIR) \\ 
    G092.11$-$34.97 & 1.0 & $1.5 \pm 0.1$ & $16.7 \pm 0.9$ & $0.58 \pm 0.06$ & 10.8 & $0 \pm 2$ & $0 \pm 1$ & $3.9 \pm 0.5$ & $16.7 \pm 1.0$ & $1.88 \pm 0.20$ & 0.9 & IRAS 23017+2048, PGCC G092.00-35.36 (FIR, MC) \\ 
    G093.02+02.76 & 1.0 & $20.3 \pm 4.0$ & $20.7 \pm 1.9$ & $0.70 \pm 0.15$ & 5.1 & $-22 \pm 21$ & $713 \pm 51$ & $84.3 \pm 4.4$ & $20.7 \pm 0.6$ & $1.47 \pm 0.05$ & 0.35 & GAL093.06+2.81 (H\textsc{ii}) \\ 
    G094.76$-$01.52 & 1.0 & $6.9 \pm 1.2$ & $21.7 \pm 2.0$ & $0.57 \pm 0.13$ & 5.8 & $-16 \pm 18$ & $299 \pm 14$ & $50.5 \pm 4.4$ & $18.8 \pm 0.8$ & $1.55 \pm 0.10$ & 0.56 & LDN 1059 \,\textsuperscript{G094.47$-$01.53}\, (DC) \\ 
    G096.89$-$29.75 & 1.2 & $1.1 \pm 0.2$ & $17.8 \pm 1.2$ & $0.47 \pm 0.11$ & 4.5 & $0 \pm 2$ & $4 \pm 3$ & $1.7 \pm 0.2$ & $19.7 \pm 1.1$ & $1.54 \pm 0.14$ & 1.5 & PGCC G096.76-29.39 (MC)\,$\dagger$ \\ 
    G098.00+01.47 & 1.0 & $3.0 \pm 0.5$ & $35.5 \pm 2.8$ & $0.53 \pm 0.15$ & 6.0 & $-5 \pm 8$ & $224 \pm 7$ & $8.5 \pm 0.6$ & $23.0 \pm 1.2$ & $1.39 \pm 0.11$ & 0.96 & GM1-12, TGU H582 (RNe, DC) \\ 
    G099.60+03.70 & 1.3 & $11.0 \pm 2.5$ & $32.6 \pm 2.9$ & $0.65 \pm 0.18$ & 4.4 & $-38 \pm 15$ & $663 \pm 21$ & $28.8 \pm 1.5$ & $23.5 \pm 0.8$ & $1.28 \pm 0.06$ & 0.46 & LDN1111 (MC) \\ 
    G100.96$-$15.48 & 1.0 & $0.8 \pm 0.3$ & $22.5 \pm 2.0$ & $0.52 \pm 0.16$ & 2.6 & $-3 \pm 4$ & $13 \pm 9$ & $3.5 \pm 0.4$ & $17.8 \pm 1.2$ & $1.59 \pm 0.20$ & 0.74 & TGU H611, LBN 462 (DC, H\textsc{ii})\,$\dagger$ \\ 
    G104.27$-$38.83 & 1.3 & $0.8 \pm 0.4$ & $24.9 \pm 4.3$ & $0.45 \pm 0.20$ & 2.3 & $-1 \pm 3$ & $1 \pm 4$ & $3.4 \pm 0.5$ & $17.0 \pm 1.2$ & $1.65 \pm 0.21$ & 0.97 & IRAS 23463+2127, TGU L73 (FIR, DC)\,$\dagger$ \\ 
    G104.60+10.75 & 1.4 & $3.4 \pm 0.6$ & $26.2 \pm 1.9$ & $0.54 \pm 0.11$ & 5.7 & $-2 \pm 4$ & $1 \pm 4$ & $11.7 \pm 0.8$ & $17.9 \pm 0.6$ & $1.56 \pm 0.09$ & 0.81 & TGU H634 (DC) \\ 
    G106.44+12.39 & 1.0 & $2.5 \pm 0.2$ & $20.2 \pm 0.6$ & $0.59 \pm 0.06$ & 10.1 & $8 \pm 3$ & $-1 \pm 3$ & $5.6 \pm 0.5$ & $17.8 \pm 0.9$ & $1.77 \pm 0.16$ & 0.78 & LDN 1199, TGU H653 (DC) \\ 
    G107.20+05.20 & 1.2 & $18.9 \pm 1.2$ & $24.1 \pm 0.9$ & $0.57 \pm 0.05$ & 15.3 & $19 \pm 10$ & $334 \pm 11$ & $36.4 \pm 1.8$ & $21.7 \pm 0.7$ & $1.54 \pm 0.06$ & 0.41 & S140 (H\textsc{ii}) \\ 
    G107.29+19.10 & 1.0 & $1.5 \pm 0.3$ & $14.7 \pm 1.4$ & $0.52 \pm 0.11$ & 5.3 & $7 \pm 3$ & $0 \pm 2$ & $6.9 \pm 2.0$ & $14.3 \pm 1.5$ & $2.16 \pm 0.39$ & 0.78 & TGU H667 (DC) \\ 
    G108.89$-$52.21 & 1.0 & $0.8 \pm 0.2$ & $23.4 \pm 3.8$ & $0.51 \pm 0.25$ & 4.1 & $-5 \pm 4$ & $8 \pm 2$ & $4.7 \pm 0.5$ & $17.3 \pm 1.0$ & $1.73 \pm 0.18$ & 1.2 & PGCC G108.79-51.98 (MC) \\ 
    G109.29+13.51 & 1.3 & $7.3 \pm 0.6$ & $16.4 \pm 0.6$ & $0.68 \pm 0.06$ & 12.8 & $1 \pm 4$ & $-8 \pm 4$ & $19.2 \pm 1.3$ & $16.3 \pm 0.5$ & $1.63 \pm 0.09$ & 0.48 & TGU H693, PGCC G109.11+13.26 (DC) \\ 
    G109.31+06.46 & 1.0 & $5.8 \pm 0.4$ & $24.4 \pm 0.7$ & $0.49 \pm 0.04$ & 15.2 & $10 \pm 6$ & $79 \pm 5$ & $17.7 \pm 1.2$ & $20.7 \pm 0.8$ & $1.64 \pm 0.09$ & 0.72 & DB 3395/3397/3401/3389, LDN 1213/14 (DC) \\ 
    G110.21+02.41 & 1.0 & $10.1 \pm 0.9$ & $31.7 \pm 1.8$ & $0.64 \pm 0.09$ & 11.7 & $2 \pm 15$ & $314 \pm 11$ & $23.3 \pm 1.6$ & $25.0 \pm 1.2$ & $1.53 \pm 0.09$ & 0.53 & Cepheus B, S155 \,\textsuperscript{G110.25$+$02.58}\, (MC, H\textsc{ii}) \\ 
    G110.72$-$00.48 & 1.0 & $9.0 \pm 1.9$ & $26.7 \pm 2.3$ & $0.64 \pm 0.15$ & 4.9 & $20 \pm 16$ & $365 \pm 25$ & $26.8 \pm 1.8$ & $22.6 \pm 1.0$ & $1.59 \pm 0.10$ & 0.92 & DB 3435/33, TGU H700 (DC) \\ 
    G111.67+00.78 & 1.0 & $9.2 \pm 1.1$ & $26.0 \pm 1.3$ & $0.47 \pm 0.07$ & 8.4 & $44 \pm 12$ & $551 \pm 15$ & $36.9 \pm 2.3$ & $22.8 \pm 0.9$ & $1.65 \pm 0.07$ & 0.32 & NGC 7538 (Open Cluster) \\ 
    G111.66+20.21 & 1.0 & $1.2 \pm 0.5$ & $11.5 \pm 2.3$ & $0.81 \pm 0.25$ & 2.3 & $7 \pm 4$ & $2 \pm 7$ & $20.3 \pm 1.2$ & $14.9 \pm 0.4$ & $1.74 \pm 0.08$ & 1.2 & LDN 1228, MBM 162 (DC, MC)\,$\dagger$ \\ 
    G116.05+19.94 & 1.0 & $1.0 \pm 0.2$ & $14.8 \pm 1.4$ & $0.78 \pm 0.14$ & 4.1 & $2 \pm 1$ & $0 \pm 3$ & $8.9 \pm 0.6$ & $14.3 \pm 0.4$ & $1.99 \pm 0.08$ & 0.4 & MBM 163/64/65, TGU H758, LBN 569 (MC, H\textsc{ii}) \\ 
    G122.67+09.73 & 1.0 & $2.9 \pm 0.6$ & $16.3 \pm 1.4$ & $0.52 \pm 0.18$ & 5.2 & $-3 \pm 6$ & $0 \pm 7$ & $10.7 \pm 1.1$ & $16.3 \pm 0.9$ & $1.67 \pm 0.16$ & 1.1 & IREC 169, TGU H814, DB 3740 (DC) \\ 
    G124.29+02.57 & 1.5 & $11.8 \pm 1.0$ & $18.4 \pm 0.7$ & $0.64 \pm 0.09$ & 12.3 & $6 \pm 6$ & $14 \pm 6$ & $34.6 \pm 2.0$ & $15.4 \pm 0.4$ & $1.70 \pm 0.08$ & 0.48 & LDN 1307, DB 3760/61/63, TGU H823 (DC) \\ 
    G124.39+30.21 & 1.4 & $2.2 \pm 0.4$ & $20.3 \pm 2.7$ & $0.60 \pm 0.11$ & 6.1 & $-1 \pm 2$ & $-1 \pm 2$ & $8.7 \pm 0.6$ & $16.2 \pm 0.6$ & $1.76 \pm 0.10$ & 0.84 & Polaris Flare (MC)\,$\dagger$ \\ 
    G124.91$-$03.81 & 1.0 & $3.4 \pm 0.4$ & $20.8 \pm 1.4$ & $0.64 \pm 0.14$ & 8.2 & $3 \pm 6$ & $25 \pm 5$ & $7.4 \pm 0.7$ & $19.3 \pm 1.1$ & $1.62 \pm 0.16$ & 0.99 & LDN 1310 (DC) \\ 
    G126.80$-$70.10 & 1.7 & $1.8 \pm 0.5$ & $25.4 \pm 5.1$ & $0.72 \pm 0.18$ & 3.9 & $-4 \pm 3$ & $3 \pm 3$ & $1.9 \pm 0.3$ & $19.9 \pm 1.9$ & $1.50 \pm 0.24$ & 0.83 & PGCC G127.36-70.07 (MC)\,$\dagger$ \\ 
    G127.67+13.88 & 1.0 & $1.3 \pm 0.2$ & $23.4 \pm 1.4$ & $0.41 \pm 0.07$ & 6.1 & $1 \pm 3$ & $4 \pm 2$ & $14.0 \pm 1.1$ & $15.4 \pm 0.6$ & $1.70 \pm 0.11$ & 0.87 & TGU H853, DB 3797/98/99/3800/01 (DC) \\ 
    G129.15$-$00.07 & 1.0 & $2.3 \pm 0.7$ & $19.6 \pm 3.9$ & $0.89 \pm 0.22$ & 3.2 & $-3 \pm 9$ & $3 \pm 11$ & $12.3 \pm 1.3$ & $17.3 \pm 1.0$ & $1.65 \pm 0.17$ & 0.84 & LDN 1332/34/37, DB 3809/12 (DC)\,$\dagger$ \\ 
    G129.17$-$04.89 & 1.0 & $1.3 \pm 0.3$ & $18.3 \pm 1.7$ & $0.43 \pm 0.15$ & 3.8 & $-4 \pm 5$ & $6 \pm 6$ & $6.1 \pm 0.8$ & $18.5 \pm 1.3$ & $1.46 \pm 0.18$ & 1.2 & PGCC G129.19-04.84, TGU H858 (DC)\,$\dagger$ \\ 
    G133.27+09.05 & 1.0 & $4.7 \pm 0.7$ & $18.8 \pm 0.9$ & $0.58 \pm 0.08$ & 7.1 & $17 \pm 8$ & $13 \pm 12$ & $49.6 \pm 2.6$ & $15.1 \pm 0.4$ & $1.75 \pm 0.07$ & 0.59 & LDN 1358/1355/1357 (MC)\,$\dagger$ \\ 
    G136.60$-$68.50 & 1.5 & $2.4 \pm 0.1$ & $22.9 \pm 1.4$ & $0.61 \pm 0.07$ & 17.9 & $-1 \pm 1$ & $0 \pm 1$ & $2.4 \pm 0.1$ & $20.2 \pm 0.7$ & $1.60 \pm 0.08$ & 1.2 & IRAS 01081-0606 (FIR) \\ 
    G136.48+12.72 & 1.0 & $2.6 \pm 0.5$ & $15.3 \pm 1.2$ & $0.57 \pm 0.12$ & 5.3 & $7 \pm 6$ & $2 \pm 9$ & $14.0 \pm 1.4$ & $15.6 \pm 0.8$ & $1.85 \pm 0.15$ & 0.53 & PGCC G136.48+12.72, TGU H894 (MC, DC)\,$\dagger$ \\ 
    G137.29+01.16 & 1.0 & $8.6 \pm 1.7$ & $29.7 \pm 2.5$ & $0.54 \pm 0.15$ & 5.0 & $8 \pm 14$ & $724 \pm 24$ & $22.8 \pm 1.4$ & $24.7 \pm 1.1$ & $1.46 \pm 0.08$ & 0.33 & S199 Soul Nebula, DB 3915/6 (H\textsc{ii}, DC) \\ 
    G139.77+21.91 & 1.2 & $1.2 \pm 0.3$ & $20.9 \pm 2.7$ & $0.60 \pm 0.18$ & 3.7 & $0 \pm 3$ & $-1 \pm 3$ & $3.7 \pm 0.4$ & $18.3 \pm 1.0$ & $1.55 \pm 0.15$ & 1.5 & IRAS 05429+7359 (FIR) \\ 
    G142.35+01.35 & 1.0 & $15.4 \pm 0.8$ & $17.5 \pm 0.4$ & $0.67 \pm 0.04$ & 18.9 & $24 \pm 7$ & $27 \pm 13$ & $73.2 \pm 2.8$ & $17.8 \pm 0.4$ & $1.63 \pm 0.04$ & 1.8 & TGU H942, DB 3984 (MC, DC)\,$\dagger$ \\ 
    G142.60+38.46 & 1.2 & $1.0 \pm 0.5$ & $19.4 \pm 4.0$ & $0.68 \pm 0.20$ & 1.9 & $-1 \pm 3$ & $0 \pm 5$ & $4.4 \pm 0.4$ & $16.8 \pm 0.8$ & $1.83 \pm 0.15$ & 1.7 & Ursa Major Complex (MC)\,$\dagger$ \\ 
    G142.94+10.02 & 1.3 & $5.5 \pm 0.4$ & $18.1 \pm 0.5$ & $0.49 \pm 0.05$ & 13.2 & $10 \pm 3$ & $3 \pm 3$ & $12.2 \pm 1.1$ & $15.5 \pm 0.7$ & $1.98 \pm 0.15$ & 0.57 & IRAS 04119+6413, TGU H951 (FIR, DC) \\ 
    G150.16+09.36 & 1.2 & $3.5 \pm 0.6$ & $19.1 \pm 1.2$ & $0.53 \pm 0.12$ & 6.4 & $-4 \pm 6$ & $7 \pm 6$ & $8.1 \pm 1.1$ & $16.4 \pm 1.2$ & $1.65 \pm 0.23$ & 1.1 & PGCC G150.43+09.39, DB 4066 (DC)\,$\dagger$ \\ 
    G151.40$-$19.37 & 1.1 & $1.4 \pm 0.2$ & $16.8 \pm 1.0$ & $0.57 \pm 0.10$ & 8.7 & $5 \pm 2$ & $3 \pm 2$ & $2.6 \pm 0.4$ & $17.7 \pm 1.3$ & $1.75 \pm 0.22$ & 0.97 & IRAS 03041+3544 (FIR) \\ 
    G154.03$-$39.80 & 1.0 & $2.1 \pm 0.4$ & $19.6 \pm 3.9$ & $0.89 \pm 0.18$ & 5.0 & $-6 \pm 7$ & $-5 \pm 5$ & $6.1 \pm 0.8$ & $17.0 \pm 1.2$ & $1.71 \pm 0.23$ & 1.1 & PGCC G154.64-39.67 (MC)\,$\dagger$ \\ 
    G154.53+02.52 & 1.0 & $2.5 \pm 0.3$ & $28.2 \pm 3.9$ & $0.71 \pm 0.17$ & 7.5 & $-12 \pm 10$ & $28 \pm 3$ & $25.0 \pm 1.8$ & $16.5 \pm 0.6$ & $1.54 \pm 0.10$ & 0.62 & B20, S211, DB 4093/96/98, TGU H1036 (H\textsc{ii}, DC) \\ 
    G155.87+05.08 & 1.0 & $2.2 \pm 0.3$ & $20.1 \pm 1.0$ & $0.59 \pm 0.09$ & 6.7 & $0 \pm 5$ & $14 \pm 9$ & $17.1 \pm 1.3$ & $15.8 \pm 0.6$ & $1.68 \pm 0.11$ & 0.87 & LDN 1436/38, DB 4110/12/13/14/15 (DC)\,$\dagger$ \\ 
    G158.40$-$20.60 & 1.0 & $5.3 \pm 0.4$ & $25.7 \pm 2.0$ & $0.61 \pm 0.08$ & 14.6 & $29 \pm 17$ & $1 \pm 2$ & $45.6 \pm 3.5$ & $16.3 \pm 0.6$ & $1.57 \pm 0.11$ & 0.37 & LDN 1450/1452, MBM 104 (DC, MC) \\ 
    G158.70+35.09 & 1.0 & $1.2 \pm 0.3$ & $12.7 \pm 1.4$ & $0.64 \pm 0.15$ & 4.2 & $2 \pm 2$ & $-2 \pm 4$ & $1.8 \pm 0.2$ & $17.8 \pm 1.4$ & $1.75 \pm 0.24$ & 1.1 & IRAS 08237+5838 (FIR) \\ 
    G159.02$-$33.88 & 1.2 & $4.0 \pm 0.3$ & $19.8 \pm 1.1$ & $0.64 \pm 0.08$ & 11.5 & $9 \pm 4$ & $1 \pm 4$ & $21.9 \pm 1.4$ & $15.3 \pm 0.5$ & $1.93 \pm 0.09$ & 0.79 & LDN 1454/53/58, DB 4162 (DC)\,$\dagger$ \\ 
    G160.26$-$18.62 & 1.3 & $19.8 \pm 0.6$ & $24.8 \pm 0.5$ & $0.48 \pm 0.02$ & 32.2 & $35 \pm 11$ & $44 \pm 3$ & $31.0 \pm 2.2$ & $19.9 \pm 0.8$ & $1.60 \pm 0.09$ & 0.45 & Perseus Molecular Cloud (MC) \\ 
    G160.27$-$12.36 & 1.5 & $9.6 \pm 2.1$ & $62.0 \pm 13.2$ & $0.84 \pm 0.20$ & 4.6 & $-1 \pm 7$ & $341 \pm 9$ & $7.2 \pm 0.4$ & $23.4 \pm 1.1$ & $1.58 \pm 0.09$ & 0.82 & California Nebula \,\textsuperscript{G160.60$-$12.05}\, (H\textsc{ii}) \\ 
    G164.48$-$05.63 & 1.0 & $1.6 \pm 0.2$ & $22.4 \pm 0.9$ & $0.47 \pm 0.05$ & 8.5 & $2 \pm 2$ & $5 \pm 2$ & $9.3 \pm 0.7$ & $15.6 \pm 0.5$ & $1.73 \pm 0.10$ & 0.72 & DB 4245/51/52/55, LDN 1481, TGU H1116 (DC) \\ 
    G169.84$-$08.99 & 1.0 & $2.5 \pm 0.4$ & $20.9 \pm 1.4$ & $0.57 \pm 0.11$ & 6.4 & $3 \pm 6$ & $1 \pm 5$ & $15.6 \pm 1.4$ & $15.9 \pm 0.7$ & $1.70 \pm 0.14$ & 0.97 & LDN 1496, TGU H1157, DB 4313 (DC) \\ 
    G170.60$-$37.30 & 2.5 & $21.8 \pm 0.6$ & $21.3 \pm 0.4$ & $0.62 \pm 0.03$ & 34.2 & $2 \pm 3$ & $1 \pm 1$ & $12.7 \pm 0.7$ & $17.1 \pm 0.5$ & $1.67 \pm 0.08$ & 1.4 & MBM 16 (MC) \\ 
    G171.48$-$41.62 & 1.1 & $2.4 \pm 0.2$ & $26.1 \pm 3.1$ & $0.69 \pm 0.11$ & 10.9 & $-9 \pm 7$ & $0 \pm 1$ & $7.4 \pm 0.8$ & $19.2 \pm 1.2$ & $1.40 \pm 0.17$ & 0.98 & IRAS 03047+0739 (FIR) \\ 
    G171.79$-$00.09 & 1.0 & $4.2 \pm 0.3$ & $26.6 \pm 1.4$ & $0.59 \pm 0.08$ & 13.1 & $4 \pm 6$ & $65 \pm 4$ & $12.7 \pm 0.7$ & $20.3 \pm 0.7$ & $1.47 \pm 0.08$ & 0.62 & TGU H1175/72/81 (DC) \\ 
    G173.56$-$01.76 & 1.0 & $5.8 \pm 0.7$ & $42.4 \pm 1.8$ & $0.44 \pm 0.05$ & 8.4 & $-34 \pm 9$ & $481 \pm 10$ & $7.8 \pm 0.5$ & $27.8 \pm 1.5$ & $1.13 \pm 0.08$ & 0.96 & IC 410 (H\textsc{ii}) \\ 
    G173.62+02.79 & 1.0 & $12.0 \pm 0.5$ & $23.4 \pm 0.6$ & $0.55 \pm 0.03$ & 24.7 & $22 \pm 8$ & $191 \pm 6$ & $53.7 \pm 2.1$ & $19.6 \pm 0.4$ & $1.52 \pm 0.04$ & 1.6 & S235 (H\textsc{ii}) \\ 
    G174.27$-$13.81 & 1.0 & $3.5 \pm 0.3$ & $19.1 \pm 0.8$ & $0.51 \pm 0.07$ & 12.0 & $14 \pm 8$ & $8 \pm 5$ & $53.3 \pm 3.7$ & $14.3 \pm 0.4$ & $1.82 \pm 0.09$ & 0.97 & LDN 1527 (DC)\,$\dagger$ \\ 
    G175.60$-$12.51 & 1.0 & $4.3 \pm 0.3$ & $18.1 \pm 0.6$ & $0.54 \pm 0.05$ & 16.8 & $4 \pm 6$ & $5 \pm 2$ & $29.9 \pm 2.1$ & $15.6 \pm 0.5$ & $1.66 \pm 0.09$ & 1.6 & DB 4466, TGU H1211 (DC) \\ 
    G176.90$-$00.41 & 1.3 & $10.8 \pm 0.9$ & $19.0 \pm 0.9$ & $0.75 \pm 0.08$ & 11.5 & $-1 \pm 8$ & $-10 \pm 7$ & $30.2 \pm 1.7$ & $16.6 \pm 0.5$ & $1.57 \pm 0.08$ & 0.63 & IRAS 05331+3115 (H\textsc{ii}) \\ 
    G178.67$-$06.61 & 1.0 & $2.1 \pm 0.2$ & $17.6 \pm 0.8$ & $0.55 \pm 0.06$ & 12.1 & $6 \pm 3$ & $0 \pm 2$ & $22.3 \pm 1.3$ & $15.0 \pm 0.4$ & $1.69 \pm 0.07$ & 0.68 & FG L1547, LDN 1548/49/52 (H\textsc{ii}, MC) \\ 
    G181.22$-$27.90 & 1.2 & $2.7 \pm 0.3$ & $22.1 \pm 2.1$ & $0.71 \pm 0.12$ & 7.9 & $-5 \pm 5$ & $7 \pm 3$ & $7.7 \pm 0.8$ & $18.0 \pm 1.0$ & $1.55 \pm 0.17$ & 1.1 & IRAS 04070+1128 (FIR) \\ 
    G181.19$-$21.51 & 1.0 & $2.9 \pm 0.4$ & $15.4 \pm 0.9$ & $0.55 \pm 0.08$ & 8.3 & $-7 \pm 4$ & $-2 \pm 5$ & $5.6 \pm 1.2$ & $16.5 \pm 1.6$ & $1.59 \pm 0.27$ & 1.0 & IRAS 04299+1524 (FIR)\,$\dagger$ \\ 
    G182.42+00.00 & 1.0 & $9.9 \pm 0.7$ & $19.1 \pm 0.6$ & $0.58 \pm 0.05$ & 13.8 & $-3 \pm 10$ & $33 \pm 13$ & $55.2 \pm 2.8$ & $16.8 \pm 0.4$ & $1.54 \pm 0.06$ & 1.3 & S242, LBN 827 \,\textsuperscript{G182.36$+$00.22}\, (H\textsc{ii}, RNe)\,$\dagger$ \\ 
    G183.81$-$20.35 & 1.3 & $2.8 \pm 0.6$ & $27.1 \pm 8.9$ & $0.96 \pm 0.29$ & 4.4 & $-4 \pm 8$ & $3 \pm 4$ & $7.2 \pm 1.1$ & $17.7 \pm 1.4$ & $1.65 \pm 0.25$ & 1.1 & TGU H1315 (DC) \\ 
    G190.00+00.46 & 1.0 & $13.2 \pm 1.5$ & $22.8 \pm 1.3$ & $0.75 \pm 0.12$ & 8.8 & $-4 \pm 18$ & $254 \pm 19$ & $41.0 \pm 2.6$ & $21.1 \pm 0.8$ & $1.45 \pm 0.08$ & 0.35 & NGC 2174/2175, S252 (H\textsc{ii} + DC) \\ 
    G191.20$-$38.67 & 1.0 & $2.4 \pm 0.3$ & $21.2 \pm 2.6$ & $0.50 \pm 0.09$ & 8.5 & $-3 \pm 4$ & $0 \pm 1$ & $9.1 \pm 0.7$ & $18.8 \pm 0.8$ & $1.52 \pm 0.11$ & 0.81 & IRAS 03547-0152 (FIR) \\ 
    G192.41$-$11.51 & 1.0 & $12.9 \pm 0.7$ & $25.2 \pm 0.6$ & $0.58 \pm 0.04$ & 17.8 & $48 \pm 11$ & $144 \pm 10$ & $25.7 \pm 2.2$ & $18.0 \pm 0.9$ & $1.90 \pm 0.16$ & 1.3 & $\lambda$ Orionis B30 \,\textsuperscript{G192.34$-$11.37}\, (DC) \\ 
    G192.60$-$00.06 & 1.0 & $7.3 \pm 0.4$ & $22.5 \pm 0.8$ & $0.70 \pm 0.07$ & 16.4 & $-2 \pm 9$ & $80 \pm 6$ & $27.4 \pm 1.6$ & $20.9 \pm 0.7$ & $1.47 \pm 0.07$ & 0.79 & S255 (H\textsc{ii}) \\ 
    G194.77$-$15.71 & 1.0 & $12.4 \pm 0.5$ & $23.5 \pm 0.3$ & $0.49 \pm 0.02$ & 26.7 & $28 \pm 7$ & $54 \pm 6$ & $21.3 \pm 1.7$ & $19.4 \pm 0.9$ & $1.82 \pm 0.12$ & 1.7 & $\lambda$ Orionis B223 (DC) \\ 
    G195.90$-$02.60 & 1.2 & $5.5 \pm 0.3$ & $23.7 \pm 0.6$ & $0.45 \pm 0.03$ & 20.8 & $-6 \pm 5$ & $50 \pm 2$ & $15.4 \pm 1.0$ & $20.4 \pm 0.7$ & $1.46 \pm 0.08$ & 0.66 & LDN 1591/92/93 (DC) \\ 
    G199.55$-$11.81 & 1.0 & $4.8 \pm 0.5$ & $25.1 \pm 3.8$ & $0.65 \pm 0.16$ & 8.8 & $3 \pm 5$ & $14 \pm 5$ & $6.9 \pm 0.6$ & $17.9 \pm 1.0$ & $1.96 \pm 0.18$ & 0.87 & DB 4695, LDN 1602/03 (DC) \\ 
    G200.30+11.90 & 1.0 & $1.4 \pm 0.3$ & $23.8 \pm 1.4$ & $0.53 \pm 0.08$ & 5.2 & $1 \pm 2$ & $-2 \pm 4$ & $7.7 \pm 0.6$ & $16.5 \pm 0.6$ & $1.84 \pm 0.11$ & 1.1 & TGU H1403 (DC + MC) \\ 
    G206.60$-$26.40 & 1.0 & $1.4 \pm 0.2$ & $23.0 \pm 4.0$ & $0.50 \pm 0.14$ & 6.0 & $8 \pm 2$ & $0 \pm 2$ & $2.8 \pm 0.3$ & $17.8 \pm 1.3$ & $2.03 \pm 0.22$ & 0.62 & PGCC G206.55-26.17, DB 4820 (MC) \\ 
    G207.30$-$23.00 & 1.0 & $1.4 \pm 0.5$ & $22.4 \pm 5.6$ & $0.64 \pm 0.21$ & 2.7 & $-5 \pm 5$ & $47 \pm 6$ & $5.9 \pm 0.6$ & $19.5 \pm 1.2$ & $1.69 \pm 0.16$ & 0.82 & LDN 1634, MBM 110, S278 (MC, H\textsc{ii}) \\ 
    G208.80$-$02.65 & 1.0 & $2.8 \pm 0.7$ & $48.6 \pm 9.2$ & $0.69 \pm 0.17$ & 4.3 & $-18 \pm 11$ & $156 \pm 5$ & $14.3 \pm 0.9$ & $20.6 \pm 0.9$ & $1.37 \pm 0.10$ & 0.65 & S280 (H\textsc{ii}) \\ 
    G216.31+09.85 & 1.0 & $1.2 \pm 0.3$ & $19.4 \pm 3.9$ & $0.52 \pm 0.15$ & 4.0 & $-3 \pm 4$ & $1 \pm 3$ & $4.8 \pm 0.7$ & $17.6 \pm 1.4$ & $1.45 \pm 0.20$ & 0.68 & TGU H1512, PGCC G216.04+09.86 (DC)\,$\dagger$ \\ 
    G217.56$-$00.34 & 1.0 & $5.2 \pm 0.4$ & $23.0 \pm 1.7$ & $0.56 \pm 0.08$ & 14.5 & $-12 \pm 11$ & $60 \pm 3$ & $37.6 \pm 2.3$ & $18.6 \pm 0.6$ & $1.44 \pm 0.07$ & 0.91 & LBN 1012, LDN 1650 \,\textsuperscript{G218.05$-$00.38}\, (H\textsc{ii}, DC) \\ 
    G219.18$-$08.93 & 1.0 & $2.5 \pm 0.2$ & $19.5 \pm 1.5$ & $0.61 \pm 0.06$ & 12.5 & $7 \pm 4$ & $0 \pm 1$ & $17.3 \pm 1.1$ & $17.6 \pm 0.6$ & $1.63 \pm 0.08$ & 0.54 & DB 5021, LBN 1015 (DC, RNe) \\ 
    G225.91$-$00.44 & 1.0 & $5.2 \pm 0.5$ & $20.7 \pm 2.3$ & $0.69 \pm 0.11$ & 9.8 & $-4 \pm 8$ & $47 \pm 4$ & $37.4 \pm 1.7$ & $16.3 \pm 0.4$ & $1.60 \pm 0.06$ & 0.39 & DB 5087/90/91/92... (DC) \\ 
    G231.83$-$02.00 & 1.0 & $4.0 \pm 0.3$ & $17.8 \pm 0.7$ & $0.47 \pm 0.06$ & 12.1 & $6 \pm 7$ & $14 \pm 2$ & $27.2 \pm 2.0$ & $16.1 \pm 0.6$ & $1.65 \pm 0.10$ & 1.4 & [K60] 201, TGU H1593, DB 5103 (DC) \\ 
    G234.20$-$00.20 & 1.0 & $8.2 \pm 0.7$ & $22.3 \pm 1.4$ & $0.68 \pm 0.11$ & 12.4 & $-3 \pm 18$ & $100 \pm 7$ & $34.0 \pm 3.0$ & $19.2 \pm 0.9$ & $1.51 \pm 0.12$ & 1.0 & S306, RCW 10 (H\textsc{ii}) \\ 
    G235.62+38.27 & 1.5 & $1.4 \pm 0.4$ & $18.4 \pm 3.5$ & $0.58 \pm 0.15$ & 3.7 & $2 \pm 2$ & $1 \pm 3$ & $2.8 \pm 0.2$ & $19.0 \pm 0.7$ & $1.72 \pm 0.10$ & 1.7 & PGCC G235.60+38.28 (MC)\,$\dagger$ \\ 
    G236.84$-$02.23 & 1.0 & $1.5 \pm 0.3$ & $25.1 \pm 3.3$ & $0.61 \pm 0.18$ & 5.6 & $-23 \pm 7$ & $57 \pm 3$ & $22.2 \pm 1.2$ & $17.0 \pm 0.5$ & $1.39 \pm 0.07$ & 1.0 & LBN 1057, DB 5114/15 (H\textsc{ii} + DC) \\ 
    G239.40$-$04.70 & 1.0 & $6.2 \pm 0.7$ & $23.1 \pm 2.0$ & $0.74 \pm 0.13$ & 9.3 & $-10 \pm 18$ & $83 \pm 8$ & $30.7 \pm 3.1$ & $18.2 \pm 0.9$ & $1.50 \pm 0.14$ & 1.2 & LDN 1667, DB 5132/33.., LBN 1059 (DC, H\textsc{ii}) \\ 
    G247.60$-$12.40 & 1.0 & $3.1 \pm 0.2$ & $20.3 \pm 1.5$ & $0.68 \pm 0.07$ & 18.9 & $-7 \pm 2$ & $24 \pm 1$ & $7.6 \pm 0.4$ & $20.0 \pm 0.6$ & $1.44 \pm 0.06$ & 0.99 & TGU H1630, DB 5147/48 (DC) \\ 
    G270.27+00.84 & 1.0 & $6.9 \pm 0.6$ & $23.5 \pm 2.2$ & $0.46 \pm 0.08$ & 11.6 & $-3 \pm 9$ & $111 \pm 6$ & $53.2 \pm 2.3$ & $19.3 \pm 0.5$ & $1.53 \pm 0.05$ & 0.92 & Vela A, RCW41 (MC, H\textsc{ii}) \\ 
    G311.55+00.02 & 1.2 & $109.7 \pm 13.7$ & $23.5 \pm 2.5$ & $0.43 \pm 0.09$ & 8.0 & $-16 \pm 30$ & $3083 \pm 90$ & $341.2 \pm 16.8$ & $21.2 \pm 0.5$ & $1.68 \pm 0.04$ & 1.3 & GAL311.89+00.10 \,\textsuperscript{G311.94$+$00.12}\, (H\textsc{ii} + DC/MC) \\ 
    G317.04+00.03 & 1.0 & $45.7 \pm 7.8$ & $24.2 \pm 3.5$ & $0.49 \pm 0.13$ & 5.9 & $-7 \pm 29$ & $2616 \pm 87$ & $220.6 \pm 12.6$ & $21.4 \pm 0.6$ & $1.67 \pm 0.05$ & 0.84 & GAL317.60-00.36, DB 6147/48/50/51 \,\textsuperscript{G317.51$-$00.11}\, (H\textsc{ii} + DC) \\ 
    G318.49$-$04.28 & 1.0 & $7.0 \pm 1.5$ & $21.7 \pm 4.9$ & $0.62 \pm 0.20$ & 4.6 & $-11 \pm 12$ & $155 \pm 16$ & $30.0 \pm 2.0$ & $19.5 \pm 0.7$ & $1.59 \pm 0.09$ & 0.54 & DB 6179/84/89/94, TGU H1978 (DC, RNe) \\ 
    G320.37$-$00.27 & 1.1 & $43.4 \pm 6.0$ & $24.5 \pm 2.7$ & $0.43 \pm 0.10$ & 7.2 & $-7 \pm 27$ & $1650 \pm 56$ & $157.2 \pm 8.8$ & $21.9 \pm 0.7$ & $1.63 \pm 0.05$ & 1.4 & GAL320.23-00.29 \,\textsuperscript{G320.27$-$00.27}\, (H\textsc{ii} + DC) \\ 
    G328.16+00.12 & 1.5 & $164.5 \pm 19.0$ & $24.5 \pm 2.1$ & $0.40 \pm 0.07$ & 8.7 & $-15 \pm 30$ & $3201 \pm 92$ & $410.1 \pm 18.4$ & $20.9 \pm 0.5$ & $1.73 \pm 0.04$ & 1.7 & IRAS 15502-5302 (H\textsc{ii} + MC/DC) \\ 
    G331.34+00.00 & 1.0 & $77.3 \pm 10.0$ & $25.2 \pm 2.4$ & $0.43 \pm 0.09$ & 7.7 & $-6 \pm 29$ & $3862 \pm 115$ & $380.7 \pm 19.4$ & $21.9 \pm 0.6$ & $1.73 \pm 0.04$ & 1.3 & IRAS 16076-5134 (H\textsc{ii} + many DC/MC) \\ 
    G333.24$-$00.12 & 1.0 & $102.5 \pm 20.7$ & $25.8 \pm 4.1$ & $0.61 \pm 0.19$ & 5.0 & $7 \pm 30$ & $6137 \pm 268$ & $336.2 \pm 20.2$ & $23.6 \pm 0.8$ & $1.70 \pm 0.05$ & 1.3 & RCW 106 / G333 Giant Molecular Complex (MC + H\textsc{ii}) \\ 
    G336.90+00.00 & 1.0 & $63.9 \pm 14.8$ & $25.2 \pm 4.3$ & $0.39 \pm 0.13$ & 4.3 & $-18 \pm 30$ & $4624 \pm 140$ & $334.1 \pm 19.7$ & $22.0 \pm 0.7$ & $1.84 \pm 0.05$ & 2.0 & IRAS 16313-4729 (H\textsc{ii} + DC, Kes39 1.5 Jy @ 1 GHz) \\ 
    G339.92+44.26 & 1.4 & $1.0 \pm 0.4$ & $21.2 \pm 5.3$ & $0.52 \pm 0.23$ & 2.6 & $-4 \pm 3$ & $1 \pm 3$ & $1.3 \pm 0.2$ & $21.6 \pm 2.3$ & $1.31 \pm 0.24$ & 0.99 & IRAS F14329-1055 (FIR)\,$\dagger$ \\ 
    G340.71$-$00.10 & 1.0 & $66.9 \pm 13.9$ & $21.4 \pm 5.1$ & $0.53 \pm 0.17$ & 4.8 & $5 \pm 29$ & $2012 \pm 99$ & $288.3 \pm 17.1$ & $21.1 \pm 0.6$ & $1.78 \pm 0.05$ & 1.1 & GAL 340.79-00.10 (H\textsc{ii} + many DC/MC) \\ 
    G342.51+08.81 & 1.0 & $3.8 \pm 0.8$ & $20.3 \pm 4.5$ & $0.57 \pm 0.17$ & 4.7 & $4 \pm 5$ & $32 \pm 7$ & $17.9 \pm 1.0$ & $16.9 \pm 0.5$ & $1.69 \pm 0.08$ & 1.2 & Lupus 5 (MC) \\ 
    G344.75+23.97 & 1.0 & $1.6 \pm 0.2$ & $28.9 \pm 1.8$ & $0.53 \pm 0.07$ & 9.3 & $-3 \pm 3$ & $14 \pm 2$ & $3.2 \pm 0.4$ & $22.4 \pm 1.7$ & $1.60 \pm 0.17$ & 1.1 & MBM 121/22 (MC) \\ 
    G348.73$-$00.55 & 1.3 & $50.9 \pm 15.7$ & $28.1 \pm 3.4$ & $0.45 \pm 0.15$ & 3.3 & $7 \pm 30$ & $2149 \pm 651$ & $174.8 \pm 10.5$ & $22.4 \pm 0.7$ & $1.73 \pm 0.06$ & 1.4 & RCW122 \,\textsuperscript{G348.73$-$00.75}\, (H\textsc{ii} + DC)\,$\dagger$ \\ 
    G351.31+17.28 & 1.0 & $7.0 \pm 0.7$ & $27.3 \pm 1.2$ & $0.44 \pm 0.06$ & 9.3 & $4 \pm 11$ & $245 \pm 9$ & $24.3 \pm 1.9$ & $24.7 \pm 1.2$ & $1.60 \pm 0.09$ & 0.93 & S9 or Gum 65, LBN1104/5/6, C130 (H\textsc{ii}, RNe) \\ 
    G351.65$-$01.23 & 1.0 & $51.1 \pm 19.7$ & $18.0 \pm 6.3$ & $0.55 \pm 0.18$ & 2.6 & $-5 \pm 29$ & $1833 \pm 89$ & $233.7 \pm 15.2$ & $21.1 \pm 0.7$ & $1.77 \pm 0.06$ & 0.98 & IRAS 17258-3637, DB 7138/39/44/56 (H\textsc{ii}, DC + MC) \\ 
    G353.05+16.90 & 1.0 & $22.6 \pm 0.8$ & $28.2 \pm 0.8$ & $0.53 \pm 0.02$ & 28.1 & $41 \pm 21$ & $20 \pm 5$ & $60.9 \pm 4.3$ & $23.4 \pm 1.0$ & $1.55 \pm 0.07$ & 1.2 & $\rho$ Ophiuchi (MC) \\ 
    G353.97+15.79 & 1.0 & $16.4 \pm 2.4$ & $24.2 \pm 1.2$ & $0.49 \pm 0.09$ & 6.7 & $42 \pm 23$ & $-7 \pm 33$ & $61.2 \pm 5.7$ & $19.8 \pm 1.0$ & $1.62 \pm 0.11$ & 0.59 & $\rho$ Ophiuchi East (DC) \\ 
    G355.44+00.11 & 1.0 & $51.4 \pm 12.8$ & $22.0 \pm 4.7$ & $0.52 \pm 0.18$ & 4.0 & $9 \pm 31$ & $1957 \pm 246$ & $318.5 \pm 18.2$ & $20.6 \pm 0.6$ & $1.78 \pm 0.05$ & 1.3 & RCW 132 (H\textsc{ii} + DC + SNR 3 Jy @ 1 GHz)\,$\dagger$ \\ 
    G355.96+20.53 & 1.0 & $7.8 \pm 0.4$ & $24.7 \pm 0.6$ & $0.49 \pm 0.03$ & 17.3 & $27 \pm 6$ & $7 \pm 5$ & $19.0 \pm 1.5$ & $17.3 \pm 0.7$ & $1.79 \pm 0.12$ & 0.75 & LDN 1719, MBM 128, LBN 1115 \,\textsuperscript{G355.63$+$20.52}\, (DC, H\textsc{ii}) \\ 
    G359.14+36.61 & 1.0 & $0.9 \pm 0.3$ & $18.2 \pm 3.4$ & $0.64 \pm 0.17$ & 3.5 & $1 \pm 2$ & $2 \pm 3$ & $3.2 \pm 0.3$ & $16.5 \pm 0.8$ & $1.73 \pm 0.14$ & 1.1 & LDN 1778/80, TGU H2202, LBN 1122 (MC, H\textsc{ii}) \\ 
    G359.43+43.98 & 1.1 & $0.5 \pm 0.2$ & $18.6 \pm 5.3$ & $0.58 \pm 0.22$ & 2.4 & $1 \pm 2$ & $-1 \pm 2$ & $1.7 \pm 0.3$ & $16.8 \pm 1.4$ & $1.74 \pm 0.25$ & 1.2 & IRAS 15179-0146 (FIR)\,$\dagger$ \\ 
	\end{longtable}
\end{landscape}
\twocolumn